# Benchmarking $CO_2$ Storage Simulations: Results from the 11th Society of Petroleum Engineers Comparative Solution Project

**Authors**:

Jan M. Nordbotten[1], Martin A. Fernø[2], Bernd Flemisch[3], Anthony R. Kovscek[4], Knut-Andreas Lie[5], Jakub W. Both[6], Olav Møyner[7], Tor Harald Sandve[8], Etienne Ahusborde[9], Sebastian Bauer[10], Zhangxing Chen[11], Holger Class[12], Chaojie Di[13], Didier Ding[14], David Element[15], Abbas Firoozabadi[16], Eric Flauraud[17], Jacques Franc[18], Firdovsi Gasanzade[19], Yousef Ghomian[20], Marie Ann Giddins[21], Christopher Green[22], Bruno R. B. Fernandes[23], George Hadjisotiriou[24], Glenn Hammond[25], Hai Huang[26], Dickson Kachuma[27], Michel Kern[28], Timo Koch[29], Prasanna Krishnamurthy[30], Kjetil Olsen Lye[31], David Landa-Marbán[32], Michael Nole[33], Paolo Orsini[34], Nicolas Ruby[35], Pablo Salinas[36], Mohammad Sayyafzadeh[37], Jakub Solovský[38], Jakob Torben[39], Adam Turner[40], Denis V. Voskov[41], Kai Wendel[42], AbdAllah A. Youssef[43]

[1] VISTA Center for Modeling of Coupled Subsurface Dynamics, Department of Mathematics, University of Bergen; NORCE Norwegian Research Center, Bergen. jan.nordbotten@uib.no
[2] Department of Physics and Technology, University of Bergen; Norwegian Research Center (NORCE), Bergen. martin.ferno@uib.no
[3] Institute for Modelling Hydraulic and Environmental Systems, University of Stuttgart. bernd.flemisch@iws.uni-stuttgart.de
[4] Energy Science & Engineering Department, Stanford University. kovscek@stanford.edu
[5] SINTEF Digital, Mathematics & Cybernetics, Oslo. knut-andreas.lie@sintef.no
[6] VISTA Center for Modeling of Coupled Subsurface Dynamics, Department of Mathematics, University of Bergen. jakub.both@uib.no
[7] SINTEF Digital, Mathematics & Cybernetics, Oslo. olav.moyner@sintef.no
[8] NORCE Norwegian Research Center, Bergen. tosa@norceresearch.no
[9] Universite de Pau et des Pays de l'Adour, E2S UPPA, CNRS, LMAP, Pau, France, etienne.ahusborde@univ-pau.fr
[10] Institute of Geosciences, Kiel University, Germany, sebastian.bauer@ifg.uni-kiel.de
[11] Department of Chemical and Petroleum Engineering, University of Calgary; Calgary, Canada; Ningbo Institute of Digital Twin, Eastern Institute of Technology; Ningbo, PR China, zhachen@ucalgary.ca
[12] Department of Hydromechanics and Modelling of Hydrosystems, Stuttgart University, Germany. holger.class@iws.uni-stuttgart.de
[13] Department of Chemical and Petroleum Engineering, University of Calgary; Calgary, Canada, chaojie.di@ucalgary.ca
[14] IFP Energies Nouvelles, France. (SPE 1605542) didier-yu.ding@ifpen.fr
[15] Tetra Tech RPS Energy, Poole, UK, david.element@tetratech.com
[16] Reservoir Engineering Research Institute and Rice University, abbas.firoozabadi@rice.edu
[17] IFP Energies Nouvelles, France. eric.flauraud@ifpen.fr
[18] Stanford University (USA), jacques.franc@univ-pau.fr
[19] Institute of Geosciences, Kiel University, Germany, firdovsi.gasanzade@ifg.uni-kiel.de
[20] CTC-Chevron Technical Center, yghomian@chevron.com
[21] SLB, UK. giddins@slb.com. SPE member 0798678, Orcid 0000-0001-7486-5013.
[22] CSIRO Energy, Australia, chris.green@csiro.au
[23] Center for Subsurface Energy and the Environment, The University of Texas at Austin, Austin, USA. brbfernandes@utexas.edu
[24] Geoscience & Engineering, Delft University of Technology, g.hadjisotiriou@tudelft.nl
[25] Earth Systems Science Division, Pacific Northwest National Laboratory, WA, USA, glenn.hammond@pnnl.gov
[26] Tetra Tech Research & Development, Lafayette, CA, USA, hai.huang@tetratech.com
[27] TotalEnergies E&P Research & Technology, USA, dick.kachuma@totalenergies.com
[28] Inria, Paris research Center, 48 rue Barrault, 75013 Paris, France; CERMICS, ENPC, 77455 Marne-la-Vallée, France, michel.kern@inria.fr
[29] Simula Research Laboratory, Oslo, Norway, timok@simula.no, ORCID 0000-0003-4776-5222
[30] ExxonMobil Technology and Engineering Company, Spring, USA. prasanna.g.krish@exxonmobil.com
[31] SINTEF Digital. kjetil.olsen.lye@sintef.no
[32] NORCE Norwegian Research Center, dmar@norceresearch.no
[33] ResFrac Corporation, michaelnole@resfrac.com
[34] OpenGoSim, 30 Nelson Street, Leicester, LE1 7BA, UK, paolo.orsini@opengosim.com
[35] CTC Chevron Technical Center, nicolas.ruby@chevron.com
[36] OpenGoSim, 30 Nelson Street, Leicester, LE1 7BA, UK, pablo.salinas@opengosim.com
[37] CSIRO Energy, Australia, mohammad.sayyafzadeh@csiro.au
[38] Reservoir Engineering Research Institute, jakub.solovsky@rerinst.org
[39] SINTEF Digital. jakob.torben@sintef.no
[40] Tetra Tech RPS Energy, Poole, UK, Adam.Turner@tetratech.com
[41] Geoscience & Engineering, Delft University of Technology; Energy Science & Engineering, Stanford, d.v.voskov@tudelft.nl
[42] Department of Hydromechanics and Modelling of Hydrosystems, Stuttgart University, Germany. kai.wendel@iws.uni-stuttgart.de
[43] Centre for Integrative Petroleum Research, King Fahd University of Petroleum & Minerals, Dhahran, Saudi Arabia. abdallah_youssef@ymail.com. SPE ID: 5373688

# Abstract

The 11th Society of Petroleum Engineers Comparative Solution Project (shortened SPE11 herein) benchmarked simulation tools for geological carbon dioxide ($CO_2$) storage. A total of 45 groups from leading research institutions and industry across the globe signed up to participate, with 18 ultimately contributing valid results that were included in the comparative study reported here.

This paper summarizes the SPE11. A comprehensive introduction and qualitative discussion of the submitted data are provided, together with an overview of online resources for accessing the full depth of data. A global metric for analyzing the relative distance between submissions is proposed and used to conduct a quantitative analysis of the submissions. This analysis attempts to statistically resolve the key aspects influencing the variability between submissions.

The study shows that the major qualitative variation between the submitted results is related to thermal effects, dissolution-driven convective mixing, and resolution of facies discontinuities. Moreover, a strong dependence on grid resolution is observed across all three versions of the SPE11. However, our quantitative analysis suggests that the observed variations are predominantly influenced by factors not documented in the technical responses provided by the participants. We therefore identify that unreported variations due to human choices within the process of setting up, conducting, and reporting on the simulations underlying each SPE11 submission are at least as impactful as the computational choices reported.

# Introduction

The 11th Society of Petroleum Engineers Comparative Solution Project (shortened SPE11 herein) benchmarked simulation tools for geological carbon dioxide ($CO_2$) storage, announced in March 2023 with submissions due in September 2024. A total of 45 groups from leading research institutions and industry across the globe signed up to participate, with 18 ultimately contributing valid results that were included in the comparative study reported here. This extensive international collaboration underscores the significance of the project as a critical step toward global advancements in $CO_2$ storage.

The processes governing geological $CO_2$ storage (Nordbotten and Celia 2011) share many similarities with those governing the extraction of subsurface energy resources such as oil and gas (Lake et al. 2014). In particular, the general geological context of a relatively deep (~1000m) and permeable host formation, overlain by a significantly less permeable caprock, is common. Moreover, the ubiquitous presence of brine in the subsurface implies that a multiphase problem of brine in addition to $CO_2$/oil/gas must always be considered. These similarities, alongside over 60 years of experience in simulating subsurface petroleum recovery, provide a strong foundation for developing $CO_2$ storage simulation tools. Many of the same modeling techniques—such as Darcy's law for multiphase flow, relative permeability functions, capillary pressure models, and numerical discretizations and solution strategies—are directly transferable to $CO_2$ storage. As a result, existing petroleum reservoir simulation frameworks may be adaptable with relatively few modifications for $CO_2$ storage applications.

On the other hand, $CO_2$ storage also presents several unique challenges that are not fully addressed by traditional petroleum industry practices (Celia et al. 2015; Ajayi et al. 2019). These include, but are not limited to, the following factors:

a) Well placement: $CO_2$ storage wells are typically placed deep in the storage formation, below the expected gas–water contact, to ensure efficient plume migration and minimize risk of upward leakage. In contrast, petroleum production wells are often located in petroleum-rich zones closer to structural highs to maximize recovery.
b) Time scales: While the active injection period of $CO_2$ storage is comparable to the operational phase of a petroleum reservoir, $CO_2$ storage also requires a more critical assessment of the post-operational phase (extending over 100 years in some jurisdictions) to ensure long-term storage security and reservoir behavior, including $CO_2$ trapping mechanisms and pressure dissipation.
c) Balance of physical processes: With the injected $CO_2$ initially out of equilibrium with the reservoir brine, processes such as dissolution, dispersion, and eventually convective mixing play a heightened role in $CO_2$ storage compared to petroleum recovery, where viscous forces often dominate.
d) Information scarcity: Petroleum production is economically profitable, and hydrocarbon reservoirs are typically characterized in detail through seismic surveys, core sampling, and well logging. In contrast, $CO_2$ storage is a cost-driven waste disposal business with small profit margins in which limited geological data availability and uncertainty can impact simulation accuracy and risk assessment.

The SPE11 project was designed to assess the applicability of state-of-the-art reservoir simulation software for $CO_2$ storage, with the aim of advancing the modeling capabilities needed for effective and safe geological carbon storage. This study builds on the longstanding traditions of comparative solution projects conducted by SPE (for a summary of SPE1 to SPE10, see Islam and Sepehrnoori, 2013) and complements previous benchmark efforts within the $CO_2$ storage research community (see, e.g., Pruess et al. 2004; Class et al. 2009; Nordbotten et al. 2012; Flemisch et al. 2024). These earlier projects, also referred to as intercomparison or benchmark studies, demonstrated that the process of stating a common reference case, even when no reference solution is defined, and carefully analyzing the proposed solutions, offers great benefits to participants as well as readers of the literature. Notably, consensus among submitted solutions provides confidence in the underlying numerical models and simulation tools, while discrepancies and insights gained from comparative assessments help identify key challenges and inform further research, methodological improvements, and software development. Moreover, establishing well-documented and common simulation cases has created baselines against which new research innovations can be discussed, extending far beyond the scope of the original studies.

The SPE11 problem definition is presented in detail elsewhere (Nordbotten et al. 2024a). This paper focuses on the outcomes of the study, starting with a brief overview of the main aspects of the SPE11 problem set. We then introduce the submitted solutions, describe how they are qualitatively compared, and present the database constructed to host them. This is followed by a comparative assessment, highlighting key learnings from the study. Finally, we present conclusions and recommendations.

# Overview of SPE11

Comprehensive details about SPE11, including problem definitions and links to collaborative resources, as well as a graphical presentation and full access to all submitted results, can be found on the official SPE CSP website (www.spe.org/csp/spe11). This section presents an overview of the process followed in the study and a brief overview of the three subcases and their common and distinguishing features.

The lessons learned from three earlier comparative solution studies (Class et al. 2009; Nordbotten et al. 2012; Flemisch et al. 2024) shaped key elements of the SPE11 process, including the following:

1. Problem definition and quality control:
    - Precise definitions of the three cases that comprise SPE11 to minimize variability in results due to subjective interpretation.
    - Early-access testing with four selected likely participants to quality-control the initial drafts of the problem descriptions.
2. Community participation:
    - An open call for participation, announced at the *2023 SPE Reservoir Simulation Conference*, including a discussion session, followed by a feedback and correction period open to all prospective participants.
    - A renewed announcement with updated details at the *2023 SPE Annual Technical Conference and Exhibition*, including a discussion session.

- Publication of the final problem description as an open-access paper in *SPE Journal* (Nordbotten et al. 2024a).
- A special issue of *SPE Journal* devoted to more in-depth technical discussions.

3. Collaborative resources:
   - A dedicated online discussion forum at *SPE Connect* to facilitate community interactions.
   - A *GitHub* repository (github.com/Simulation-Benchmarks/11thSPE-CSP/) offering pre-processing tools, sample input decks, post-processing tools, and quality-control utilities—supported by regular development meetings to assist participants, foster collaboration, and improve transparency.
4. Workshops and deadlines:
   - Two pre-submission checkpoints with submission deadlines and intercomparison workshops to encourage early sharing of challenges and learnings, resolve ambiguities, and consolidate results.
   - A post-submission workshop dedicated to reducing mistakes in data processing and reporting.

The timeline of the full SPE11 project, illustrating key milestones, deadlines, and community engagement points, is shown in Figure 1.

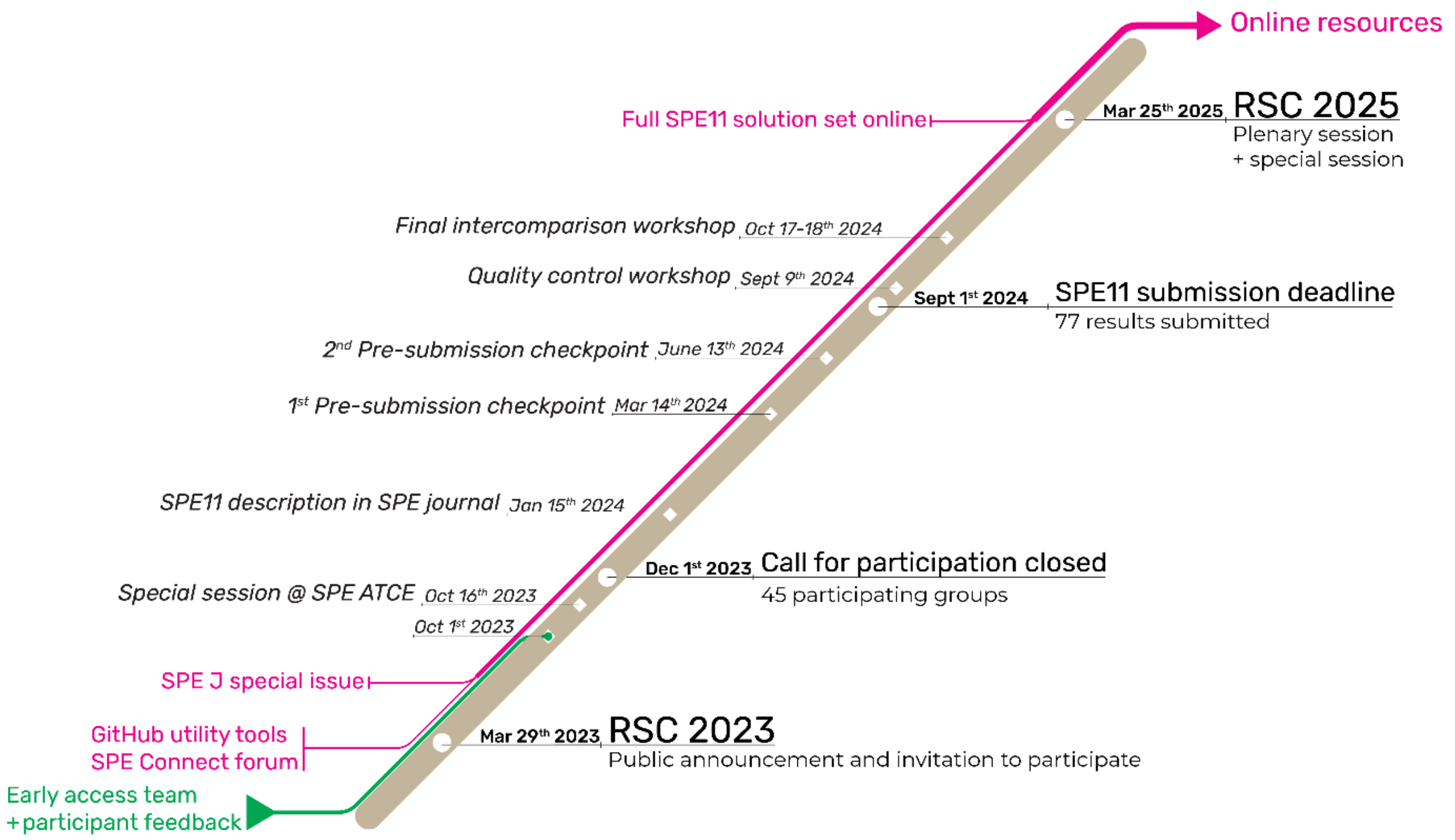

**Figure 1**: Timeline of the SPE11 comparative solution project.

The SPE11 problem definition itself comprises three subcases representing a laboratory experimental setting (SPE11A), a vertical 2D field transect (SPE11B), and a 3D field model (SPE11C), as illustrated in Figure 2. These subcases share a common geometric framework that reduces the overhead for participants engaged in the full study. Differences in scaling and placement within pressure-temperature space influence how the various physical processes balance each other. Consequently, each subcase poses different computational challenges.

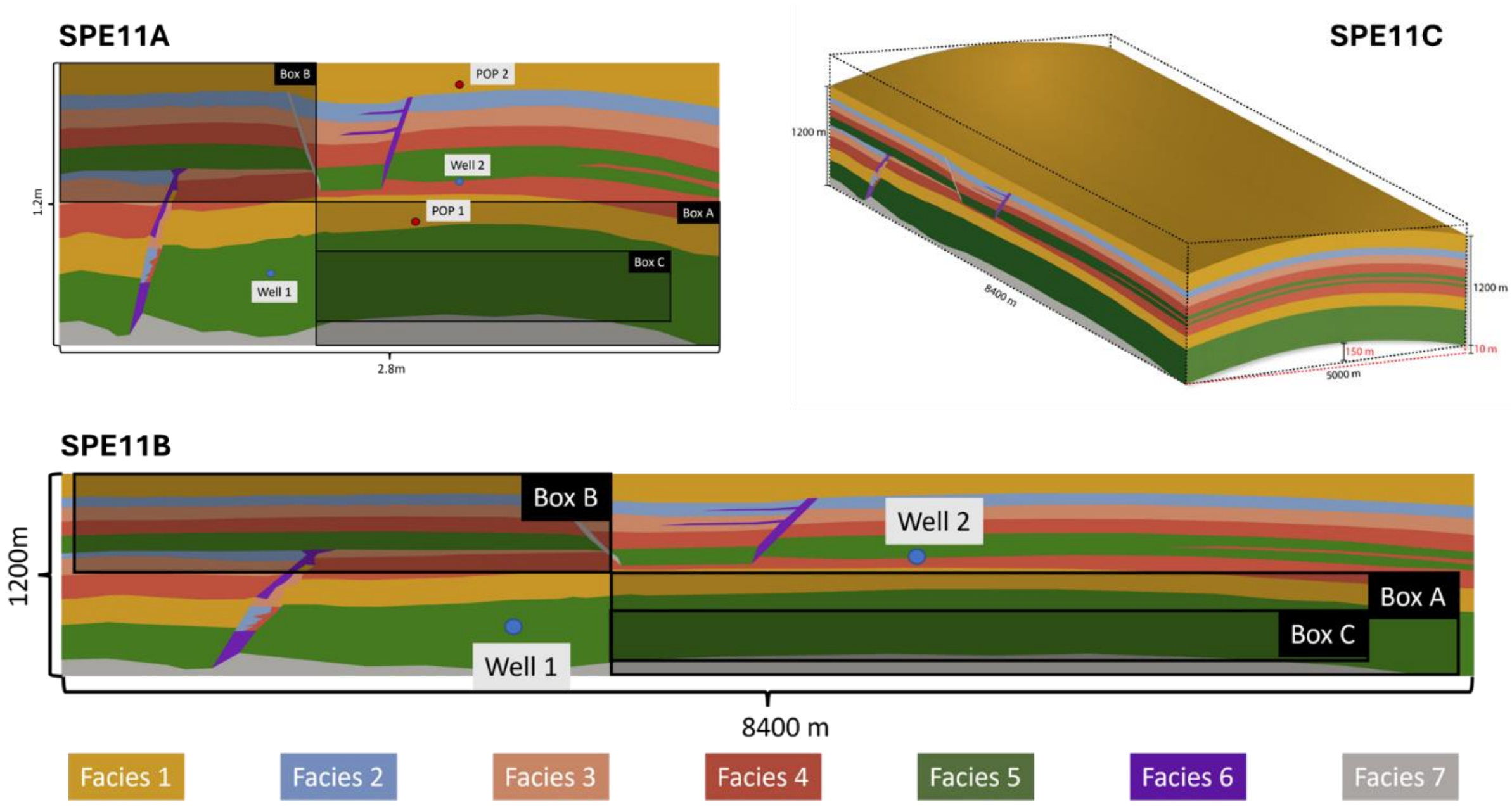

**Figure 2:** The SPE11 subcases are derived from the same geometry but are set respectively as a 2D experiment at the laboratory scale and atmospheric conditions (SPE11A), as a 2D transect at field scale and reservoir conditions (SPE11B), and as a synthetic 3D storage reservoir at field scale and reservoir conditions (SPE11C). Figure from Nordbotten et al. (2024a).

## Common features of all three SPE11 subcases

The official description includes a summary of the governing equations for thermal, two-phase, two-component flows in heterogeneous porous media (Nordbotten et al. 2024a). Moreover, constitutive laws such as capillary pressure, relative permeability and phase behavior are provided and are common for the three subcases. On the other hand, some processes are omitted, including geochemical reactions and geomechanics.

The geometry of the SPE11 subcases features consistent elements, including facies enumeration, well placements, pressure observation points (POP 1 and POP 2 in Figure 2), and designated volumes of interest for detailed reporting of simulation results (Boxes A, B, and C in Figure 2). The injection schedule is also standardized and prescribes constant-rate injection in Well 1 for the duration of the whole injection period, and constant-rate injection in Well 2 during only the second half of the injection period.

The structure of the reporting requirements is also similar across the subcases. The first thirteen “sparse data” types are reported as scalar quantities derived from the simulation with high resolution in time (summarized in Table 1). In contrast, the subsequent eight “dense data” types are field variables reported on a prescribed reporting grid at a low temporal resolution. These include pressure, gas saturation, mass fraction of $CO_2$ in the liquid phase, mass fraction of $H_2O$ in the $CO_2$-rich phase, phase mass density of the $CO_2$-rich phase, phase mass density of the water phase, total mass of $CO_2$, and temperature. To supplement the reporting data, participants were required to complete questionnaires detailing simulator choices, numerical discretization, solution strategies and solvers, possible deviations from the official description, etc. This additional information is essential for interpreting variations in the simulation results.

| | # | Type | Specification | Units |
|---|---|---|---|---|
| Sparse data | 1 | Pressure | At pressure observation point POP 1 | Pa |
| | 2 | | At pressure observation point POP 2 | |
| | 3 | Phase Composition Box A | Mobile gas-phase $CO_2$ | kg |
| | 4 | | Immobile gas-phase $CO_2$ | |
| | 5 | | $CO_2$ in liquid phase | |
| | 6 | | $CO_2$ in seal facies 1 | |
| | 7 | Phase Composition Box B | Mobile gas-phase $CO_2$ | kg |
| | 8 | | Immobile gas-phase $CO_2$ | |
| | 9 | | $CO_2$ in liquid phase | |
| | 10 | | $CO_2$ in seal facies 1 | |
| | 11 | Convective mixing in Box C | Evaluation of a specific integral | m |
| | 12 | $CO_2$ in seal | All $CO_2$ in facies 1 | kg |
| | 13 | Boundary $CO_2$ | All $CO_2$ in boundary volumes | kg |
| Dense data | 14 | Pressure | Pressure at center of reporting grid cells | Pa |
| | 15 | Saturation | Gas saturation in reporting grid cells | - |
| | 16 | Mass fraction | $CO_2$ in liquid phase in reporting grid cells | - |
| | 17 | | $H_2O$ in vapor phase in reporting grid cells | |
| | 18 | Density | Liquid phase density in reporting grid cells | kg/m3 |
| | 19 | | Vapor phase density in reporting grid cells | |
| | 20 | Mass | Mass of $CO_2$ in reporting grid cells | kg |
| | 21 | Temperature | Temperature at center of reporting-grid cells | C |

**Table 1**: Reporting quantities as defined in the SPE11 description (Nordbotten et al. 2024a). Reporting data number 13 and 21 are applicable to SPE11B and SPE11C only. At any given time, the sparse data are scalar quantities while the dense data are spatially varying quantities.

The description also requested the reporting of various performance data such as solver time, number of linear and nonlinear iterations, time steps, residuals, and error measures. These requirements, however, were treated as optional, and were only reported in about half of the submissions. For completeness, plots of the performance data are included in Appendix B as Figures B.1, B.2 and B.3.

## Particular features separating the three SPE11 subcases

The following features differentiate the three subcases. This description provides a concise overview. For an exhaustive description, please refer to Nordbotten et al. (2024a).

**SPE11A:** This subcase is motivated by the application of reservoir simulators to analyze laboratory-scale experiments (Fernø et al. 2024), specifically inspired by the recent *FluidFlower* validation study (Flemisch et al. 2024). It is characterized by the following features:

- **Domain and geometry:** The domain is 2.8 meters wide and 1.2 m high, and it is treated as a two-dimensional system with a nominal thickness of 1 cm used for converting to volumetric quantities.
- **Boundary conditions:** The top boundary is set to atmospheric pressure, while all other boundaries are defined as no-flow.

- **Temperature:** The system is assumed to be isothermal at 20 °C, consistent with the experimental setup.
- **Facies and permeability:** Facies properties are derived from unconsolidated sands, with a permeability of 4000 D in the main reservoir (facies 5) and 40 D in the seal (facies 1).
- **Initial conditions:** The initial condition is pure water at hydrostatic pressure.
- **Injection conditions:** The injection period lasts for 5 h, with a total simulation time of 120 h. The injection rate is set to 5 $cm^3$/min for each well.
- **Reporting schedule:** Sparse data are reported every 10 min, while dense data are recorded hourly. The reporting grid for the dense data is a 1 cm by 1 cm Cartesian grid, containing approximately 33,000 cells.

**SPE11B:** This subcase is motivated by subsea geological $CO_2$ storage and reflects conditions typical of the Norwegian Continental Shelf (Halland et al. 2013). A hypothetical two-dimensional transect is considered for this case. It is characterized by the following features:

- **Domain and geometry:** The domain is 8.4 km wide and 1.2 km high, treated as a two-dimensional system with a nominal thickness of 1 m used for conversion to volumetric quantities. The well placement and pressure observation points are the same as for SPE11A, relative to the geometry.
- **Boundary conditions:** The boundary conditions include a buffer volume to emulate embedding the domain into a larger aquifer structure. The top and bottom boundaries are set to constant temperature, consistent with a geothermal gradient of 25°C/km and a depth of approximately 2 km to the top of the reservoir.
- **Facies and permeability:** Facies properties are typical of a prime $CO_2$ storage location, with a permeability of 1 D in the main reservoir (facies 5) and 0.1 mD in the seal (facies 1).
- **Initial conditions:** The simulation is initialized with pure water at hydrostatic pressure and geothermal gradient pressure, set 1000 years before injection begins.
- **Injection conditions:** The injection period lasts for 50 y, with a total simulation time of 1000 years. The injection rate is approximately 1100 tons of $CO_2$ per year for each well.
- **Reporting schedule:** Sparse data are reported as 10 data points per year, while dense data are recorded every five years, both starting at the time of injection. The reporting grid for dense data is a 10 m by 10 m Cartesian grid, with a total of approximately 100,000 cells.

**SPE11C**: This subcase is a three-dimensional extension obtained by extruding the geometry of SPE11B along a parabolic shape. It is characterized by the following features:

- **Domain and geometry:** The areal footprint of the domain is 8.4 km by 5 km, with a uniform vertical thickness of 1.2 km. The deformation follows a parabolic shape, with the maximum elevation occurring near the center at approximately 150 m. To accommodate the 3D geometry, Well 1 was designated as a straight well, while Well 2 was designated to follow the parabolic extrusion of the geometry. Each well is placed in the central 3 km of the domain.
- **Boundary conditions:** As for SPE11B, the boundary conditions implement a buffer volume to mimic the embedding of the domain into a larger aquifer structure. The top and bottom boundaries are set to constant temperature, consistent with a geothermal gradient of 25°C/km and a reservoir depth of approximately 2 km.

- **Facies and permeability:** Same as in SPE11B, but with minor adjustments for the three-dimensional extension of the domain.
- **Initial conditions:** Same as in SPE11B.
- **Injection conditions:** Same as in SPE11B, except the injection rate is approximately 1.5 megatons $CO_2$ per year for each well.
- **Reporting schedule:** Sparse data are reported as 10 data points per year, while dense data are collected at selected time intervals. The reporting grid for dense data is a 50 m by 50 m by 10 m Cartesian grid, containing approximately 2 million cells.

# Qualitative overview of submitted solutions

In a comparative solution project like SPE11, where no benchmark solutions exist, ranking the submitted solutions by correctness is neither practical nor meaningful. In this section, we will therefore provide a descriptive introduction to the submitted solutions and a qualitative discussion of the most visually notable differences. A quantitative analysis is deferred to the next section.

## Overview of submitted solutions

A total of 45 groups signed the participation agreement for the SPE11. Of these, 18 groups submitted valid data of sufficient quality for inclusion in the comparative study; these groups will be referred to hereafter as the "participants" of SPE11. An additional two groups submitted data that were deemed unphysical and were subsequently excluded from the study.

Each participant was allowed to submit up to four sets of results, and partial submissions—including only one or two of the subcases—were also accepted. As a result, 13 participants submitted a total of 21 results to SPE11A, all 18 participants submitted a total of 34 results to SPE11B, while 14 participants submitted a total of 22 results to SPE11C. Table 2 provides an overview of the participating groups.

| Participants | | | # of submitted results | | |
|---|---|---|---|---|---|
| *Main institution(s)* | *Short name* | *Simulator name(s)* | *SPE11A* | *SPE11B* | *SPE11C* |
| University of Calgary | Calgary | PRSI-CGCS | 1 | 1 | 1 |
| University of Kiel | CAU-Kiel | OPM Flow 2024.04 | 1 | 1 | 1 |
| Commonwealth Scientific and Industrial Research Organisation | CSIRO | MOOSE | 2 | 1 | 1 |
| Chevron | CTC-CNE | SLB IX 2023.4 | 1 | 1 | 1 |
| TU Delft | DARTS | openDARTS 1.1.4 | - | 1 | - |
| LLNL, Stanford, TotalEnergies, Chevron | GEOS | GEOS v1.0.1 | 2 | 2 | 2 |
| IFP Energies nouvelles | IFPEN | Geoxim | 1 | 2 | 1 |
| King Fahd University of Petroleum and Minerals | KFUPM | Eclipse-300, 2024 | - | 1 | - |
| OpenGoSim | OpenGoSim | PFLOTRAN-OGS 1.8 | 1 | 3 | 2 |
| NORCE, SINTEF | OPM | OPM Flow 2024.10 | 4 | 4 | 4 |
| University of Pau | Pau-Inria | DuMux 3.8 | 1 | 1 | 1 |
| Pacific Northwest National Laboratory | PFLOTRAN | PFLOTRAN | 1 | 1 | 1 |
| Rice University | Rice | Higher-Order Reservoir Simulation Engine | - | 2 | - |
| SINTEF | SINTEF | JutulDarcy.jl v 0.2.28 | - | 4 | 3 |
| SLB | SLB | SLB IX 2024.2 and Eclipse Comp. 2024.2 | 2 | 1 | 1 |
| University of Stuttgart | Stuttgart | DuMux 3.9 | - | 4 | - |
| TetraTech | TetraTech | STOMP and tNavigator 24.2 | 1 | 2 | 2 |
| University of Texas at Austin | UT-CSEE | SLB IX 2023.1 and CMG GEM 2023.30 | 3 | 2 | 1 |
| | | **Total results per case** | **21** | **34** | **22** |

**Table 2**: Overview of participating groups and submissions. For detailed descriptions, see Appendix A.

As illustrated in Figure 3, the total amount of data submitted is on the order of 500 GB, precluding any substantial manual processing. In anticipation of this, the SPE11 description contains strict formatting guidelines. In addition, scripts for automatic prechecks of the formatting were made available to the participants. To ensure the highest possible quality of the submitted data, two intermediate submission deadlines, each with corresponding workshops, were offered to participants, as outlined in the timeline of Figure 1. Nevertheless, some errors in such a large dataset are unavoidable and ultimately require a combination of manual and automatic correction (Nordbotten, 1963; De Waal et al, 2011). The organizers, in collaboration with the participants, have strived to eliminate clerical errors, such as issues with units, post-processing errors, reporting misunderstandings, and so forth. This process was limited by the following constraints, in fairness to all participants: New simulation results were not accepted after September 20th, 2024; and the data correction phase of the SPE11 was concluded by January 1st 2025, thus no results were updated after this date.

Despite the efforts to minimize and correct errors, some errors undoubtedly will have escaped our attention. Indeed, in the process of analyzing the submissions, some errors were identified after the cutoff date for finalizing the data correction, yet before the completion of this report. These are reported where relevant in the text below. Moreover, some outliers may be reasonably suspected of being the result of errors, but have not been possible to ascertain within the scope of the study. We are therefore faced with a persistent challenge within the analysis of complex systems, wherein we must acknowledge the presence of known, suspected, and unknown errors, several of which are more an expression of the users, rather than the simulation code (for an extended discussion, see Roache, 1998).

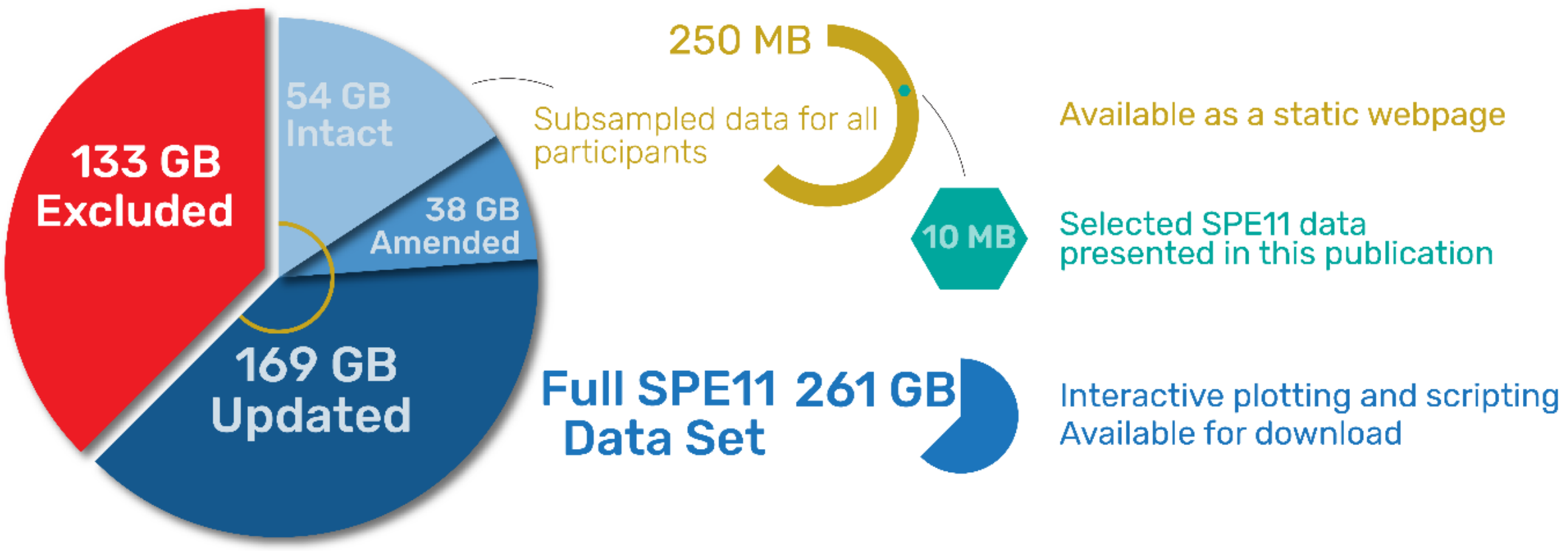

**Figure 3**: Overview of submitted data, and data availability.

We give some impression of the error correction efforts as follows: Of the data submitted by the participants by the original deadline, the two excluded groups and one withdrawn SPE11C result accounted for 9% of the total submissions (see Table 3). Additionally, 24% of the data was excluded because it was reported at time steps not requested in SPE11C. Approximately 43% was updated by the participants themselves by the corrections deadline, accounting for 65% of the remaining data. Following the start of the detailed analysis, an additional 10% required corrections by the organizers, based on bilateral exchanges with the respective participants, before being included in the results published openly and analyzed herein. In addition to the above, the organizers discovered and remedied two instances of data being corrupted during data transfer. In a final step, the format of all file headers, column delimiters, and reported numbers was harmonized.

| | SPE11A | | | SPE11B | | | SPE11C | | | Total | |
|---|---|---|---|---|---|---|---|---|---|---|---|
| | sparse | dense | GB | sparse | dense | GB | sparse | dense | GB | GB | % |
| ***Submitted*** | ***21*** | ***21*** | ***10*** | ***36*** | ***36*** | ***87*** | ***23*** | ***23*** | ***297*** | ***394*** | ***100*** |
| Excluded<br>- *Unphysical*<br>- *Not requested* | | | | 2<br>*2* | 2<br>*2* | 5<br>*5* | 1<br>*1* | 4<br>*1*<br>*3* | 128<br>*32*<br>*96* | 133<br>*37*<br>*96* | 34<br>*9*<br>*24* |
| Updated | 13 | 12 | 6 | 21 | 23 | 55 | 13 | 14 | 108 | 169 | 43 |
| Amended | 3 | 2 | 1 | 12 | 6 | 14 | 1 | 3 | 23 | 38 | 10 |

**Table 3**: Numbers and sizes of submitted, excluded, updated, and amended results.

While these figures indicate the overall precision of submissions, they also obscure some systematic deficiencies. Notably, not all groups reported all sparse quantities #6, 10 and 11, and for these groups, these quantities were estimated based on the submitted dense data. Similarly, a few groups used an alternative notion of "immobile gas-phase $CO_2$," and for these groups, sparse data #3, 4, 7, and 8 were estimated based on the dense data (see Appendix C for details). Substantial deviations were observed in the submitted questionnaires, in terms of missing or imprecise information, necessitating considerable manual corrections.

The following subsections introduce the results submitted to the three subcases, starting with SPE11B, for which all participants submitted results. For each subcase, we have designated a "median submission." This submission can be considered as the most representative of all submissions, in the sense of being the least outlier. The precise definition and identification of these median submissions are deferred to the section on quantitative comparisons.

Given the extensive nature of the submitted data, it is not possible to present fully the results within the format of a journal paper. Therefore, as illustrated in Figure 3, three online portals have been created to provide access to the data. First, the "SPE11-at-a-glance" webpage (www.spe.org/csp/spe11) offers comprehensive visualizations of the submitted data from all participating groups. Secondly, a full SPE11 data repository enables full dataset download (doi.org/10.18419/DARUS-4750). Finally, an online repository of plotting and analysis tools allows direct access to the full dataset without the need for downloading (github.com/Simulation-Benchmarks/11thSPE-CSP). Full details about these three online portals are provided at the end of this section.

## An introduction to the SPE11B solutions

The SPE11 submissions format included questionnaires for self-reporting various numerical, computational and solver-related aspects of the underlying simulations. These questionnaires have formed the basis for an informal "taxonomy" of the SPE11 submission for each of the three subcases. In Figure 4, this taxonomy is used to illustrate some of the main qualitative differences and similarities among the 34 individual submissions from the 18 participating groups for SPE11B. Notably, nearly half of the submissions (indicated by a blue background) belong to the main branch of the tree, which consists of standard finite-volume simulations on the reporting grid. Fourteen different groups contributed 16 submissions in this category, employing 13 different simulators, with DuMuX and SLB IX being the only simulators used in more than one submission. These 16 submissions are collectively interpreted as the baseline submissions for the SPE11B. We remark that while we treat all these as "baseline" simulations, being methodologically similar, they do have distinct variability in setup, which to some extent was reported in the questionnaires. The baseline simulations contain simulations based both on a pseudo black-oil formulation as well as a fully compositional formulation. Some of the simulations include an explicit representation of physical dispersion, while others do not.

The most common variation from the baseline submissions was altering the computational grid. Seven submissions used refined Cartesian grids while retaining the standard simulation methodology (marked in green in Figure 4). Additionally, five submissions employed non-orthogonal grids to more accurately represent the geological structure (marked in orange). In two of these (by SINTEF), the standard two-point discretization scheme was replaced with a consistent multipoint method to reduce the effect of inconsistency errors.

Finally, three groups provided submissions that are methodologically different from the others (indicated by a white background in Figure 4). Specifically, Rice submitted two simulations employing high-order numerical methods, OPM3 introduced a model for upscaled dissolution due to convective mixing, while Stuttgart2 to Stuttgart4 explored the trade-off between accuracy and reduced computational cost by using a simplified immiscible model during the first part of the simulation period. These six submissions provide valuable context to the SPE11B data set and indicate the impact of fundamental design choices, such as high-order numerical methods, upscaling strategies, or simplified physical representations.

Further details of all submissions are provided in Appendix A, while Figure 5 gives a visual impression of the variability between the simulations, in terms of the distribution of $CO_2$ at the end of the simulation period.

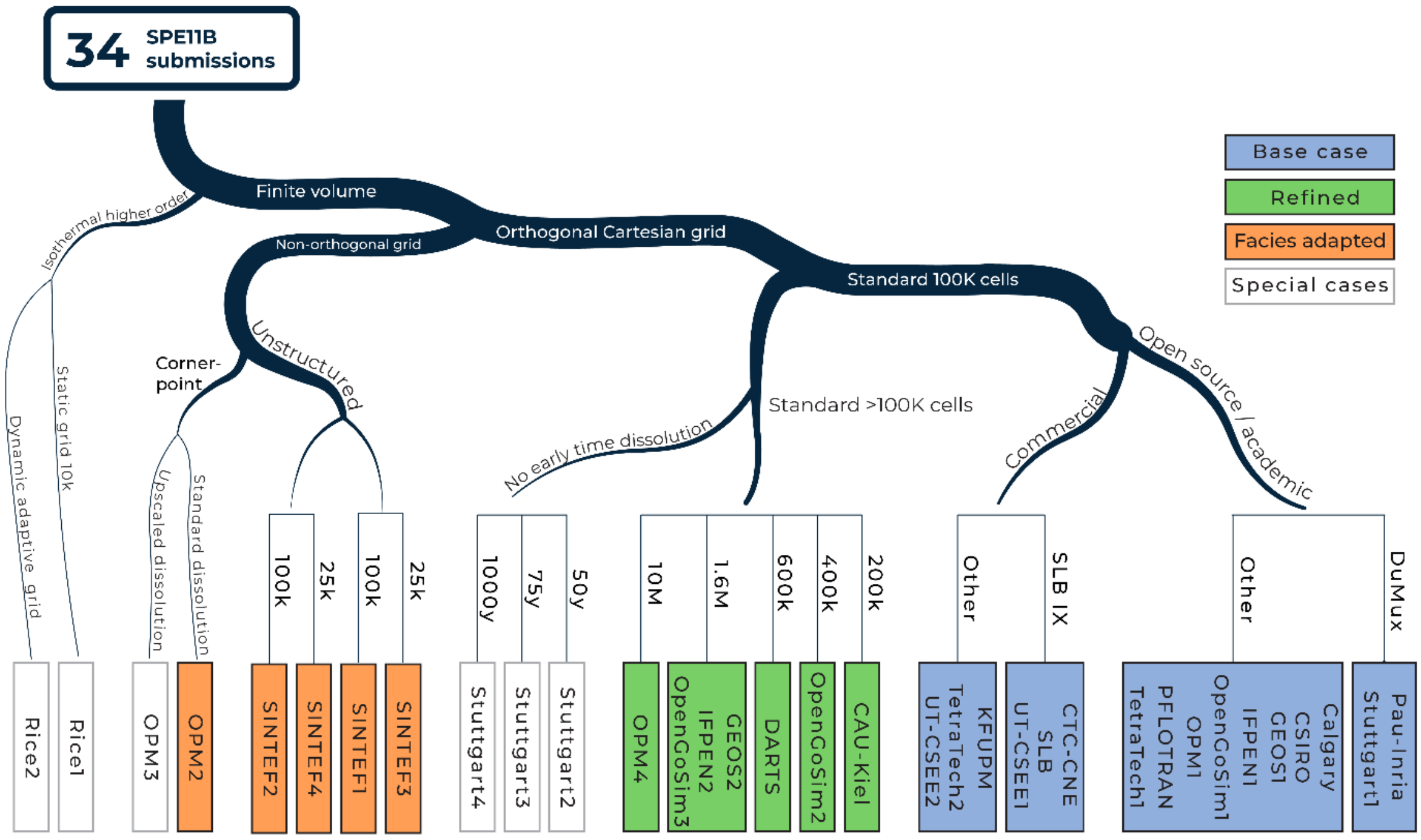

**Figure 4**: An overview of the SPE11B submissions. Curved branches indicate methodological choices, while angled brackets indicate grid resolution and choice of simulator.

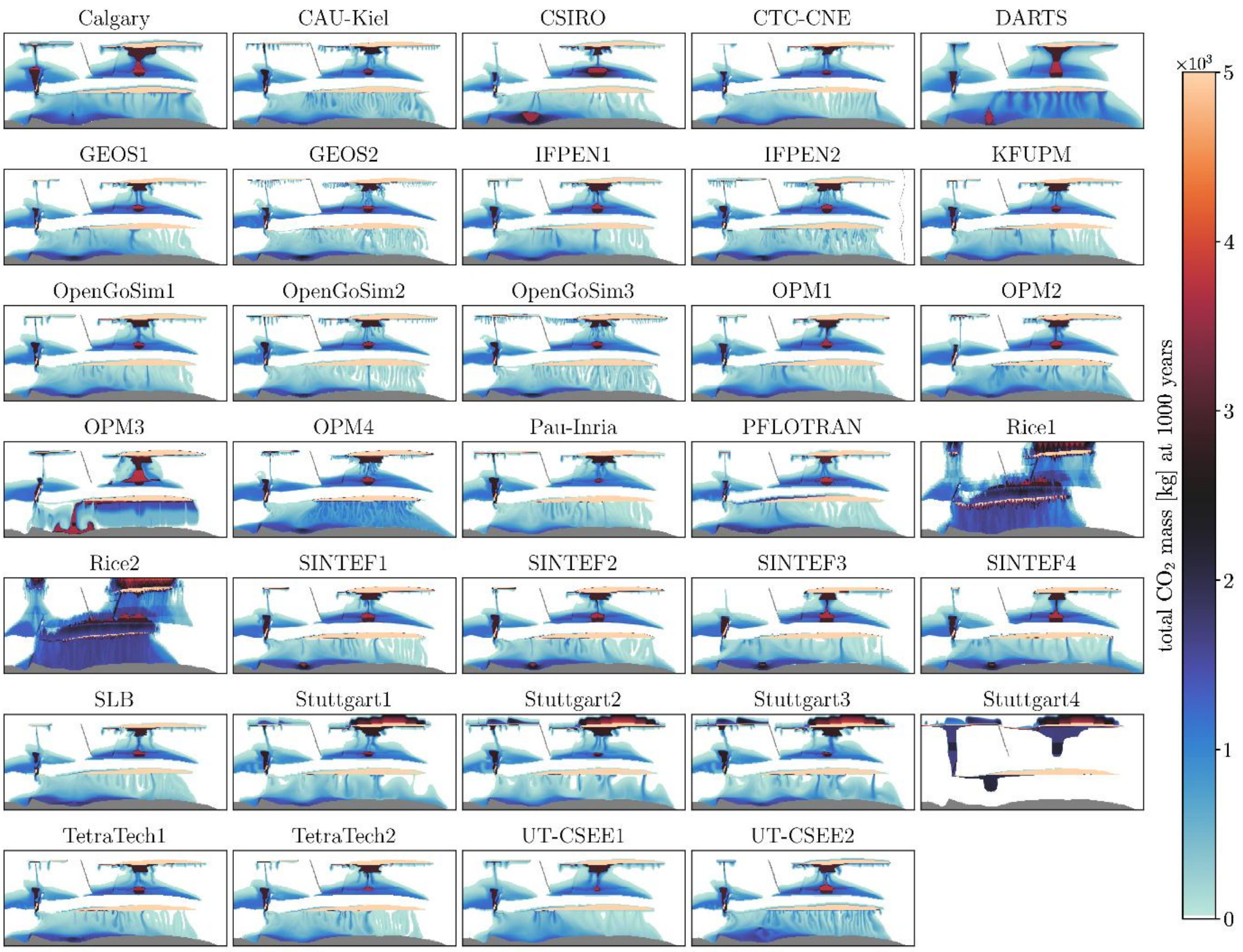

**Figure 5**: An overview of the SPE11B submissions, in terms of distribution of $CO_2$ at the end of the simulation period. Note that the figures are based on the submitted data, thus for the higher resolution simulations, some details are lost relative to the underlying simulation data.

To gain a qualitative understanding of the variation within the submitted solutions, we compare example results at both the end of the injection period and the end of the simulation. We chose to visualize the total $CO_2$ mass (Figures 6 and 7) and temperature (Figures 8 and 9) in each cell of the reporting grid, as these provide a visual guide to the full simulation results. These plots are available for all groups at the SPE11 website (www.spe.org/csp/spe11), together with additional plots of field variables such as pressure, saturation, and mass fractions.

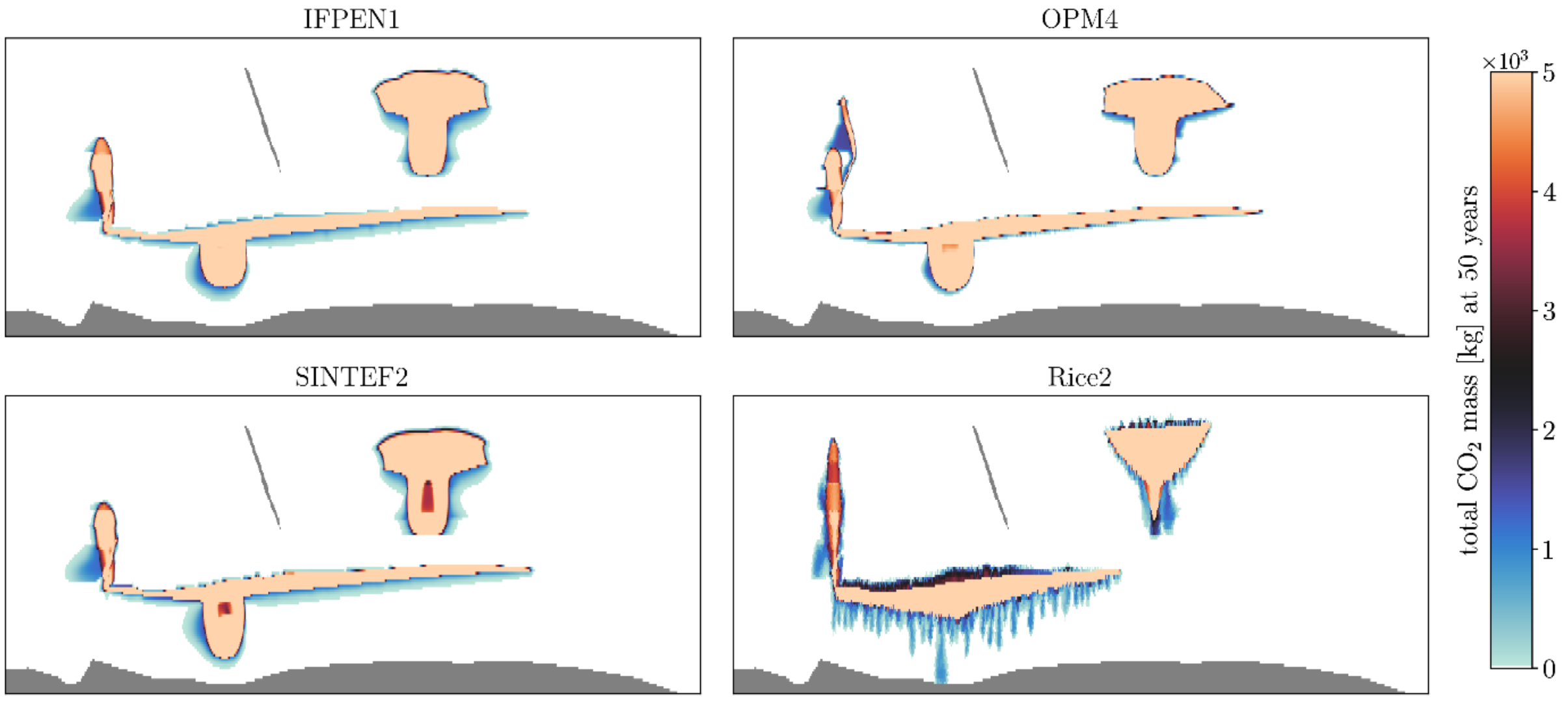

**Figure 6**: Total mass of $CO_2$ in each cell of the reporting grid at the end of injection (50 yr) for SPE11B. The median submission (IFPEN1, upper left), the simulation with highest grid resolution (OPM4, upper right), a relatively coarse simulation on a facies-adapted grid (SINTEF2, lower left), and a simulation on a time-dynamic adaptive grid (Rice2, lower right). Note that the color-scale is capped at 5 tons to better visualize the dissolution dynamics.

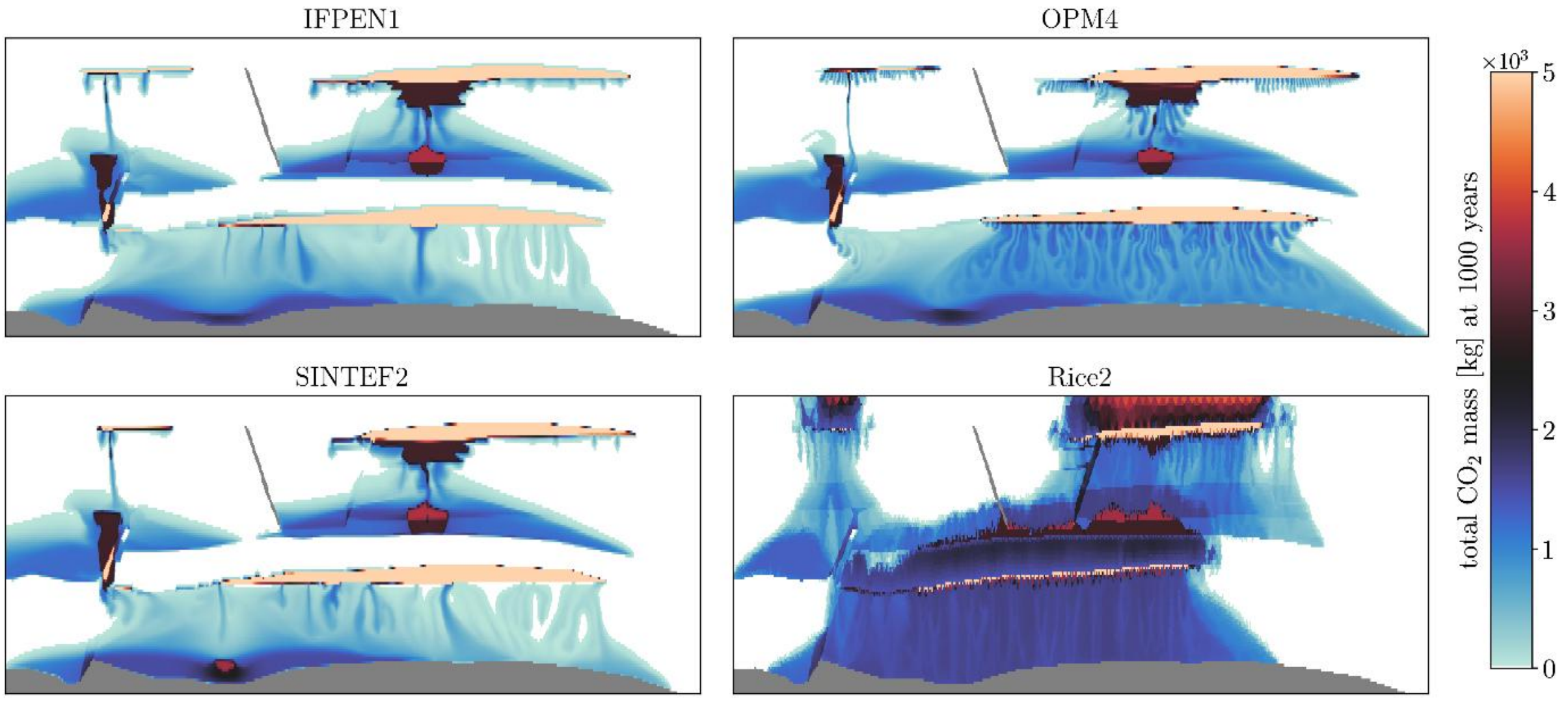

**Figure 7**: Total mass of $CO_2$ in each cell of the reporting grid at the end of simulation (1000 yr) for SPE11B.

We first consider the total $CO_2$ mass distribution at the end of injection, shown in Figure 6, illustrating the footprint of the $CO_2$ plume as it migrates toward the primary sealing units for each injection well. The injection rate for the lowest well was intentionally chosen to ensure that some $CO_2$ migrates to the permeable faulting structure on the domain's left side. The IFPEN1 submission is identified as the "median submission" for SPE11B, with several other submissions being visually indistinguishable from it (see the next section on quantitative results for a detailed discussion). The simulation captures this migration clearly, showing $CO_2$ that rises toward the secondary storage unit through the fault.

A comparison between IFPEN1 and the highly resolved OPM4 simulation reveals some moderate differences already during the injection phase, although the overall visual impression is that the solutions are qualitatively similar. Moderate differences are also observed when compared to SINTEF2. On the other hand, the high-order, time-adaptive simulation of Rice2 exhibits more pronounced deviation from the other submissions in the along-slope migration of $CO_2$, the influence of heterogeneity above the upper well, and the early onset of convective mixing.

Figure 7 shows the total $CO_2$ mass at the end of the simulation, where the minor differences observed in Figure 6 have become more pronounced. The median submission, IFPEN1, shows three distinct accumulations of free-phase $CO_2$, with convective mixing occurring below each accumulation due to dissolution of $CO_2$ into the water phase. Notably, a substantial amount of $CO_2$-enriched water accumulates at the base of the storage reservoirs. The high-resolution OPM4 simulation shows much the same pattern of free-phase $CO_2$ accumulation, but with qualitative and quantitative differences in the convective mixing. The SINTEF2 simulation, conducted on a facies-adapted grid, likewise agrees on the free-phase $CO_2$ accumulation, but provides a third suggestion for the convective mixing. These differences arise from the formation of self-reinforcing dissolution fingers that form in a grid-dependent manner at points where the discrete $CO_2$–brine interface deviates from the true geological boundary; see Holme et al. (2025) for further discussion. Finally, Rice2 constitutes an outlier, in general suggesting a much smaller impact of material heterogeneity and predicting substantial migration of $CO_2$ into the sealing facies.

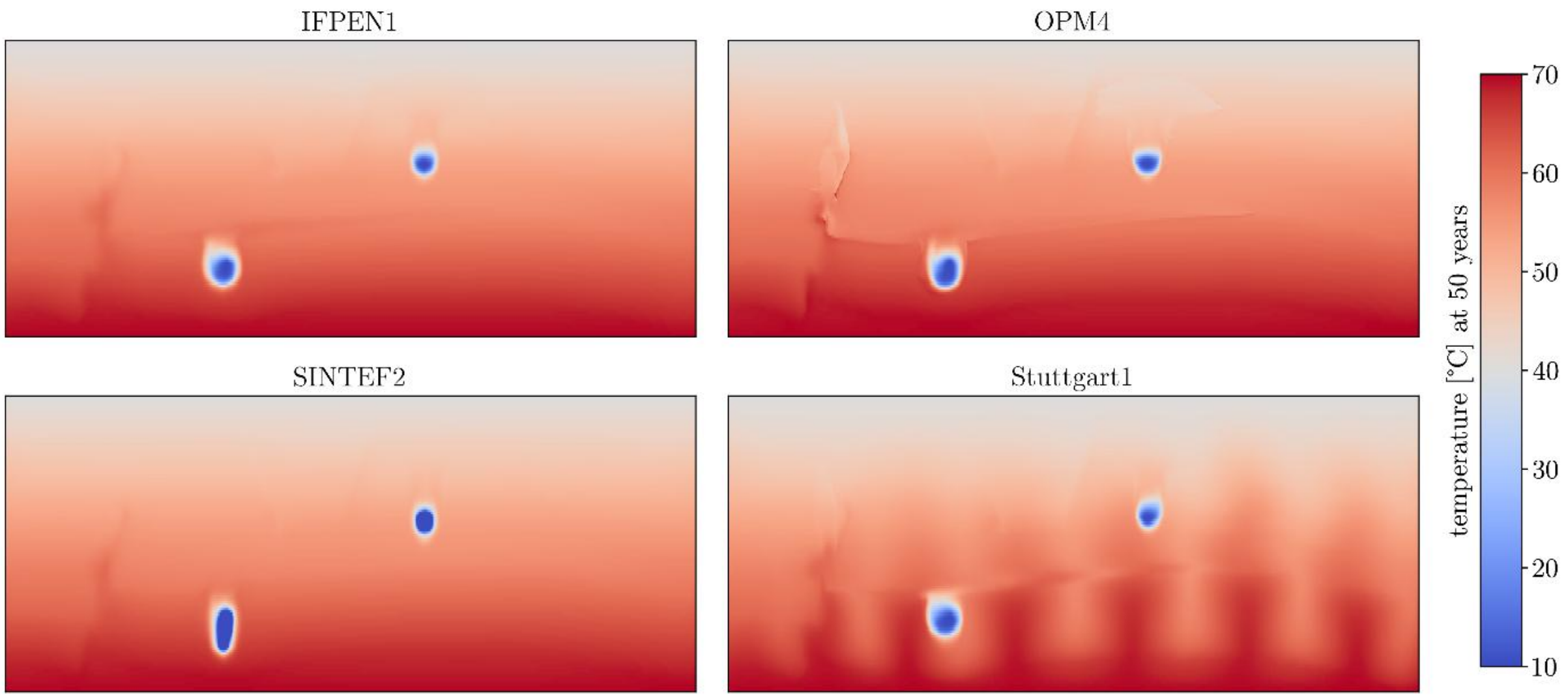

**Figure 8**: Temperature at the end of injection (50 yr) for SPE11B. The median submission (IFPEN1, upper left), the simulation with highest grid resolution (OPM4, upper right), a simulation on a facies-adapted grid (SINTEF2, lower left), and the Stuttgart1 submission (lower right). The color scale is capped below at 10 ºC, which is the minimum temperature reported by IFPEN1, OPM4, and Stuttgart1. The SINTEF2 submission has a near-well cooling down to -50 ºC.

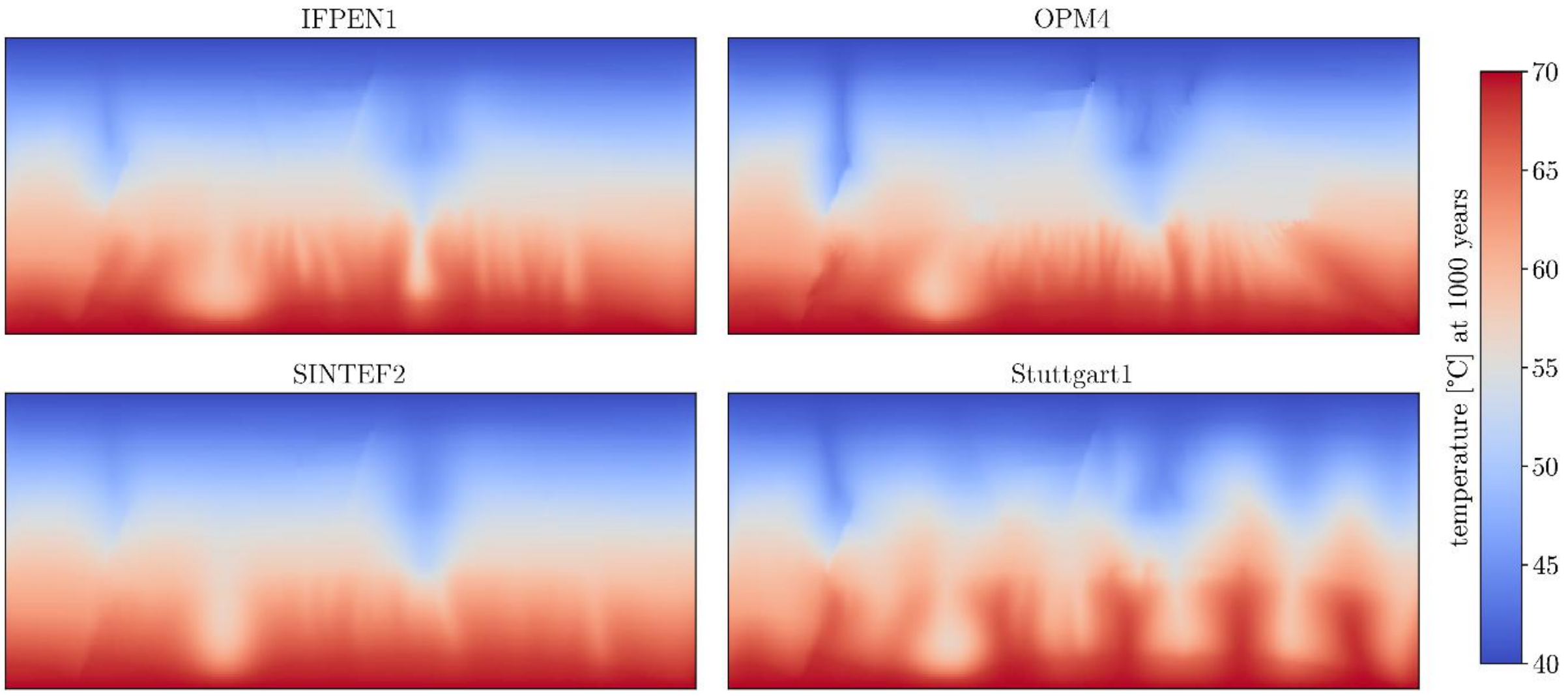

**Figure 9**: Temperature at the end of simulation (1000 yr) for SPE11B.

Figure 8 shows the evolution of temperature after 50 years. In the median simulation, IFPEN1, the most notable effect is the near-well cooling, driven by the lower injection temperature of 10 °C. The high-resolution OPM4 simulation captures more clearly the localized cooling linked to the Joule–Thomson effect as $CO_2$ rises from the injection wells. The unstructured SINTEF2 simulation is broadly speaking comparable to IFPEN1 but features a more compact near-well cooling zone, with temperatures dropping to -50 °C near the well. This, however, appears to be not a grid effect, but rather differences in treatment of injection enthalpy (i.e., using the provided NIST table at reservoir pressure rather than computing it from heat capacity). Because Rice2 conducted an isothermal simulation and did not report temperature values, Stuttgart1 is included for comparison instead. The Stuttgart team's submissions were the only ones indicating the presence of natural thermal convection of water in the system, evident in the figure (note that all simulations were initiated 1000 years before injection to allow for the possibility of this phenomenon.)

Figure 9 shows the temperature after 1000 years. There is minimal visual deviation among IFPEN1, OPM4, and SINTEF2, with the primary mechanism being energy transport driven by the convective mixing seen in Figure 7. In contrast, the Stuttgart1 simulation exhibits a combined effect of thermal convection and energy transport due to convective mixing of $CO_2$.

By examining the time series of the sparse data summarized in Table 1, we gain insight into how the full set of submitted results compares to the selected representative cases in Figures 6 to 9. From the 13 measured quantities, we present pressure at POP 2, convection in Box C, the amount of mobile $CO_2$ in Boxes A and B, and the total amount of $CO_2$ in seal facies and boundaries (quantities 2, 11, 3, 7, 12, and 13 in Table 1); see Figure 10. Plots for all quantities are available from the SPE11 website (www.spe.org/csp/spe11).

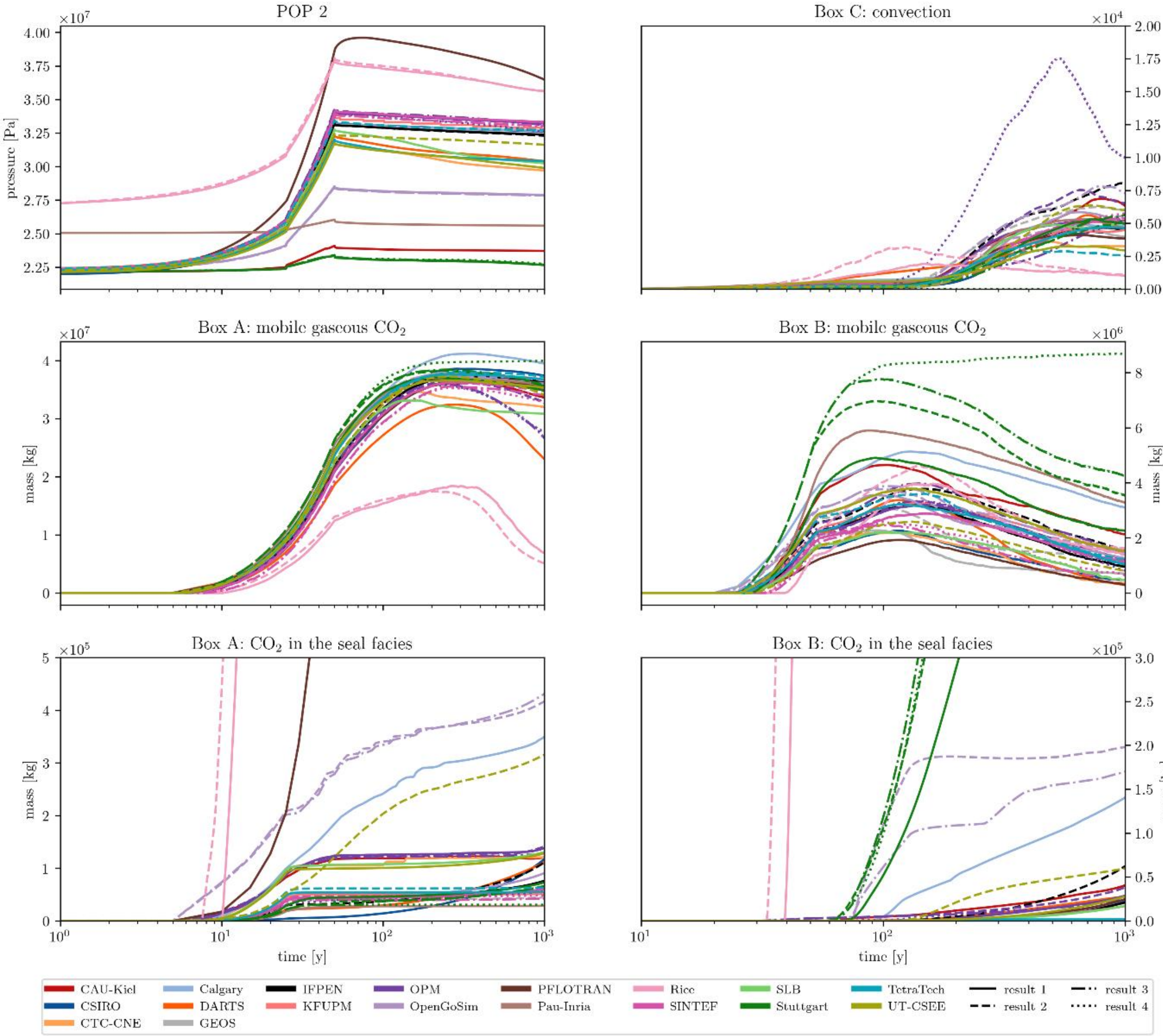

**Figure 10**: Plots of selected sparse data for SPE11B based on all submissions.

Figure 10 broadly supports observations from the dense data. Notably, the two Rice submissions consistently appear as outliers across most plots. Moreover, the Stuttgart submissions stand out in terms of $CO_2$ entering sealing units and the migration of mobile $CO_2$ past the spill point and into Box B. This is due to an erroneous (constant pressure) top boundary condition. The high-resolution OPM4 simulation is also a clear outlier in terms of the measure of convective mixing.

The remaining submissions display a stronger consensus. We specifically emphasize that they agree that only negligible amounts of $CO_2$ enter the sealing facies or reach the domain boundaries. There is also broad agreement on the accumulation of free-phase $CO_2$ within the primary storage unit in Box A. Pressure dynamics at observation point POP2 exhibit some variability, with most simulations falling within a span of 10–20 bars, which is arguably a substantial variability for this metric. Even greater variability is observed for the amount of free-phase $CO_2$ in Box B, where the main group spans a factor of 2 to 3. This variability is comparable to that seen in the measure of convective mixing.

Overall, the submissions to SPE11B display a strong level of internal consistency, suggesting that participants largely interpreted the problem definition similarly and submitted data of high quality. This is particularly noteworthy given the diverse backgrounds of the participating groups and the variety of simulation software used. Moreover, most submissions agree on important questions, such as the minor amounts of $CO_2$ that enter the sealing facies or reach the domain boundaries. On the other hand, substantial differences remain in finer details, particularly concerning the volume of $CO_2$ migrating into the secondary storage unit within Box B and the extent of convective mixing.

## An introduction to the SPE11C solutions

Of the 18 participating groups, 14 submitted results to SPE11C, resulting in a total of 22 individual submissions. An overview of these submissions is given in Figure 11, and they are color-coded in the same way as in Figure 4.

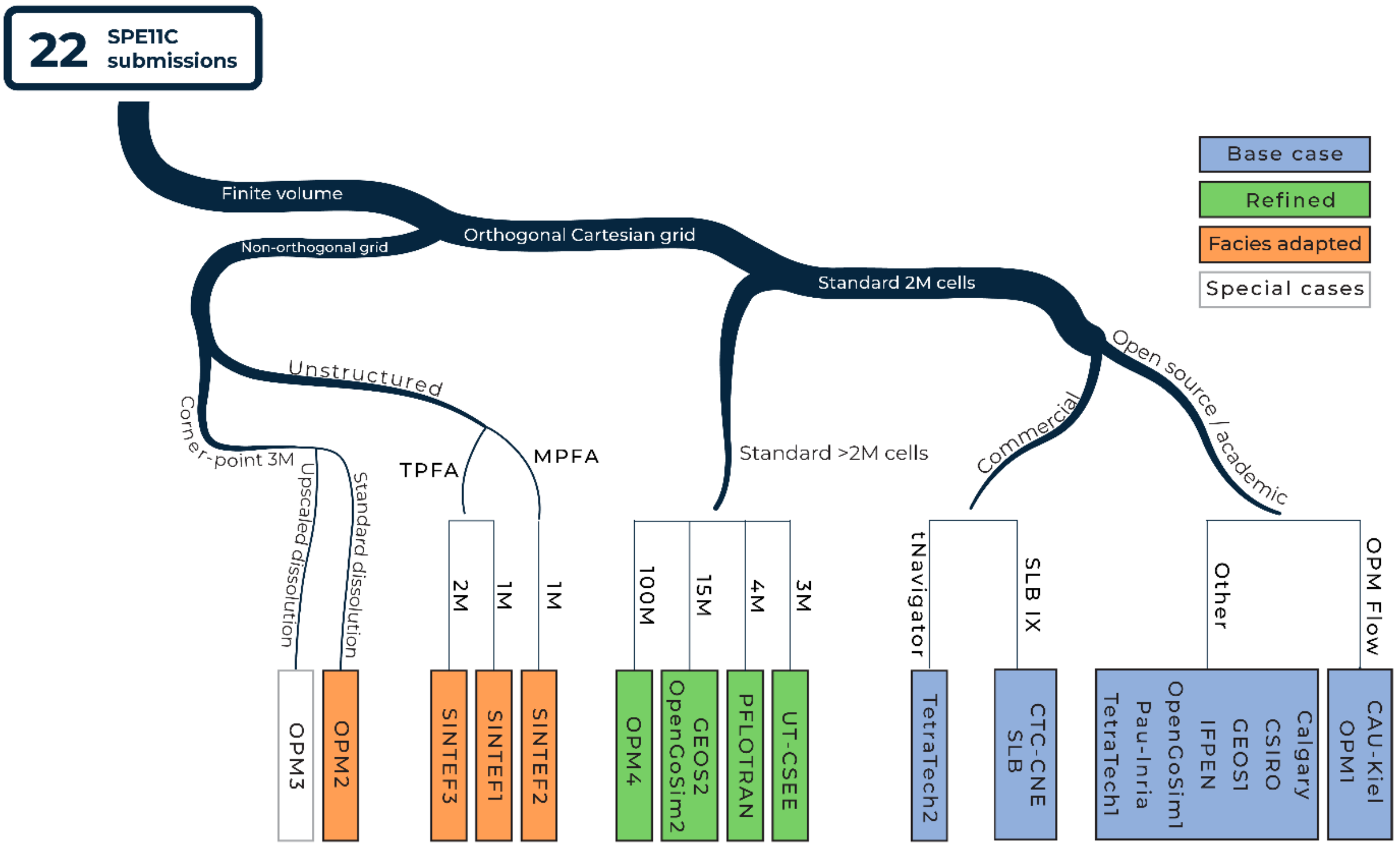

**Figure 11**: An overview of the SPE11C submissions. Curved branches indicate methodological choices, while angled branches indicate grid resolution and choice of simulator.

For SPE11C, 11 groups provided 12 submissions based on simulations using the reporting grid with "standard" methodological choices, represented in blue in Figure 11. Additionally, four submissions used refined Cartesian grids, while five submissions employed grids that were to some extent adapted to the facies. Only one submission deviated from standard simulation choices: OPM3, which utilized a model for resolving the convective mixing through upscaled dissolution rates.

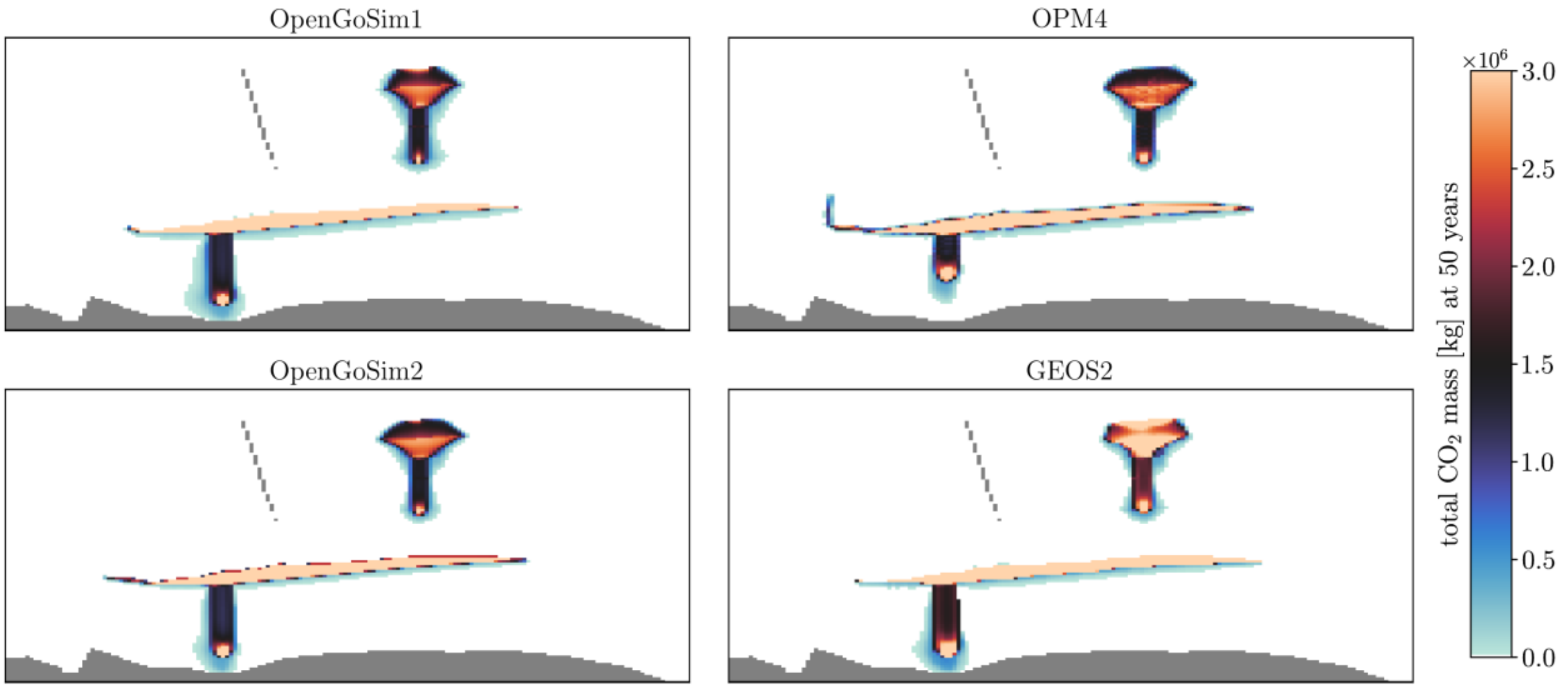

**Figure 12**: Total mass of $CO_2$ within each cell of the reporting grid at the end of injection (50 yr) for SPE11C, plotted on an xz-plane located at y=2500m. The median submission (OpenGoSim1, upper left), the highest-resolution simulation (OPM4, upper right), and two more moderately resolved simulations (OpenGoSim2, lower left, and GEOS2, lower right).

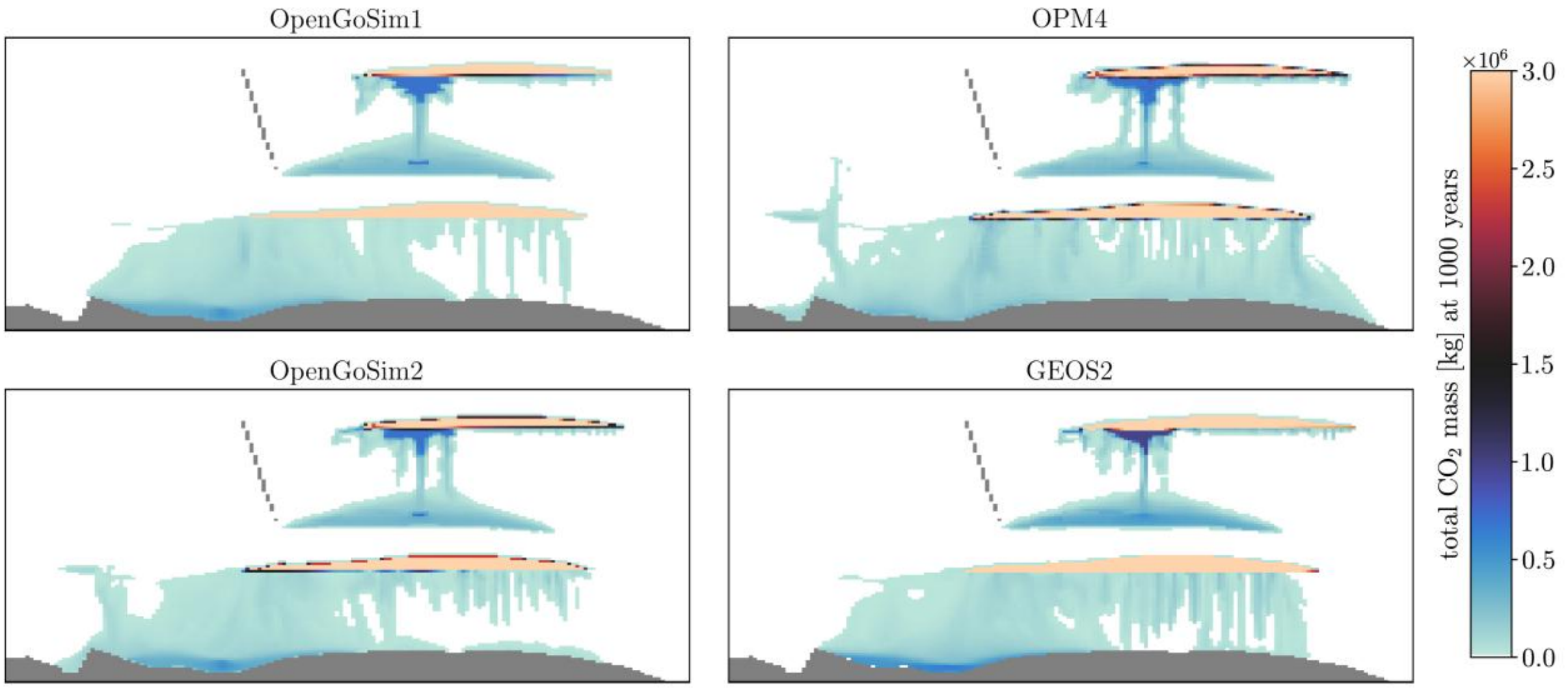

**Figure 13**: Total mass of $CO_2$ within each cell of the reporting grid at the end of simulation (1000 yr) for SPE11C, plotted on an xz-plane located at y=2500m.

The total mass of $CO_2$ within each cell of the reporting grid after 50 and 1000 years is plotted in Figures 12 and 13, respectively, for four selected groups. For SPE11C, OpenGoSim1 is identified as the median simulation in the quantitative analysis. Overall, there is good qualitative agreement between the submissions. The most notable visual deviations are in the amount of $CO_2$ that migrates from the primary storage unit to the fault, which increases with grid refinement, and in the amount and structure of $CO_2$ that dissolves. Regarding the latter, there is a noticeable difference in the amount of dissolved $CO_2$ beneath Well 1. This becomes even more pronounced when considering a 3D rendering of the solutions, as in Figure 14. Here we note a substantial variability in the amount and structure of convective mixing, even between the two submissions based on 15M cell simulations (OpenGoSim2 and GEOS2).

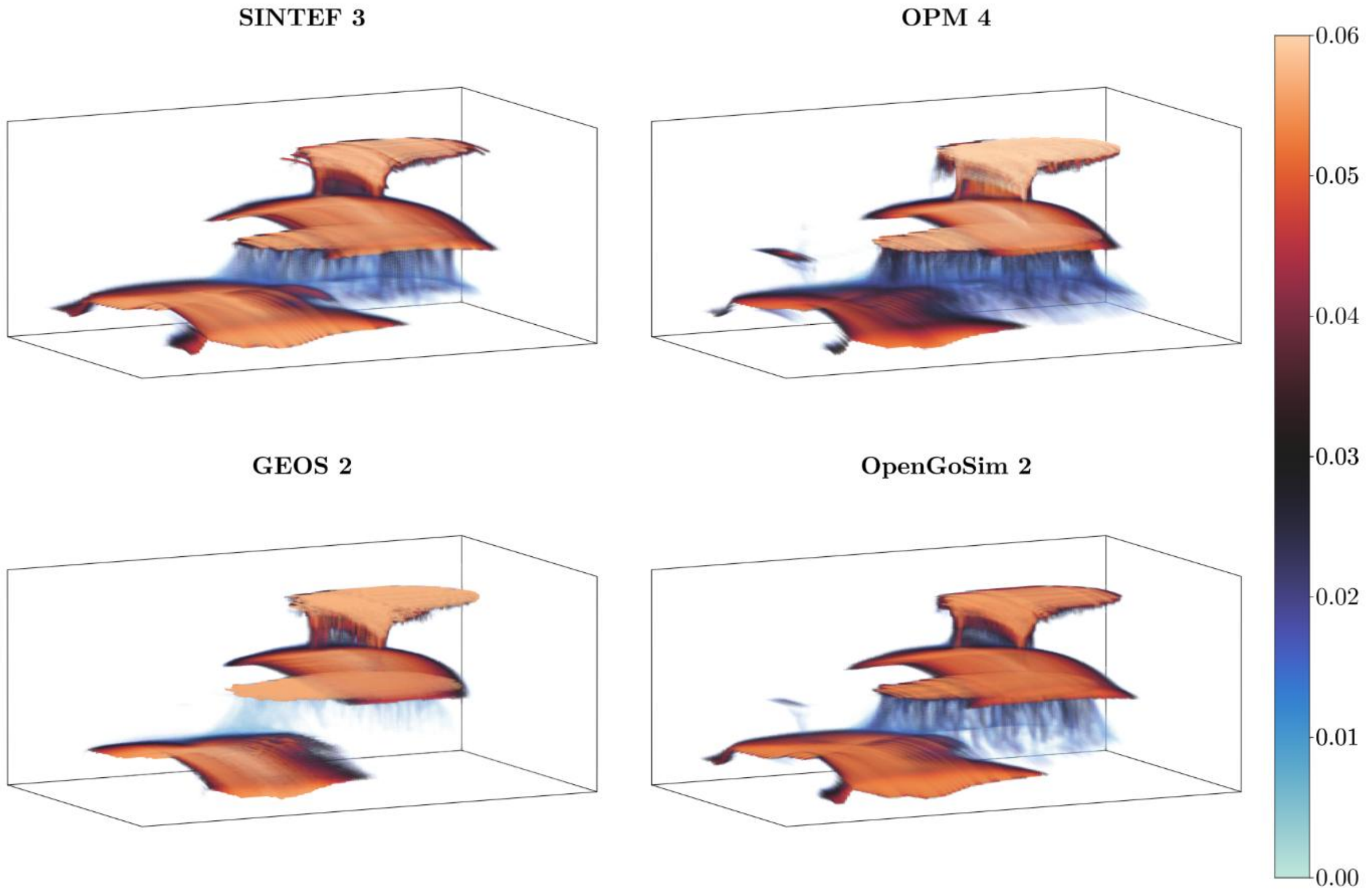

**Figure 14**: Distribution of $CO_2$ at the end of simulation (1000 yr) for SPE11C, rendered in 3D. For visualization purposes, the data is set to be transparent inversely proportional to the $CO_2$ content. Same submissions as in Figure 12.

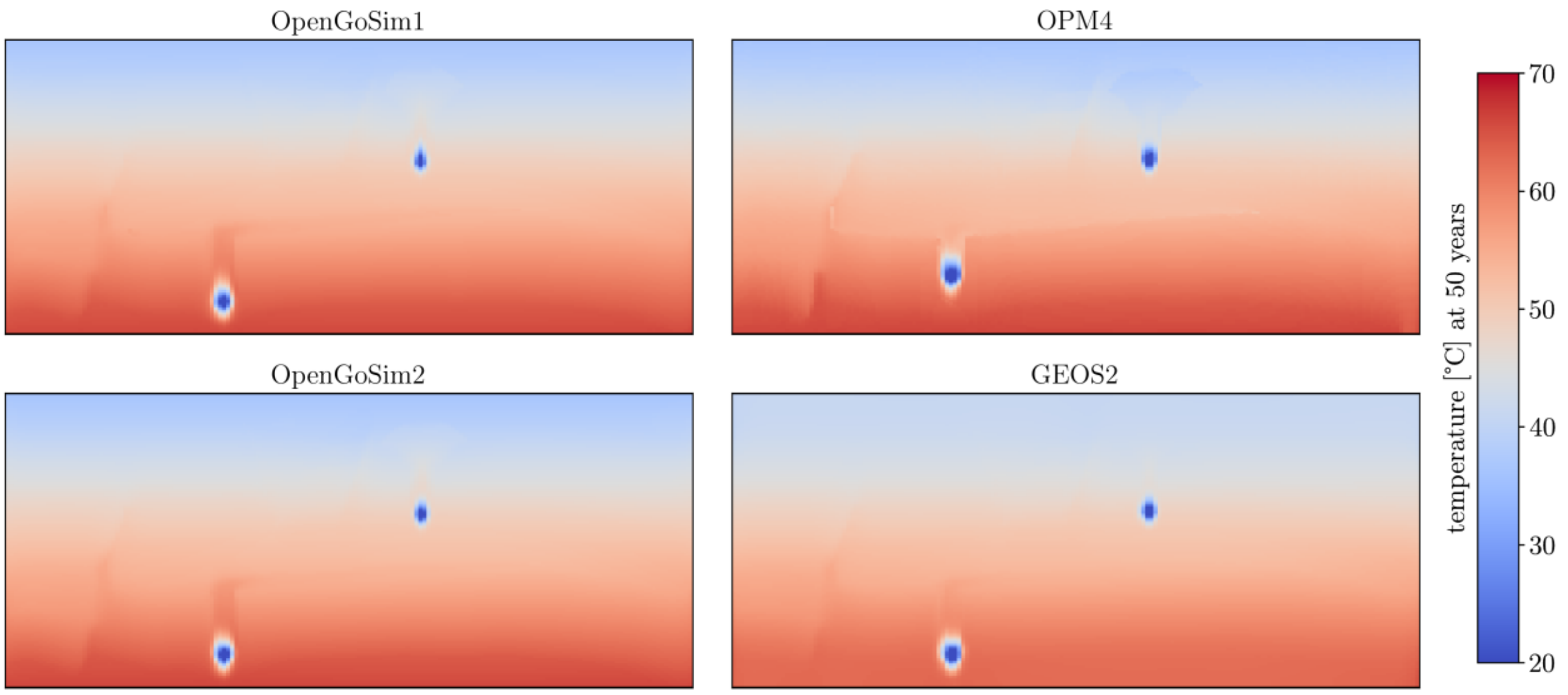

**Figure 15**: Temperature at the end of injection (50 yr) for SPE11C, plotted on an xz-plane located at y=2500m.

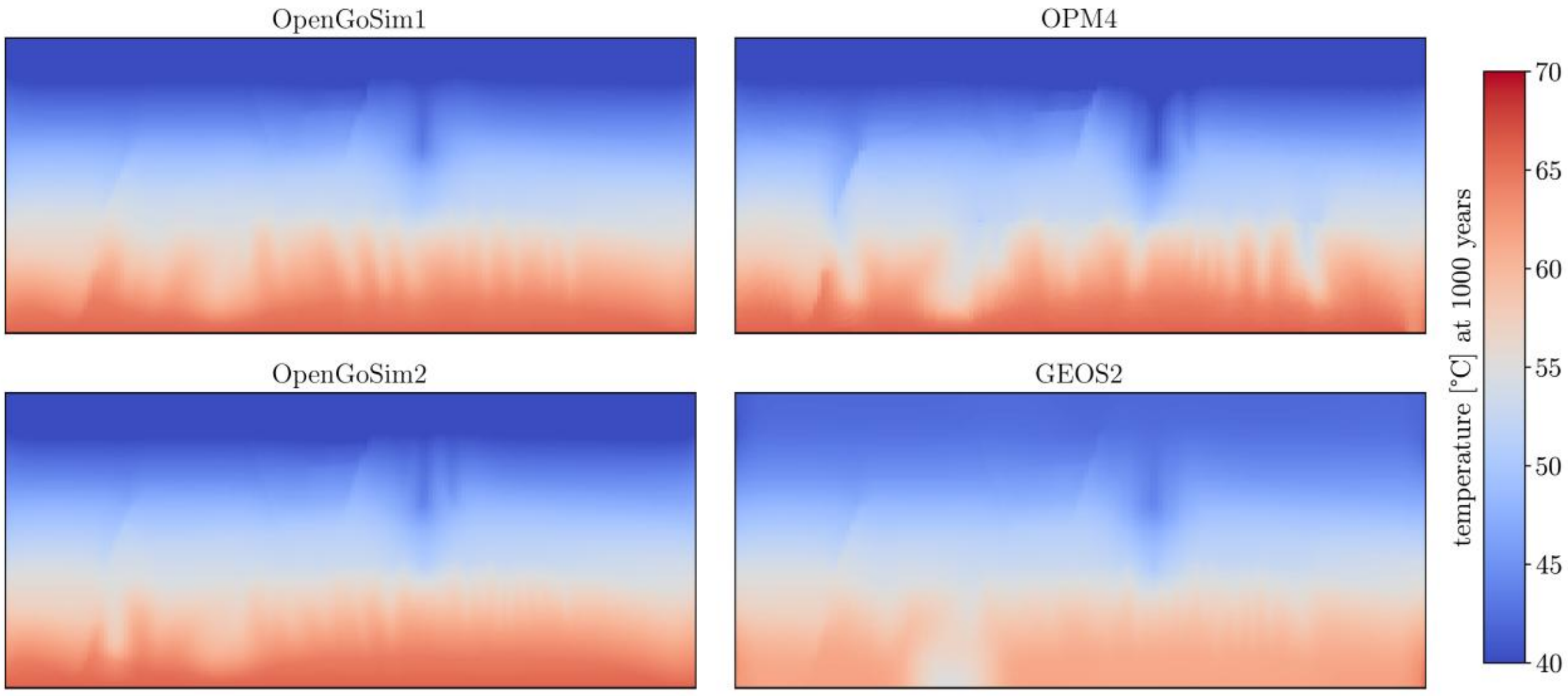

**Figure 16**: Temperature at the end of simulation (1000 yr) for SPE11C, plotted on an xz-plane located at y=2500m.

Figures 15 and 16 report the temperature data for the same submissions as shown in Figures 12 and 13. At the end of injection, near-well cooling can be seen in all four submissions. However, the amount varies, with OpenGoSim2 reporting significantly less near-well cooling compared to the other three. Moreover, near Well 1, where the cold injected $CO_2$ is heavier than the water, the extent of downward migration of $CO_2$—and thus a lowering of the footprint of near-well cooling—is reported quite differently between OpenGoSim1, GEOS2 and OpenGoSim2 on the one hand, and OPM4 on the other. This discrepancy can be explained by the OPM simulations having placed Well 1 at an elevation 100 m higher than stated in the SPE11 description, impacting all their submissions for SPE11C.

Indeed, a surprising aspect of the SPE11C submissions is the variability in the implementation of well paths and injections. To highlight this issue, we show near-well saturation for all groups in Figure 17. The figure clearly reveals that several groups have well paths that deviate from the official description. Moreover, the near-well saturations also suggest that there is substantial variability in the distribution of injected $CO_2$ along the well and thus in the initial dynamics of $CO_2$ rising from the injection wells.

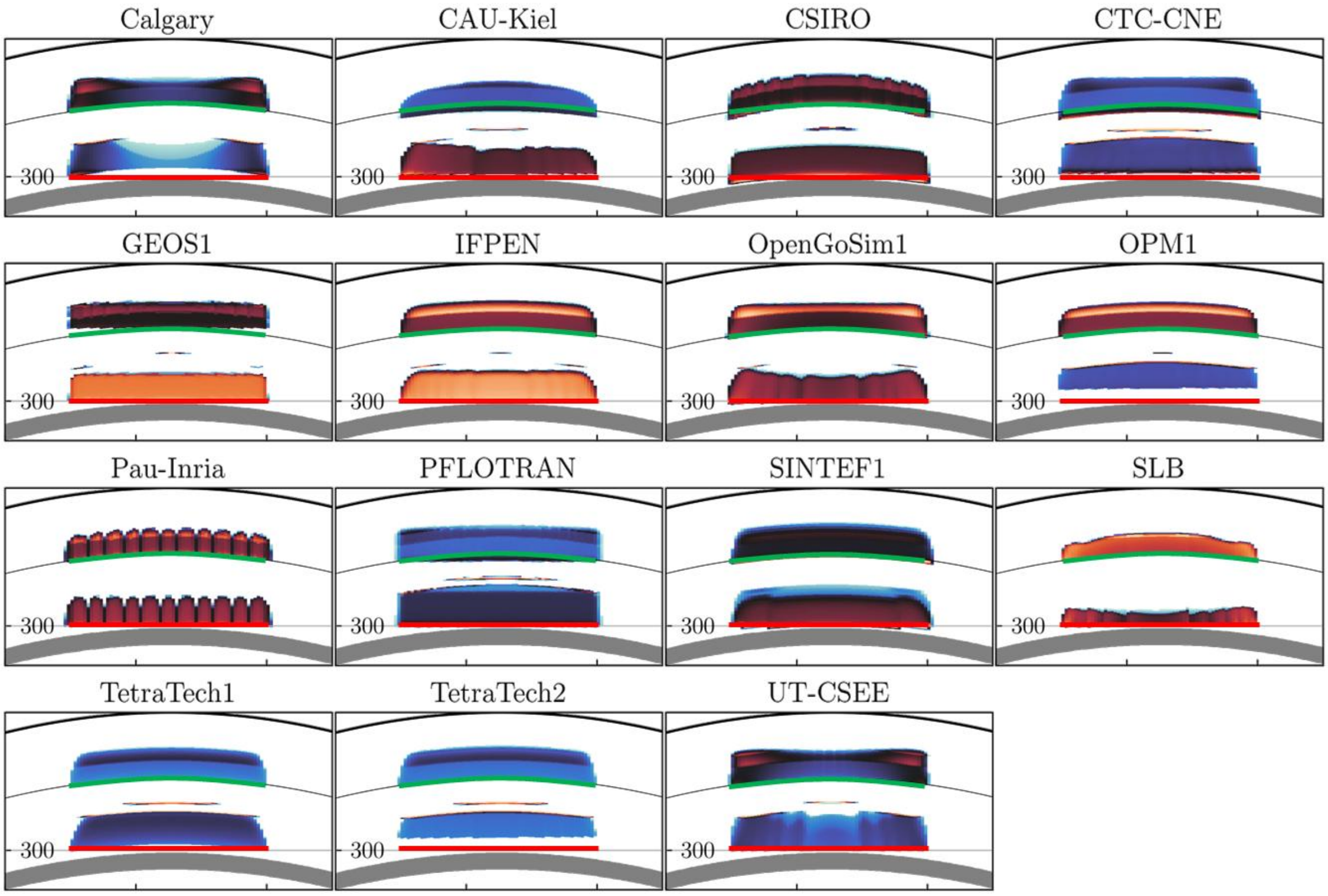

**Figure 17**: Near-well saturations for all groups. The lower part of each figure shows saturation after 5 years along the plane defined by x=2700m, and the upper part of each figure shows saturation after 30 years (5 years after injection starts for Well 2) at x=5400m. Red and green lines indicate prescribed well-paths for Well 1 and Well 2, respectively.

Selected sparse data for all groups are presented in Figure 18. The reported data indicate substantial discrepancies in pressure, already at the initial time, with Pau-Inria and CTC-CNE being the major outliers (dense data reported by these groups confirm the trends seen in the sparse data). The remaining groups report a more modest disagreement in initial pressure on the order of 15 bars. All groups agree on the general dynamics of accumulation of mobile $CO_2$ in Box A. However, only the OPM submissions and TetraTech1 report that any $CO_2$ will migrate into Box B. While the precise dynamics vary, all groups agree that 1% or less of the total injected amount of $CO_2$ will enter the sealing facies. Similarly, all groups report only a minor amount of $CO_2$ reaching the boundaries of the domain.

Finally, we note that there is a substantial disparity between the reported convective mixing within Box C. Notably, the OPM submissions generally report higher values than the remaining submissions, with the high-resolution OPM4 simulation being by far the outlier. Other groups with multiple submissions of varying resolution (GEOS and OpenGoSim) also report an increase in convection with grid refinement, suggesting that this process is not resolved on the reporting grid. On the other hand, within the groups submitting results based on the reporting grid, there is generally greater agreement in this metric, apart from the mentioned OPM1 result.

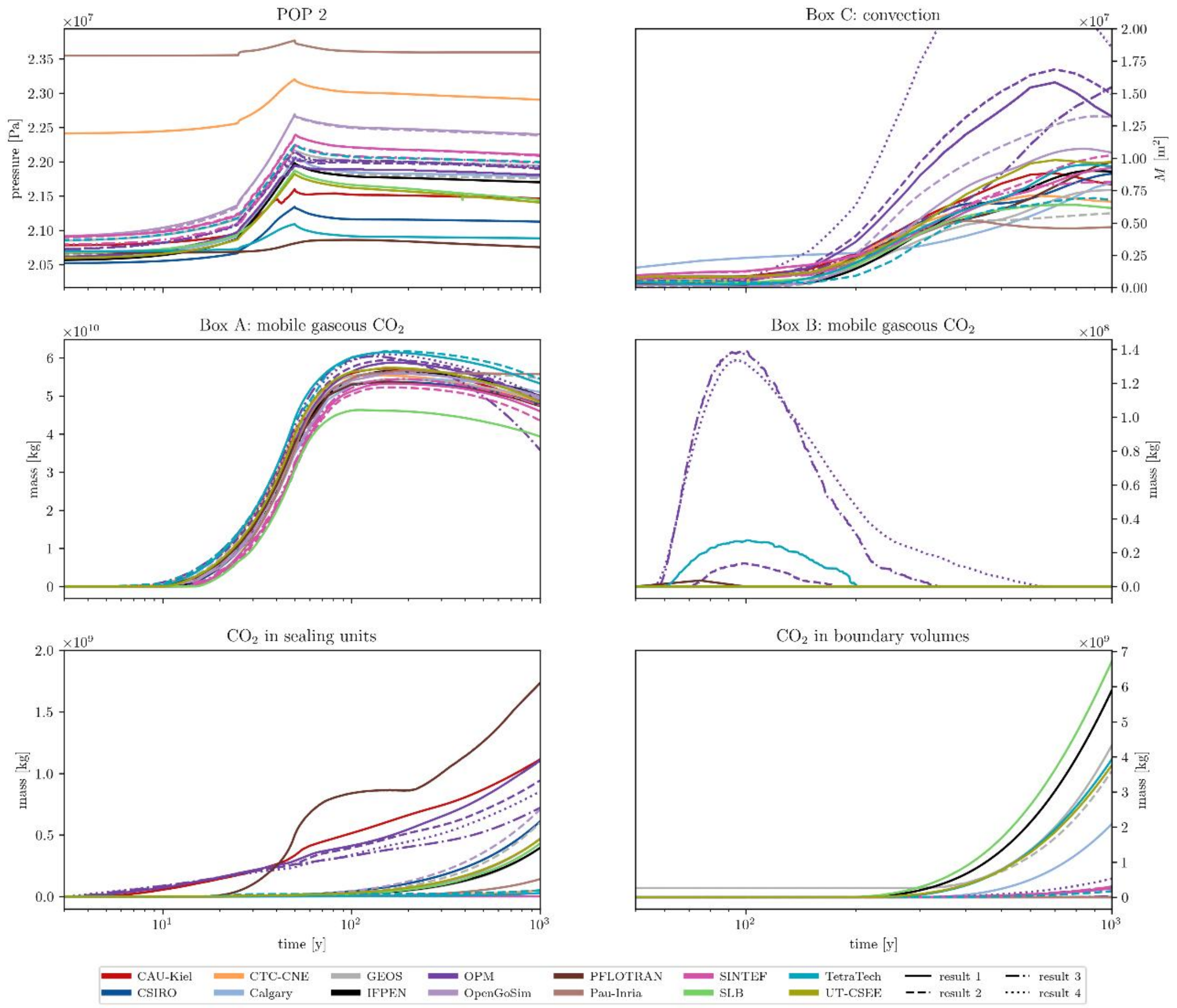

**Figure 18**: Plots of selected sparse data submissions for SPE11C.

# An introduction to the SPE11A solutions

Of the 18 participating groups, 13 submitted results to SPE11A, resulting in a total of 21 individual submissions. An overview of these submissions is given in Figure 19, color-coded in the same fashion as in Figure 4.

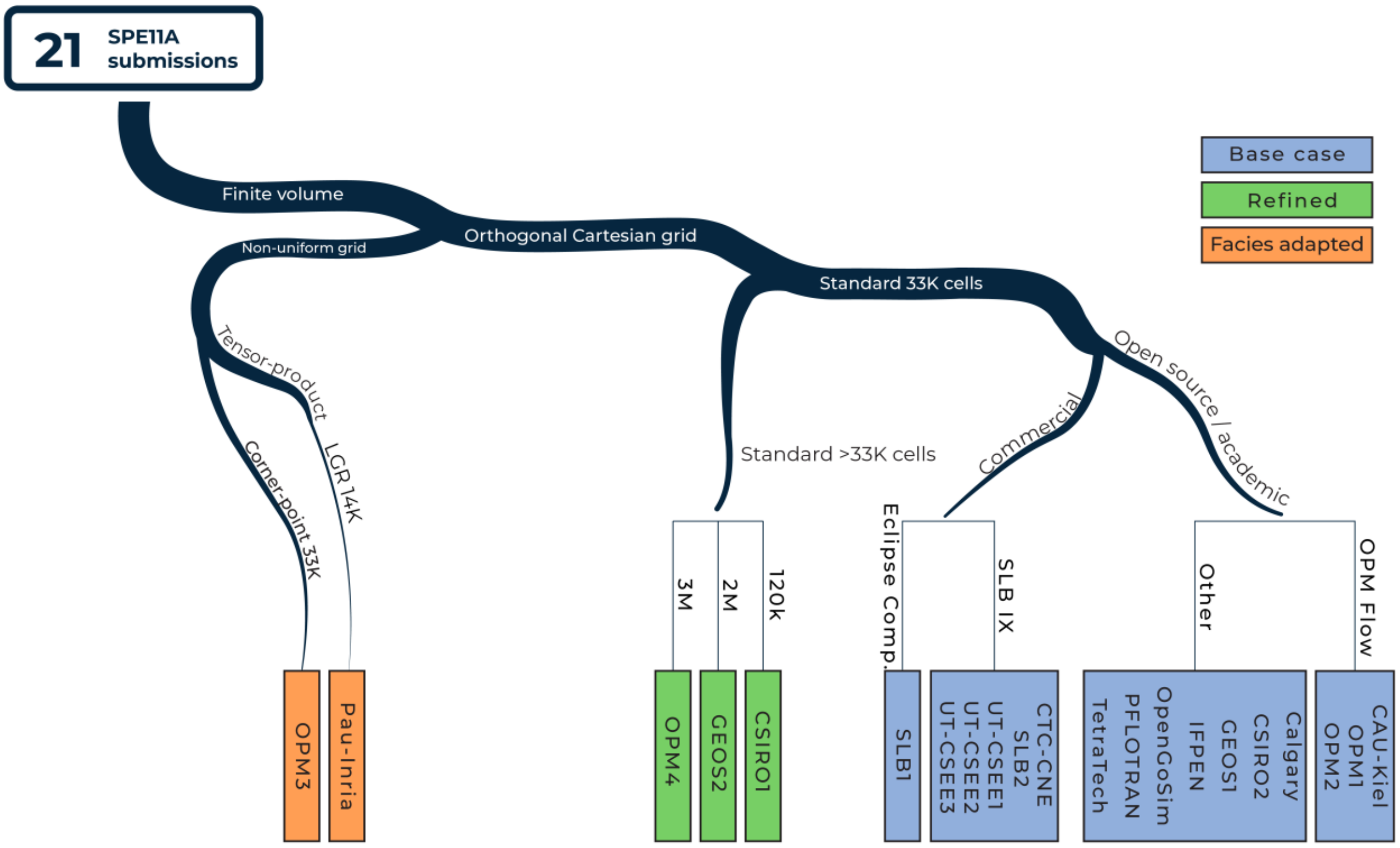

**Figure 19**: An overview of the SPE11A submissions. Curved branches indicate methodological choices, while angled branches indicate grid resolution and choice of simulator.

SPE11A is representative of an experiment at laboratory conditions and is modeled as isothermal. We therefore only visualize the $CO_2$ mass distribution to provide an impression of the submitted results, as shown in Figures 20 and 21.

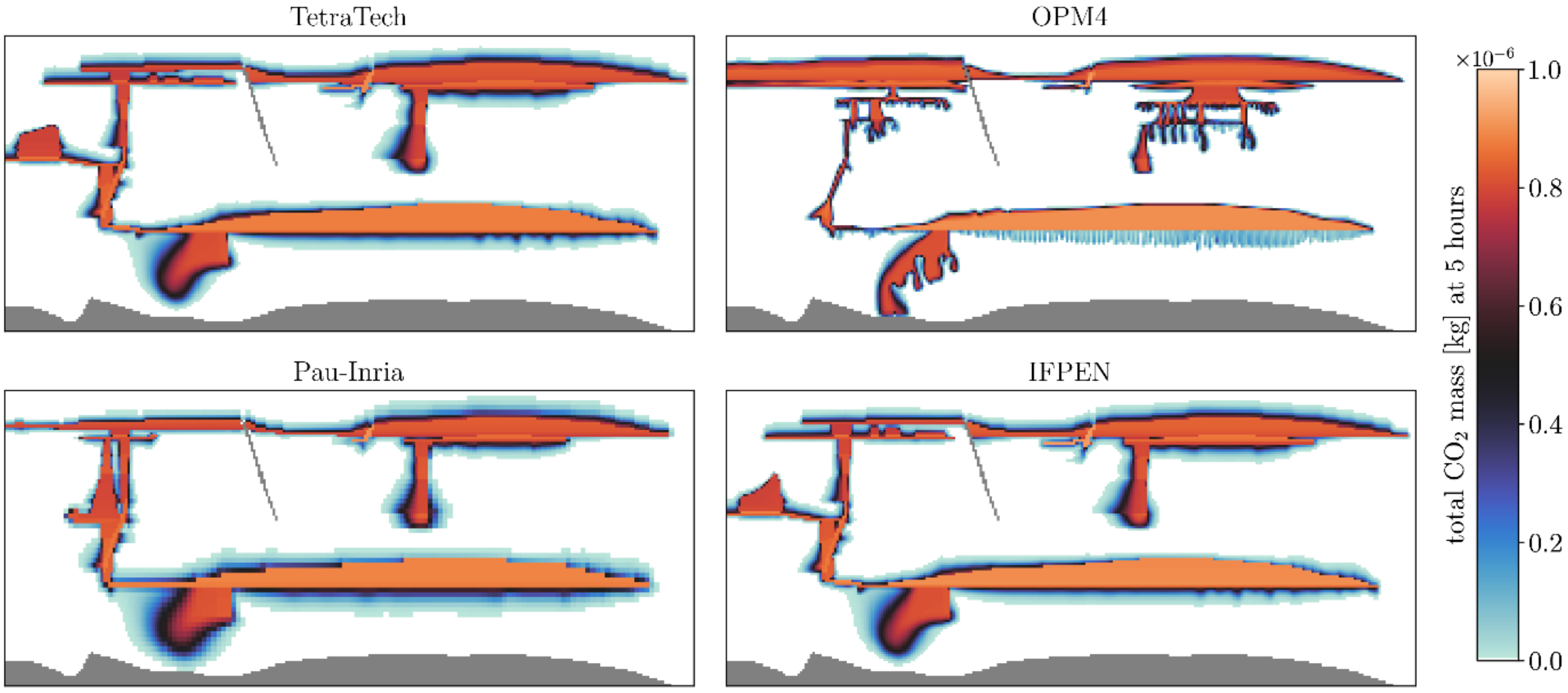

**Figure 20**: Total $CO_2$ mass in each cell of the reporting grid at the end of injection (5 hr) for SPE11A. The median submission (TetraTech, upper left), the simulation with highest grid resolution (OPM4, upper right), a relatively coarse simulation on a locally refined tensor-product grid (Pau-Inria, lower left), and a representative of the baseline simulations (IFPEN, lower right).

For SPE11A, TetraTech is identified as the median submission. After the injection period, we clearly see in Figure 20 the accumulation of free-phase $CO_2$ above each of the two injection wells, as well as some migration of $CO_2$ through the fault structure into the secondary storage unit within Box B. The TetraTech submission shows some effects of the heterogeneous layering in the upper formations and also indicates some flow dynamics associated with migration through the fault leading to lateral spreading of $CO_2$. Already at this relatively early time, the onset of convection is visible at the right-hand side of the primary storage site.

The results of IFPEN closely match those of TetraTech, indicating the similarity between many of the submissions. On the other hand, the submission from Pau-Inria, using a locally refined grid, differs in the migration pattern through the fault structure. Moreover, several substantial differences in migration patterns appear in the more geologically complex top storage units. Finally, the high-resolution OPM4 submission diverges significantly from the preceding three submissions. In particular, the effects of heterogeneity are much more pronounced in the upper reservoir, and the onset of convection is observed throughout almost all regions of the domain where $CO_2$ is in contact with water. This underscores the importance of sufficiently resolving heterogeneous structures, indicating that the reporting grid size may not be sufficient to capture the initial onset of convection.

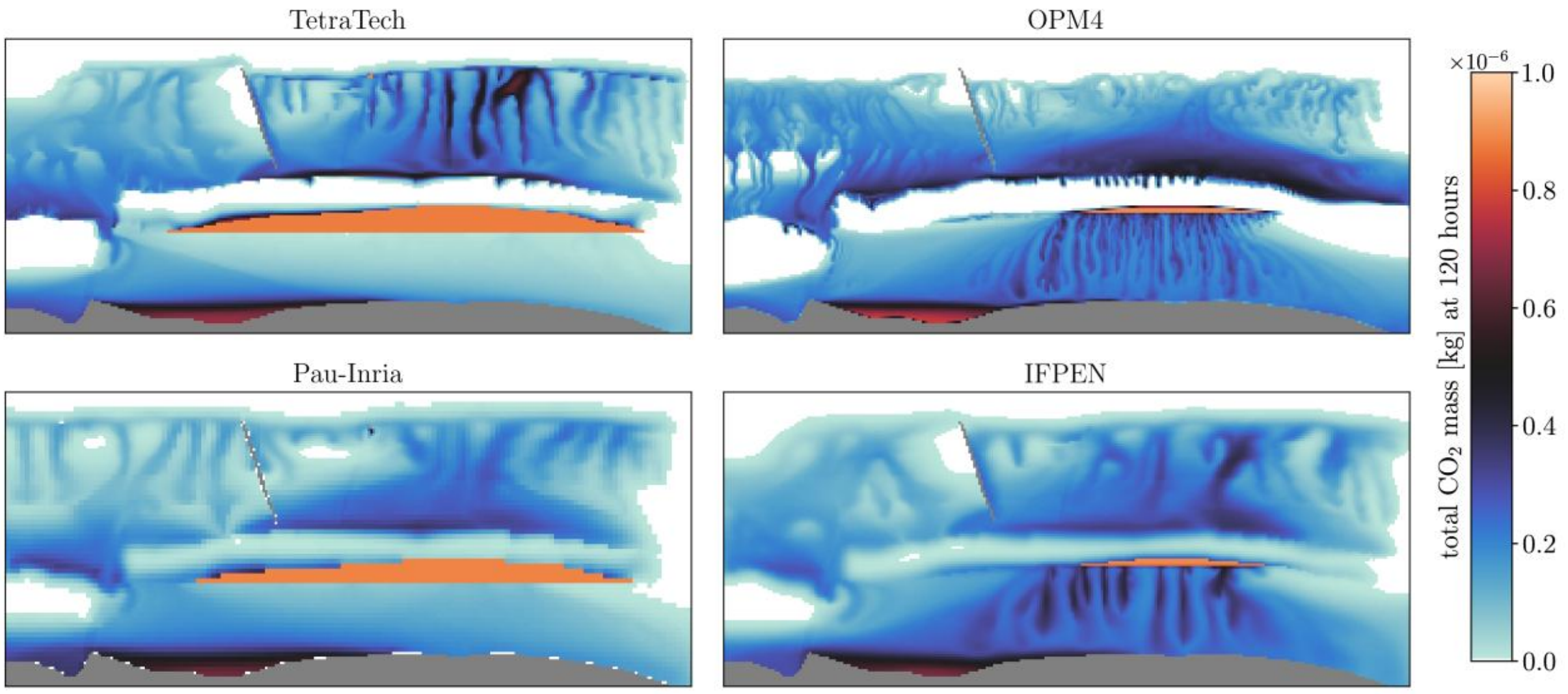

**Figure 21**: Total $CO_2$ mass per reporting cell at end of simulation (120 hr) for SPE11A. Same submissions as in Figure 20.

After 120 hours, as shown in Figure 21, the simulations have diverged significantly. While TetraTech and IFPEN were in close agreement after 5 hours, they now report substantially different amounts of free-phase $CO_2$ (orange color in Figure 21). Both submissions agree that the convective mixing will dissolve all free-phase $CO_2$ in the upper storage units, but the TetraTech submission indicates no convective mixing in the lower storage unit above Well 1. This is likely due to an artifact associated with Cartesian grids, as discussed previously (Flemisch et al, 2024). The Pau-Inria submission closely aligns with the TetraTech submission, although notable differences exist in the precise structure of the convective mixing in the upper storage units. Finally, as expected, the OPM4 submission resolves significantly more of the convective mixing patterns, even reporting some convective fingers penetrating down into the relatively low-permeability seals.

Selected sparse data for all groups are reported in Figure 22. As for SPE11B and SPE11C, there is some disagreement regarding pressure, even at the initialization of the simulation. The strong compressibility of gaseous $CO_2$ at laboratory conditions impacts the subsequent migration patterns, and groups reporting the highest pressures at POP1 tend to observe lower amounts of $CO_2$ migrating into Box B. Within Box A, most groups report a similar buildup, followed by a decline, in mobile gaseous $CO_2$. However, the rate of decline varies strongly among the groups, as it depends on resolving convective patterns. In particular, we note that most groups exhibit a "staircase-like" decline in mobile $CO_2$ caused by the same Cartesian grid phenomenon mentioned above. These staircases seem to diminish with grid refinement and are not noticeable in the highly refined GEOS2 and OPM4 submissions. The submitted results of the CAU-Kiel group appear as outliers in all panels of Figure 22, with their results shifted to larger times. This is due to a scripting error from the participating team when preparing their sparse data for submission and only discovered after the cutoff date for updating results. Their dense results are not affected by this issue.

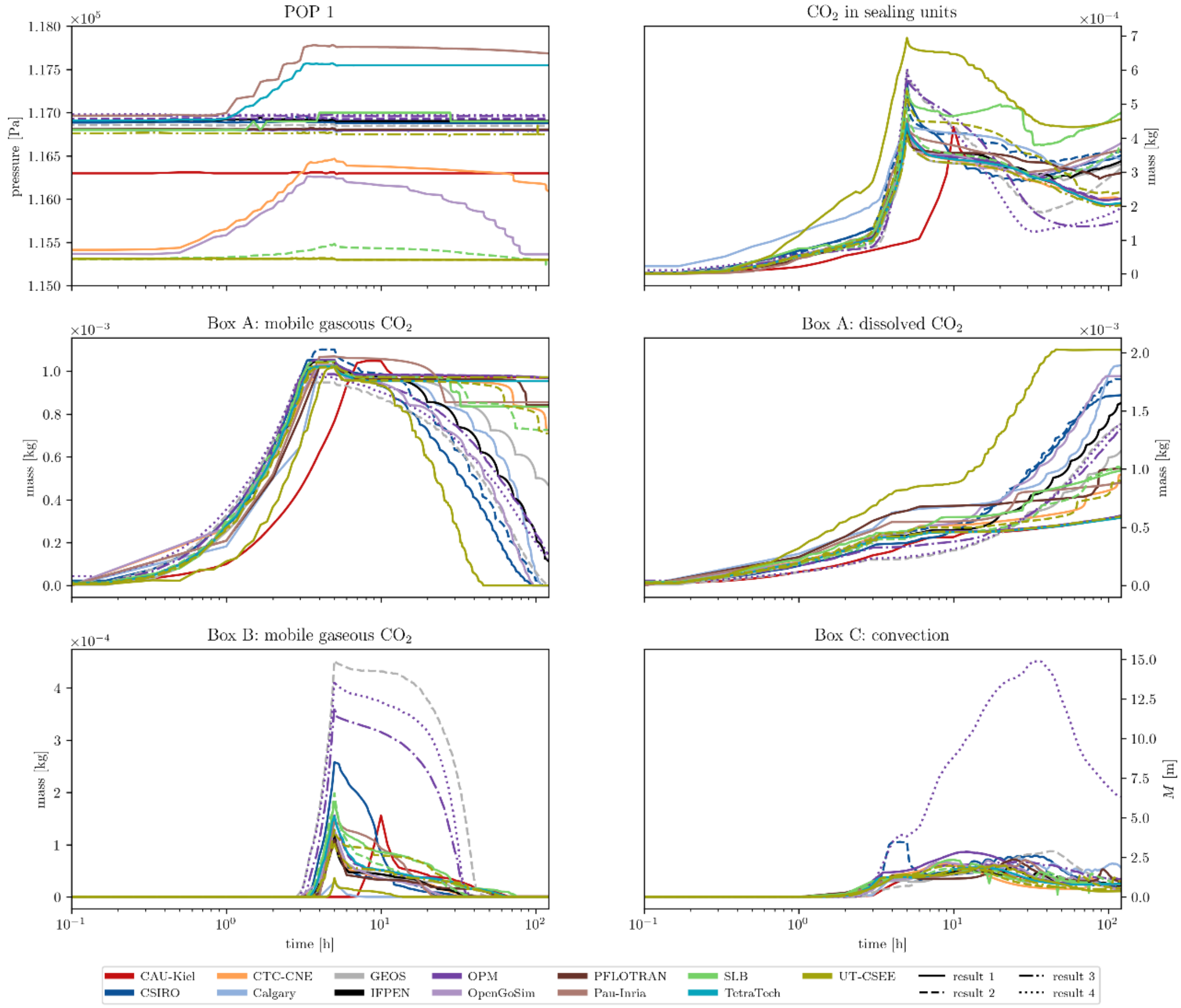

**Figure 22**: Plots of selected sparse data submissions for SPE11A.

## The full SPE11 solution set

The full dataset consisting of all valid submissions to SPE11 has been made available online through three different online resources. These are as follows:

**SPE11-at-a-glance**: A static webpage, hosted at spe.org (spe.org/csp/spe11), has been created to display sparse-data plots from all submissions, as well as plots of the dense data at selected time-steps. The webpage additionally contains illustrative movies for all submissions, intending to give the page visitor a quick understanding of the main characteristics of each submission. Although static, the webpage is generated from the full SPE11 dataset using an openly available script (github.com/moyner/CSP11Visualizer.jl), making its contents reproducible and, in principle, regenerable with additional data from future researchers.

**Full SPE11 dataset**: A set of 77 zip-files containing the contribution from each team to each of the three subcases is available for download at doi.org/10.18419/DARUS-4750

**Interactive online data access**: All analysis and plotting scripts used to generate the results and the figures in this paper are available on GitHub in the dedicated repository github.com/Simulation-Benchmarks/11thSPE-CSP. This repository also includes scripts for generating PVT tables, the precise geometry definitions compatible with the Gmsh mesh generator (gmsh.info), and exemplary input decks for some of the simulators used in the study. Moreover, a JupyterLab environment is directly accessible from the GitHub repository, where guest users have direct access to the full SPE11 dataset, giving them the opportunity to recreate and modify the plots of this paper, and generate virtually any other plot from the dataset. A particularly important feature of the interactive access is that it has functionality for guest users to upload their own data and compare their results to the original SPE11 dataset.

# Quantitative comparative assessment

The fundamental building block of our comparative analysis is the pairwise differences between solutions. We use these to develop a holistic quantitative analysis of the submission for each of the three SPE11 subcases. This approach enables us to address the following questions:

- Q1. Which solution(s) can be considered the “median” in the sense of being the least outlying among all submissions?
- Q2. Are the baseline submissions - based on similar qualitative choices - more closely aligned than the full set of submissions (blue background in Figures 4, 11, and 19)?
- Q3. Which has a greater influence on the results: the choice of simulator or the participating group?
- Q4. What is the impact of grid refinement (green background in Figures 4, 11, and 19)?
- Q5. What is the impact of adapting the grid to the geometry (orange background in Figures 4, 11, and 19)?

Given the larger number of submissions for SPE11B, this case provides the most statistically meaningful basis for analysis. Therefore, we first analyze SPE11B in detail in the next three subsections before substantiating the findings from this case through analogous analyses for SPE11A and SPE11C.

To maintain interpretive validity of our analysis, we avoid drawing conclusions from individual submissions that represent unique choices without comparable counterparts, as indicated by the white background in Figures 4 and 11.

## The SPE11 distance and median simulations

As highlighted earlier, the official SPE11 description requested the submission of time series for 13 scalar quantities (referred to as sparse data) and 8 field variables (dense data). Each quantity or variable can, in principle, be compared (pairwise) across submissions at any point during the simulation period. Selected examples of such comparisons were presented as part of the qualitative discussion in the previous section.

To provide a quantitative comparison between submissions, we introduce the SPE11 distance metric that reflects the overall characteristics of the submitted data. For each analyzed data type $\gamma$, and at any time $t$, we specify the distance $D_\gamma$ between two data fields indicated by $\gamma_i$ and $\gamma_j$, where $i$ and $j$ refer to different submissions. Concretely, for scalar quantities, the distance is defined as the absolute difference, $D_\gamma^{i,j}(t) = |\gamma_i(t) - \gamma_j(t)|$. For field variables, we choose either the $L^2$-norm, the $L^2$-seminorm, or the weighted Wasserstein distance, as detailed in Appendix D. The choice of distance measures is specified in Table 4.

| | # | $D_\gamma$ | $\bar{\mathcal{D}}_\gamma$ | PCC |
|---|---|---|---|---|
| Sparse data | 3 | Absolute value of the difference | 1.61e11 | 0.88 |
| | 4 | Absolute value of the difference | 2.69e9 | 0.51 |
| | 5 | Absolute value of the difference | 5.57e10 | 0.77 |
| | 7 | Absolute value of the difference | 5.29e10 | 0.32 |
| | 8 | Absolute value of the difference | 1.88e10 | -0.05 |
| | 9 | Absolute value of the difference | 2.72e10 | 0.79 |
| | 11 | Absolute value of the difference | 6.97e7 | 0.32 |
| | 12 | Absolute value of the difference | 1.36e10 | 0.93 |
| | 13 | Absolute value of the difference | 6.04e8 | 0.29 |
| Dense data | 14.e | $L^2$-seminorm for $t \le t_{\mathrm{IS}}$; zero for $t > t_{\mathrm{IS}}$ | 3.79e9 | 0.87 |
| | 14.l | Zero for $t \le t_{\mathrm{IS}}$; $L^2$-norm for $t > t_{\mathrm{IS}}$ | 9.21e10 | 0.30 |
| | 20.e | Weighted Wasserstein distance for $t \le t_{\mathrm{IS}}$; zero for $t > t_{\mathrm{IS}}$ | 4.48e12 | 0.79 |
| | 20.l | Zero for $t \le t_{\mathrm{IS}}$; weighted Wasserstein distance for $t > t_{\mathrm{IS}}$ | 2.03e13 | 0.95 |
| | 21.e | $L^2$-seminorm for $t \le t_{\mathrm{IS}}$; zero for $t > t_{\mathrm{IS}}$ | 4.06e4 | 0.86 |
| | 21.l | Zero for $t \le t_{\mathrm{IS}}$; $L^2$-seminorm for $t > t_{\mathrm{IS}}$ | 5.03e4 | 0.91 |

**Table 4**: Selection of quantities included in the calculation of the SPE11 distance. Numbers refer to the enumeration of submitted data as summarized in Table 1. For dense data, the dataset is split into the injection period (denoted by e for early), and post-injection (denoted by l for late). For each quantity, the median distance $\bar{\mathcal{D}}_\gamma$, as defined in Equation (3), is also given. Finally, with respect to the overall SPE11 distance $\mathcal{D}$ defined in Equation (4), the Pearson correlation coefficient (PCC) is provided in the last column (for detailed discussion, see the end of this section).

To balance the relative importance between the injection and post-injection periods, where the latter is an order of magnitude longer, we define a weight factor:

$$\tau(t) = \begin{cases} 1 & \text{if} \quad t \leq t_{\mathrm{IS}} \\ 0.1 & \text{if} \quad t > t_{\mathrm{IS.}} \end{cases} \tag{1}$$

Here, $t_{IS}$ represents the time the injection stops, which is 5 hours for SPE11A and 50 years for SPE11B and SPE11C. With this weight, we calculate the weighted-square distance over time $\mathcal{D}_\gamma$ as:

$$\mathcal{D}_\gamma^{i,j} = \sqrt{\int_0^{t_{\mathrm{final}}} \tau(t) D_\gamma^{i,j}(t)^2 \, dt}. \tag{2}$$

The integral is approximated using all reported values of the data.

We denote the set of quantities listed in Table 4 as $\Gamma$. For any two quantities $\beta$ and $\gamma$ in $\Gamma$, the respective distances $\mathcal{D}_\beta$ and $\mathcal{D}_\gamma$ may have different units and scalings and therefore not be directly comparable. To address this, we define

$$\overline{\mathcal{D}}_\gamma = \underset{i,j\in\Gamma}{\mathrm{median}}\left(\mathcal{D}_\gamma\right). \tag{3}$$

as the median value of all non-zero $\mathcal{D}_\gamma^{i,j}$. The total "***SPE11 distance***" between any two submissions for SPE11A, SPE11B, or SPE11C is then defined as the mean value of the rescaled and transformed individual distance metrics:

$$\mathcal{D}^{i,j} = \underset{\gamma\in\Gamma}{\mathrm{mean}_{\mathrm{AG}}}\left(\frac{\mathcal{D}_\gamma^{i,j}}{\overline{\mathcal{D}}_\gamma}\right). \tag{4}$$

Here, we have used a combination of the arithmetic and the harmonic mean, denoted AG-mean for short:

$$\mathrm{mean}_{\mathrm{AG}}(\{x_i\}_{i\in I}) = \mathrm{symln}^{-1}\left(\frac{1}{|I|}\sum_{i\in I} \mathrm{symln}(x_i)\right). \tag{5}$$

Here the nonlinear transformation $\mathrm{symln}$ is defined for positive values based on the natural logarithm $\ln$ as:

$$\mathrm{symln}(x) = \begin{cases} x & \text{if} \quad x \leq 1, \\ 1 + \ln(x) & \text{if} \quad x > 1. \end{cases} \tag{6}$$

Thus, the $\mathrm{symln}$ transformation leaves values less than 1 unchanged, while values greater than 1 are adjusted on an "order-of-magnitude" basis. When combined with the mean calculation, this approach effectively applies the arithmetic mean to values less than 1 and the geometric mean (i.e., the average on a logarithmic scale) to values greater than 1. Importantly, this combination avoids penalizing groups that have submitted outlier data in only a few categories too severely, ensuring a more balanced comparison across all submissions.

Note that the SPE11 distance has the interpretation that if two submissions $i$ and $j$ are median-distant from each other across all individual metrics, then $\mathcal{D}^{i,j} = 1$. A table of all relative distances for the submissions to SPE11B is presented in Figure 23. This table also shows that the mean distance between all submissions is 1.31 $\pm$0.07. (Here, and later in the text, the sampling error is always given relative to a 95% confidence interval.) On the main diagonal, we have indicated the mean distance of each submission to all other submissions. We note that IFPEN1 has the smallest mean distance of 0.92, making it the "median submission." This answers question Q1 from the introduction of this section and justifies the inclusion of IFPEN1 as the most representative simulation in Figures 6 to 9.

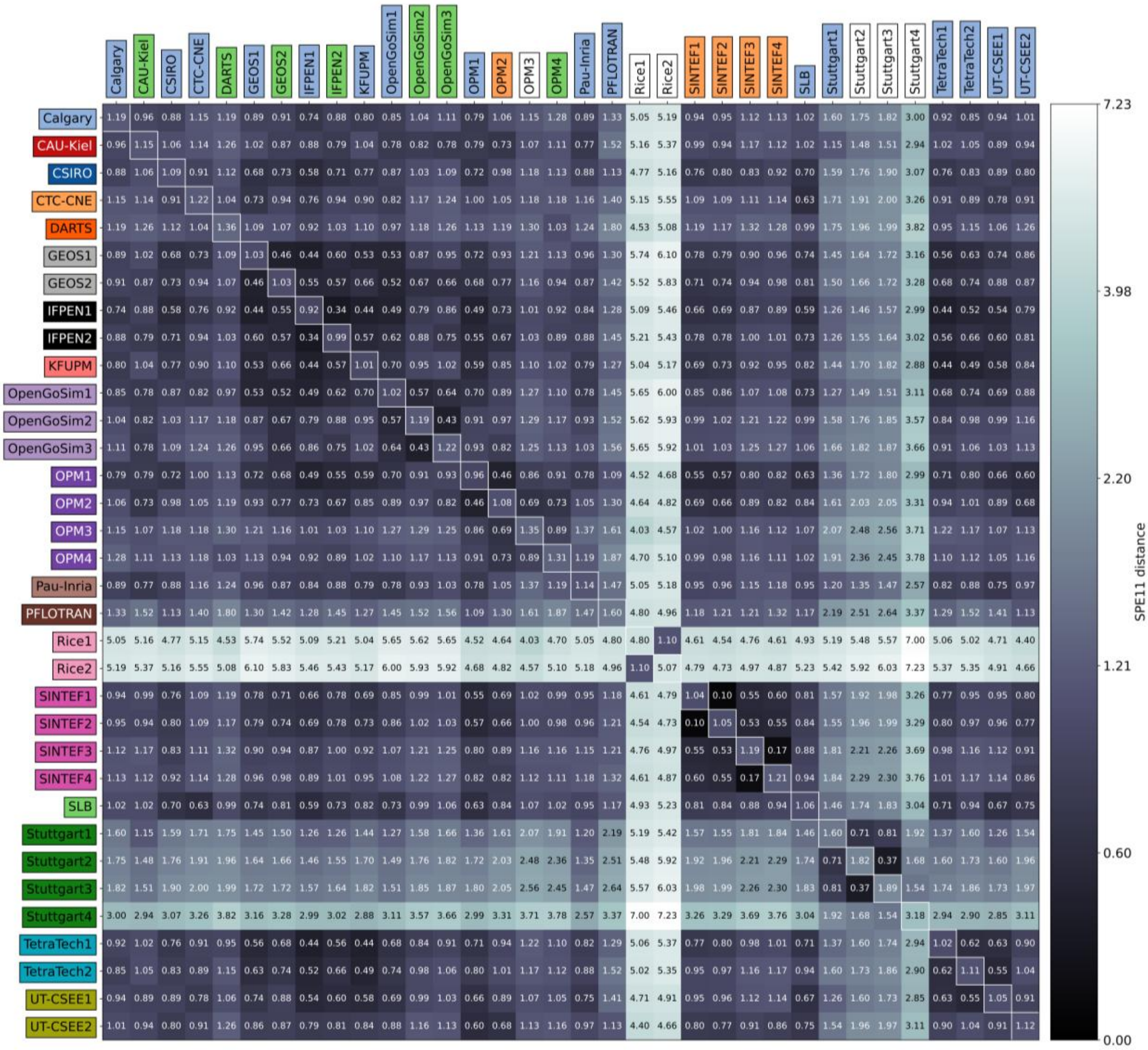

**Figure 23**: Pairwise distances for all SPE11B submissions, measured in the SPE11 distance. On the diagonal, the mean distance of the respective group towards all submissions is provided. Colors on top conform to category colors in Figure 4, colors on left conform to group colors in Figure 10.

The relative distances between submissions are used to cluster them hierarchically. This provides a simplified visualization of the submission structure, akin to the methodological classification presented in Figure 4. The result is a so-called dendrogram, shown in Figure 24, which is widely used in evolutionary biology. For this clustering, we apply average linkage clustering, where the distance between two sub-clusters is defined as the average distance of all pairwise comparisons of all submissions from the two sub-clusters (Sokal et al., 1958). Using agglomerative clustering, sub-clusters are merged sequentially, starting with individual submissions and continuing until all form a single cluster.

Comparing Figure 4 with Figure 24, we observe a high degree of correlation between the two diagrams, with the most deviating design choices (represented by Rice1 and Rice2) also being outliers in terms of the SPE11 distance. Another notable observation is that submissions from the same group tend to be relatively close in the SPE11 distance, even if they contain qualitatively different design choices. Finally, it is also striking that no two submissions conducted by different groups can be considered very close in the SPE11 distance, with the smallest cross-group distance being $\mathcal{D} = 0.44$ between IFPEN1 and TetraTech1, GEOS1, and KFUPM.

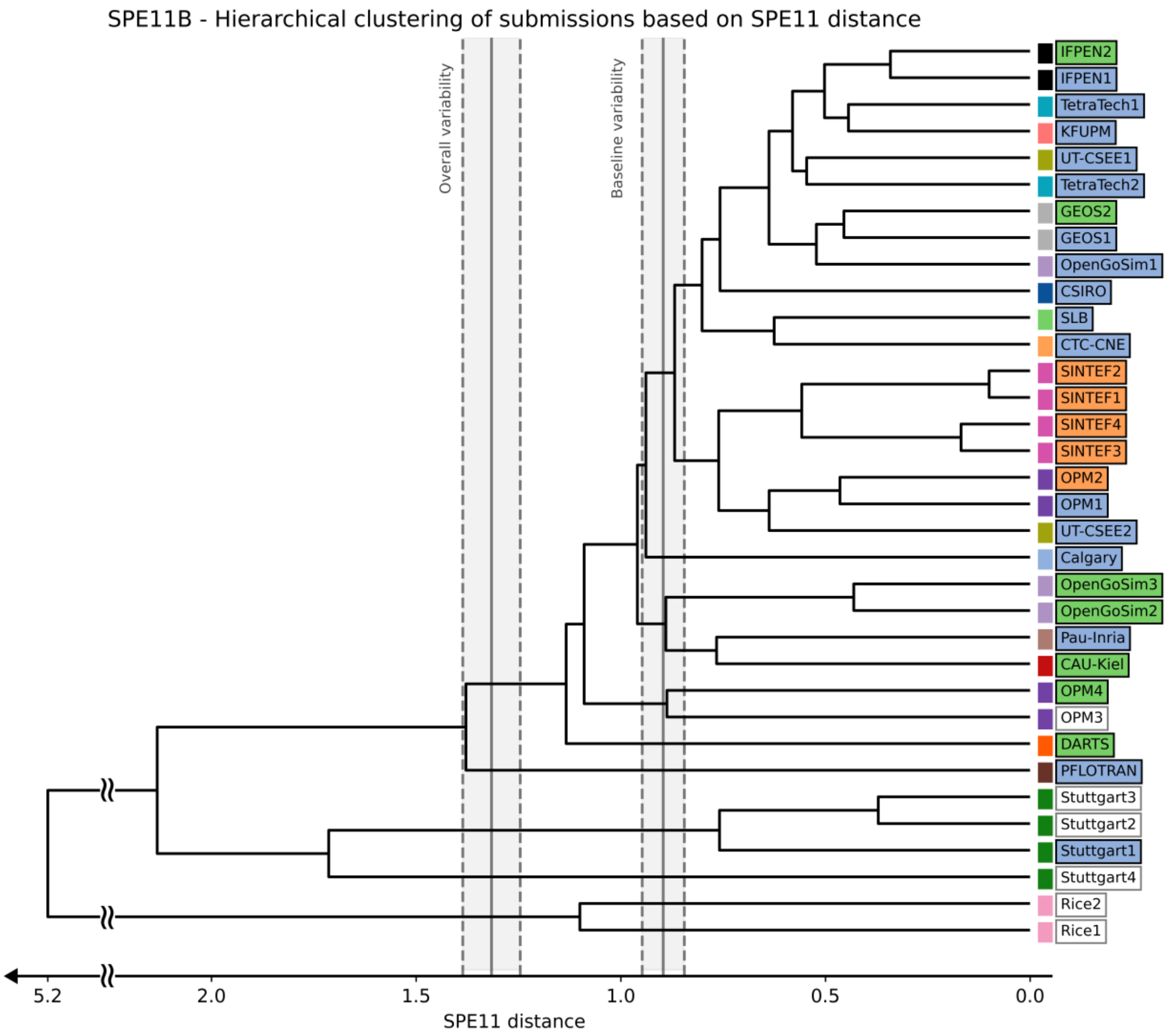

**Figure 24**: A dendrogram illustrating the hierarchical clustering of the SPE11B submissions based on the SPE11 distance. The submissions are colored based on the respective team (small box) and their category (label background) as defined in the taxonomy in Figure 4. Based on average linkage clustering, the distance value of each connection is equal to the average distance between all pairs of submissions from the two associated sub-clusters. The branches are sorted with respect to the count of associated leaves highlighting a median character.

We acknowledge that some of the choices made in defining the SPE11 distance are somewhat arbitrary, such as the selection of submitted metrics and the use of the AG-mean values. To better understand the impact of these choices, we have included information on how the individual distances correlate to the full SPE11 distance in the last column of Table 4. The Pearson correlation coefficient (PCC), ranging from -1 to 1, measures the linear relationship between each sub-metric and the global metric. A value of 0 indicates no correlation, while 1 indicates perfect correlation. We observe that most quantities in Table 4 are well represented by the overall SPE11 distance, with 60% having a PCC value above 0.75. The notable quantity that does not correlate with the overall SPE11 distance is the amount of $CO_2$ in the immobile gas phase in Box B (quantity #8).

## Variability between baseline simulations

Next, we aim to identify how much of the observed variability among submissions stems from technical factors and how much is influenced by human aspects, such as unintentional ambiguities in the SPE11 description, or simplifications, misinterpretations, and errors by the participants. This analysis addresses questions Q2 and Q3.

In this analysis, we focus on the submissions marked with a blue background in Figure 4. As a working hypothesis, we postulate that all these submissions are qualitatively "identical," meaning that the choices of grids and numerical discretization are consistent across simulations and that all participants use well-established, extensively verified software. We thus treat these simulations as the "*baseline simulations*." The variability observed within this group can arguably be attributed to a combination of numerical factors (such as tolerances on iterative methods, timesteps, etc.) and human aspects. Theoretically, the numerical errors within this baseline group should be controllable through user-defined tolerances, making the variability observed a proxy for the impact of human factors.

The mean SPE11 distance within the baseline simulations is 0.89±0.05, with a 95% confidence interval. We refer to this quantity as the "*baseline variability*." Although comparable, this value is lower than the variability for the whole dataset reported in the previous subsection, with a p-value less than .001. (Note that the use of p-values only addresses the uncertainty associated with the sampling error in the data, and is not representative of the full uncertainty introduced by the analysis choices made herein). Therefore, the initial answer to Q2 is that the baseline simulations are indeed more similar to each other than the full dataset.

As a follow-up question to Q2, we now consider the use of commercial software as a design choice. Within the subset of the baseline submissions using commercial software, the variability is 0.77±0.08, which is lower than the baseline variability (p-value equal 0.01).

To answer Q3, we contextualize the baseline variability considering the distance between the submissions from TetraTech and UT-CSEE. Both groups provided baseline submissions with two different simulators. While the responsibility for the submissions was to some extent delegated to "sub-teams" within the teams, the variability between these submissions should to a greater extent reflect differences between simulators rather than human factors. For TetraTech, the distance between STOMP and tNavigator submissions is $\mathcal{D} = 0.62$, while for UT-CSEE, the distance between SLB IX and CMG GEM is $\mathcal{D} = 0.91$. While the average of these values is smaller than the baseline variability, the sample size is too small to conclude that the difference due to different simulators used by the same simulation team is smaller than the baseline variability.

The baseline variability is also contextualized by examining the distance between different groups using the same simulator. For DuMuX, the distance is $\mathcal{D} = 1.20$ between the Pau-Inria and the Stuttgart1 submissions (this in part due to an incorrect top boundary condition in the Stuttgart1 submission). For groups using SLB IX, the distances are $\mathcal{D} = 0.62$ (CTC-CNE to SLB), $\mathcal{D} = 0.78$ (CTC-CNE to UT-CSEE1), and $\mathcal{D} = 0.67$ (SLB to UT-CSEE1). The sample size is again small, and the three last values are not independent, and it is not possible to conclude that the variability between the different groups using the same simulator is smaller than the baseline variability.

Overall, the analysis of the baseline submissions confirms that their variability is of similar magnitude, but distinctly smaller, than the mean variability observed across all submissions. This strongly supports the expectation that the choices related to numerical methods and simulation setup influence results. Nevertheless, when comparing the variability within groups using different simulators to that between groups using the same simulator, we find that while simulation setup and numerical methods are important, the specific choice of numerical simulator cannot be ascribed to a leading-order effect. To put it explicitly: who conducts the study appears to matter as much as what software they use.

## Impact of grid refinement and adaptivity

The impact of grid quality and refinement is a key consideration in both engineering practice and the verification of simulation software. The SPE11 dataset, and in particular the simulations marked in green and orange in Figure 4, provides insight into how grid type and resolution influence results relative to the baseline variability defined in the previous subsection.

To investigate this, we reformulate question Q4 as two more specific subquestions: Q4a - does grid refinement significantly change the overall submitted results? Q4b - does grid refinement reduce the variability between submitted results?

Question Q4a is addressed in Figure 25, which plots the distance between submissions from the same group as a function of the mesh refinement factor (defined as the square root of the ratio of the number of grid cells for this 2D case). We observe that small to moderate grid refinement (up to a refinement factor of 4, which corresponds to a factor 2 in each dimension) results in a mean SPE11 distance change of 0.50 ± 0.06. This change is significantly smaller than the baseline variability, suggesting that while moderate grid refinement influences the results, its effect is considerably less than the variability observed among the baseline simulations (p-value less than 0.001).

Only one submission involves an order-of-magnitude grid refinement, provided by the OPM group. The distance between OPM1 and OPM4 is $\mathcal{D} = 0.91$. This difference is comparable to the baseline variability. Indeed, the highly refined OPM4 submission is relatively far away from the baseline simulations, indicating that the chosen reporting grid may not be sufficiently fine to obtain a “grid-converged” solution to SPE11B.

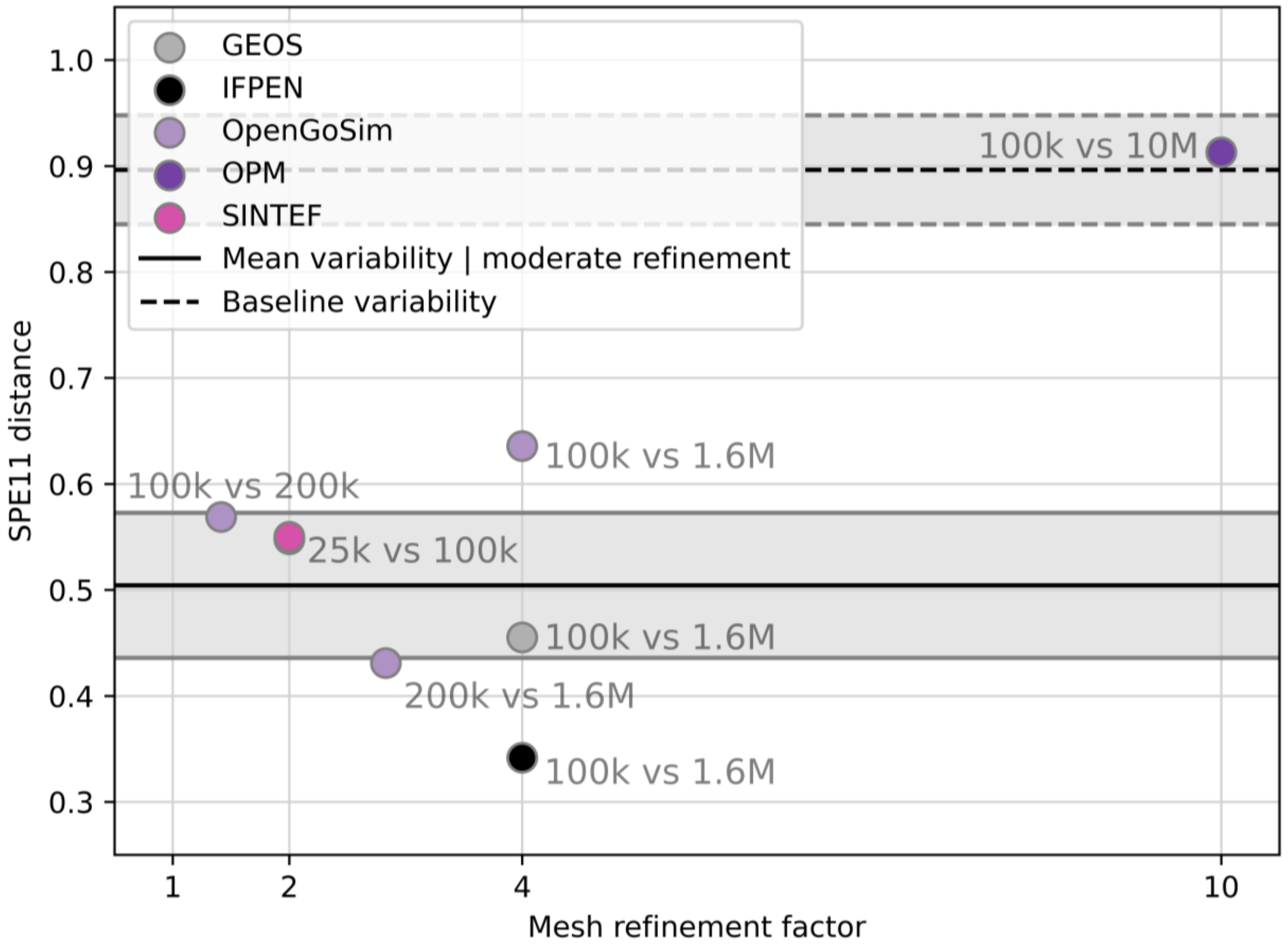

**Figure 25**: SPE11 distance between two submissions of the same team with refined mesh. The mesh refinement factor (per axis) is computed by the square root of the ratio of number of cells. The mean variability with its margin of error and respectively the baseline variability are illustrated by the gray regions.

Question Q4b is addressed by analyzing the variability among the three groups using relatively similar grids of 1.6M cells, namely GEOS2, IFPEN2, and OpenGoSim3. The mean variability between these moderately refined submissions is 0.66 ± 0.08, which is statistically smaller than the baseline variability (p value 0.005). On the other hand, the mean variability between the baseline simulations by the same groups, namely GEOS1, IFPEN1 and OpenGoSim1 is 0.48 ± 0.04. Therefore, while the 1.6M refined simulations are closer together than the baseline variability, they have actually diverged relative to the baseline simulations by the same groups. Thus, we do not have support for stating that grid refinement reduces variability between submitted results.

Only five submissions used non-orthogonal grids: all four SINTEF submissions and the OPM2 submission. The dendrogram in Figure 24 shows that these five submissions are not only relatively close to each other, but also very close to the baseline OPM1 submission. In particular, the distance $\mathcal{D} = 0.46$ between OPM1 and OPM2 indicates that, for the OPM team, the choice of grid had a relatively minor impact on the submitted results, comparable to a moderate grid refinement. Similarly, 10 of the 16 baseline submissions, including the median simulation (IFPEN1), are less than a mean distance of 0.8 from the four SINTEF submissions. Thus, it would not be reasonable to consider the SINTEF submissions as outliers relative to the baseline. Based on this data, there is no evidence to suggest that facies-adapted grids substantially impact the SPE11B submissions as evaluated by the SPE11 distance. We encourage interested readers to consult Holme et al. (2025) for a more in-depth discussion of the impact across a wider variety of non-orthogonal grid types.

## SPE11A and SPE11C

The analysis conducted for SPE11B above has also been performed for SPE11A and SPE11C. The full results are available as part of the online data resources. Here, we present a summary of the most relevant results, with complete figures of pairwise SPE11 distances, as well as the correlation of the various submetrics to the full SPE distance, included in Appendix E as Figures E.1, E.2 and table E.1, respectively.

**Summary of quantitative comparative assessment of SPE11C**

The OpenGoSim1 submission has the lowest mean SPE11 distance to the other submissions and is thus identified as the "median submission" for SPE11C (ref. Q1). Figure 26 presents the dendrogram of all submissions, constructed using the approach outlined above.

The initial impression from the dendrogram is that the submissions to SPE11C are significantly more uniformly distributed compared to SPE11B, with fewer extreme outliers. Notably, the main outliers are the high-resolution (100M cells) OPM4 simulation and the simulation with upscaled convective mixing (OPM3). Moreover, as in the case of SPE11B, the dendrogram suggests that the central part of the hierarchical clustering is dominated by baseline simulations.

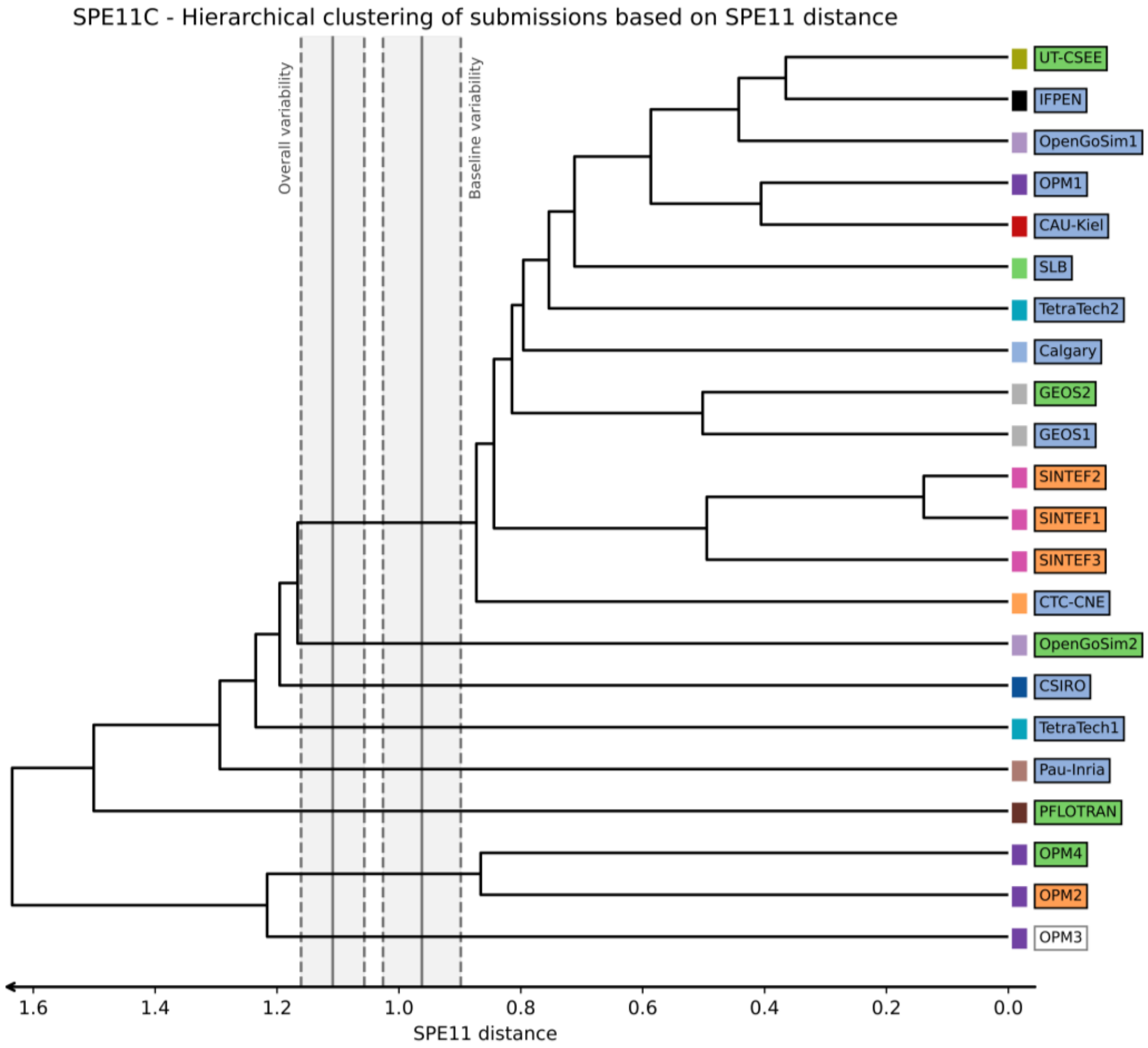

**Figure 26**: Dendrogram for SPE11C based on the pairwise SPE11 distances.

Quantifying the variability of the submissions, the overall variability is 1.10 ±0.05 and the baseline variability is 0.96 ±0.06. While the baseline variability is smaller than the overall variability (Q2, p-value less than 0.001), supporting the observation from SPE11B. Similarly, the variability within the commercial sub-group is 0.87 ±0.08, which provides weak support for the notion that the commercial variability is smaller than the overall baseline variability. Interestingly, the distance between the two groups using OPM Flow in the baseline group (OPM1 and CAU-Kiel) is only $\mathcal{D} = 0.40$, closely following IFPEN and UT-CSEE being the closest submissions in the SPE11 distance from distinct groups with $\mathcal{D} = 0.36$, while the difference between the two groups using SLB IX (SLB and CTC-CNE) is $\mathcal{D} = 0.95$. In contrast, the difference between the two submissions by TetraTech using tNavigator and STOMP is $\mathcal{D} = 1.05$. Although the sample size remains too small to draw definite conclusions, the SPE11C results suggest that simulator choice plays a more important role than in SPE11B (Q3).

Regarding grid refinement, both OpenGoSim and GEOS provide simulations on the baseline grid as well as on a refined grid with 15M cells. The distance between the OpenGoSim simulations is $\mathcal{D} = 0.83$, while the distance between the GEOS simulations is $\mathcal{D} = 0.50$. This supports the observation made above that moderate grid refinement matters less than the baseline variability (Q4a). Similarly, the difference between the refined simulations (OpenGoSim2 and GEOS2) is $\mathcal{D} = 1.29$, indicating that the refined simulations are not closer to each other than the baseline variability (Q4b), and again further away from each other than the baseline submissions by the same groups ($\mathcal{D} = 0.76$ between OpenGoSim1 and GEOS1)

Five submissions were based on unstructured grids, represented by OPM2, OPM3 and the SINTEF submissions. Interestingly, the SINTEF submissions are all closely grouped together and linear the center of the hierarchical clustering shown in the dendrogram. On average, the SINTEF submissions have a mean distance of 0.98 from the full set of submissions. Within the OPM submissions, OPM2 is most closely related to OPM1, with a distance of $\mathcal{D} = 0.48$. Overall, these findings suggest that while grid types may matter for SPE11C and possibly have larger effect than moderate grid refinement, grid type does not have a dominating impact on the SPE11 distance (Q5).

**Summary of quantitative comparative assessment of SPE11A**

The TetraTech submission has the lowest mean SPE11 distance to the remaining submissions and is therefore identified as the "median submission" for SPE11A, in the sense of Q1. Figure 27 presents the dendrogram of the full set of submissions.

Once again, the initial impression is that SPE11A contains fewer extreme outliers than SPE11B and that the submissions are more uniformly distributed. The exceptions are OPM1 and OPM2, which are not only very close in the SPE11 distance but also qualitatively very similar (see Figure 19). It is interesting that the TetraTech submission is also very similar to this pair of OPM submissions. The OPM1 and TetraTech submissions also identify the two closest submissions, measured in the SPE11 distance, originating from distinct groups, with a distance being only $\mathcal{D} = 0.17$. The closeness TetraTech and OPM1 thus confirms that it is possible for two different groups, using disparate simulators (STOPM and OPM Flow, respectively) to provide very similar results. As also observed in SPE11B and SPE11C, simulations with grid refinement tend to be outliers within the dataset, again reinforcing the impression that the baseline grid is not sufficiently refined for a grid-converged solution.

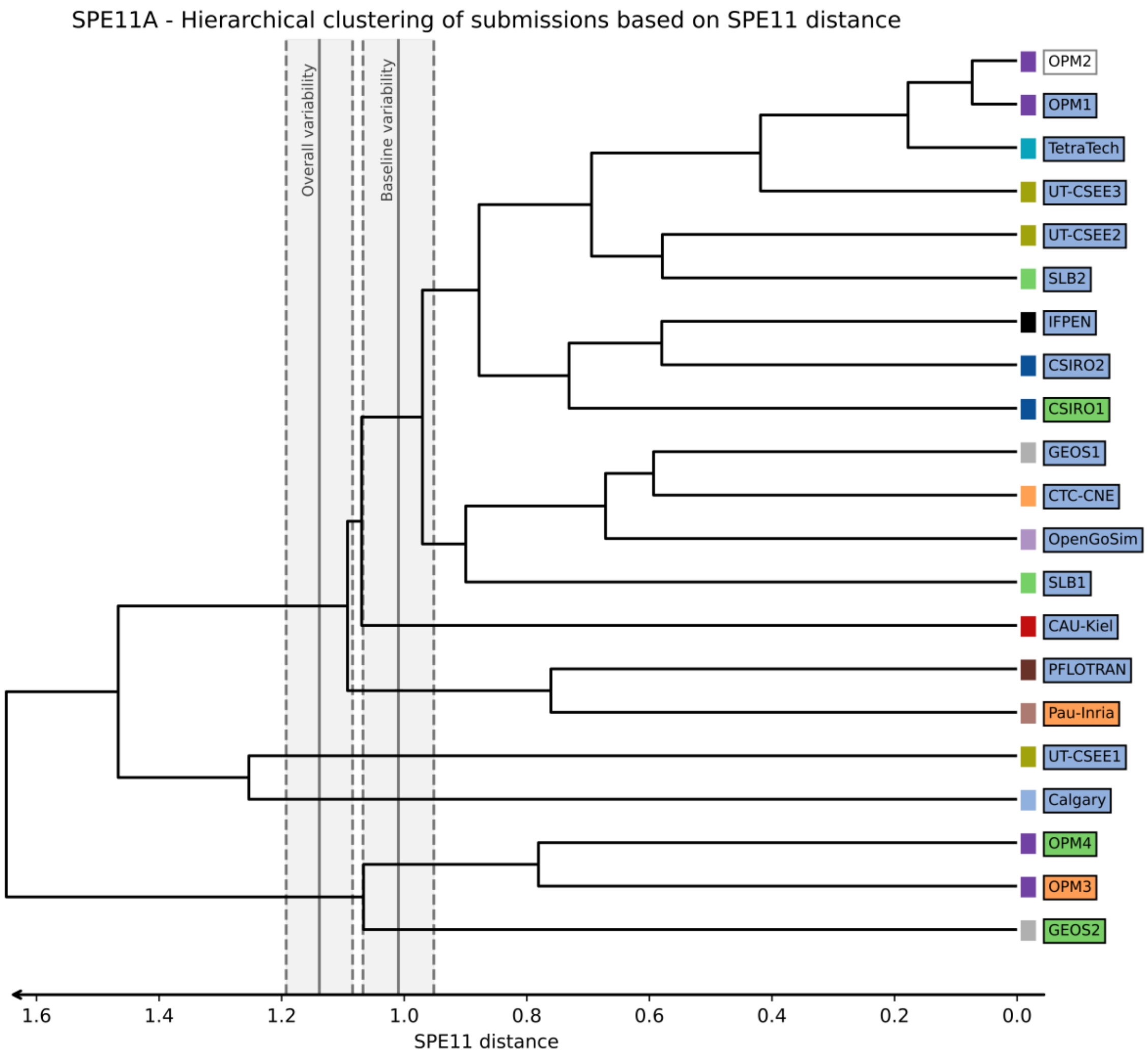

**Figure 27**: Dendrogram for SPE11A based on the SPE11 distances.

The mean variability across all submissions is 1.13 ±0.05, while the baseline variability is 1.01 ±0.05. This confirms that, as in previous cases, the baseline variability is smaller than the overall variability of the submissions (Q2, p-value less than 0.001), whereas the variability within the subgroup using commercial simulators is 1.03±0.17 and thus higher than the baseline variability.

Within the baseline simulations, three groups used SLB IX and two groups used OPM Flow. The pairwise differences between the SLB IX submissions are $\mathcal{D} = 0.84$ (SLB2 to CTC-CNE), $\mathcal{D} = 1.54$ (CTC-CNE to UT-CSEE1) and $\mathcal{D} = 1.35$ (UT-CSEE1 to SLB2). The distance between the two OPM Flow submissions is $\mathcal{D} = 0.82$ (CAU-Kiel to OPM1). As previously noted, these values are not sufficient to draw any strong conclusions, but they do not suggest that groups using the same simulator are closer together than the baseline variability (Q3). For SPE11A, only SLB used two different simulators within the baseline submissions. The difference between the Eclipse Compositional submission (SLB1) and the SLB IX (SLB2) is $\mathcal{D} = 0.87$, giving a data point in support of the statement that the variability when changing simulators within the same team is smaller than both the mean variability and the baseline variability.

Finally, two groups conducted relatively highly refined simulations: OPM4 used 3M cells and GEOS2 used 2M cells. These grid refinement results in a distance of $\mathcal{D} = 1.57$ for OPM (OPM1 to OPM4) and $\mathcal{D} = 1.25$ for GEOS. These distances are relatively large compared to SPE11B and SPE11C and suggest a greater impact of grid refinement in SPE11A relative to the two other cases (Q4a). The distance between the two refined simulations is $\mathcal{D} = 1.10$ (OPM4 vs GEOS2), and the refined simulations are thus not notably closer to each other than the general variability within the submissions (Q4b).

There is not much data supporting a discussion on the effect of unstructured grids, with only two relevant submissions from the OPM group. The distance between their simulations on Cartesian and corner-point grids is $\mathcal{D} = 1.08$ (OPM1 vs. OPM3). For comparison, the submission of Pau-Inria based on a tensor-product grid with local refinement has an average distance to the remaining submissions of $1.23$. Overall, these values suggest that for SPE11A, grid type may have a greater impact than for SPE11B and SPE11C (Q5).

# Reflections and conclusions

The 11$^{th}$ SPE Comparative Solution Project was made possible through the substantial commitment of human and computational resources from the participating groups. In this section, we offer some reflections on the entire SPE11 project, along with conclusions drawn from the results and analysis presented herein.

## Reflections

The organizers invested substantial effort in developing both a process and a description of the three subcases and the requested deliverables from participants, that were as mature and clear as possible. As a result, the final submissions could be meaningfully compared, which in turn enabled the organizers to enhance the overall impact of the effort of the participants by establishing a publicly available and thoroughly analyzed baseline for simulation technology in modelling $CO_2$ storage.

Nevertheless, learning is continuous, and through the analysis of the submissions, the organizers became aware of issues that could have been better addressed in the case definitions of the comparative solution project. We briefly summarize these below:

1. The temperature of the injected $CO_2$ was specified, but it might have been more robust to specify enthalpy. However, this in turn may have led to new problems for some of the simulators.
2. The organizers attempted to define a relatively simple set of thermodynamic relationships so that these could be represented consistently across submissions. Many participants, however, were limited to selecting among the thermodynamic relationships available within their own codes, making a unified approach to phase behavior difficult to achieve.
3. The SPE11 description called for the simulation to be initialized with pure water. While mathematically well defined, this initial condition does not exist in nature and introduced unnecessary numerical difficulties for some participants. A small initial $CO_2$ concentration in the water phase might have been a better initial condition.

To lower the barriers for participation, the organizers encouraged the sharing of scripts and input files. This was objectively successful, as evidenced by the fact that SPE11 attracted a larger number of participants and submissions than any previous comparative solution project overseen by SPE (Islam & Sepehrnoori, 2013). Nevertheless, more than half of the groups that signed participation agreements ultimately did not submit valid results. While there may be individual reasons behind each group's decisions to drop out, further efforts could have been made to help more of these groups submit valid results.

All data from the SPE11 study, together with all analysis scripts, have been made openly available and easily accessible through the development of a set of online resources. This was done in part to ensure scientific rigor and reproducibility of the analysis and conclusions presented herein. More importantly, however, this was done to facilitate further use of the data by future researchers. We envision two primary avenues for these online resources. First, by allowing the SPE11 submissions to form a reference dataset against which future developments in reservoir simulation can be benchmarked. Second, to allow more in-depth analysis of the SPE11 submission than what has been possible within the scope of this review paper.

## Conclusions

The SPE11 study contains clear learning outcomes, as presented in the preceding sections. We summarize and discuss some of the most important ones here.

Measured by the participation rate, SPE11B clearly represents the easiest subcase, being in 2D and at field conditions. Importantly, this implies that when applying reservoir simulators to analyze laboratory experiments, one should be aware that the performance of many simulators is optimized for field conditions.

Qualitative analysis of the submissions indicates that differences in reporting quantities of a factor of 2 to 3 are common for several metrics, even within the subgroup of submissions based on simulations on the reporting grid. Particularly notable are differences associated with the thermal equation, dissolution-driven convective mixing, and material heterogeneities.

A quantitative analysis, based on the "SPE11 distance," provides insight into the root factors for disparities among submitted results. A clear dependence on even modest grid refinement is observed for all three subcases, suggesting that the majority, if not all, submissions are based on computational grids that are not sufficiently refined to obtain convergent results within the SPE11 distance. Moreover, a striking observation is that the choice of simulator, grid type, and a two-times grid refinement in each direction generally matter less (or at least not more) than the variability among participating groups. This holds true despite all efforts to reduce errors, and the fact that a full 80% of the submitted data were updated or corrected after the submission deadline. Thus, even within a project with substantial quality control measures, the largest variability in the results appears to be ascribable to the humans participating in the study, rather than any numerical or technical aspects.

# Acknowledgements (people)

The authors thank: Chad Timken at SPE for technical support through the SPE11 process; Iain George Johnston and Jan Bulla (U. of Bergen), Jesus Carrera (CSIC-IDAEA) and Hoshin Gupta (U. of Arizona, Tucson) for discussions on the comparative assessments of the solutions; Enrico Facca (U. of Bergen), for discussions on efficient implementation of Wasserstein distance calculations; Sarah E. Gasda (NORCE Norwegian Research Center) for hosting the final intercomparison workshop, Daniel Patel (Rapid Geology AS) for developing 3D graphics for the SPE11 online presence, and Geir Terje Eigestad (Harbour Energy Norge AS) for general discussions on the scope and content of the SPE11.

**Calgary:** additional contributions by Yizheng Wei, Kun Wang, Hui Liu and Lihua Shen, all from the University of Calgary.

**CSIRO:** additional contributions by James Gunning at CSIRO Energy.

**CTC-CNE:** additional contributions by Ian Benson, Choongyong Han, Ichiro Osako, and Qiang Xu (all from Chevron Technical Center), and Matthew Flett (Chevron Australia).

**DARTS:** additional contributions by John Sass (Equinor ASA) and Aleks Novikov, Michiel Wapperom, and Ilshat Saifullin (Delft University of Technology).

**GEOS:** additional contributions by Thomas J. Byer, Nicola Castelletto, Matteo Cusini, Victor A. P. Magri, Randolph R. Settgast, and Joshua A. White (Lawrence Livermore National Laboratory, USA); Herve Gross and Francois Hamon (TotalEnergies E&P Research & Technology); Pavel Tomin (Chevron Technical Center); and Mohammad Karimi-Fard (Stanford University)

**IFPEN:** additional contributions by Anthony Michel and Isabelle Faille (IFP Energies Nouvelles).

**KFUPM:** additional contributions by Mohamed Abdalla (KFUPM) in selecting appropriate simulator keywords.

**OPM:** additional contributions by Atgeirr Rasmussen, Bård Skaflestad, Andreas Brostrøm, Kai Bao, Halvor Møll Nilsen, Olav Møyner, Elyes Ahmed (SINTEF Digital); Lisa J. Nebel and Markus Blatt (OPM-OP); Eduardo Barros and Negar Khoshnevis Gargar (TNO); Alf Birger Rustad (Equinor); and Trine Mykkeltvedt (NORCE).

**PFLOTRAN:** additional contributions by Katherine Muller and Xiaoliang He (Earth Systems Science Division, Pacific Northwest National Laboratory).

**SINTEF:** additional contributions by Kristian Holme, Knut-Andreas Lie, Halvor Nilsen, Odd Andersen (SINTEF Digital).

**SLB:** additional contributions by Jarle Haukås (SLB Norge) and Marat Shaykhattarov (SLB UK).

**Stuttgart:** additional contributions by Dennis Gläser (University of Stuttgart).

**TetraTech:** additional contributions by Ali Bahrami (Tetra Tech RPS Energy), Mark White (Pacific Northwest Modeling, LLC).

**UT-CSEE:** additional contributions by Mojdeh Delshad and Kamy Sepehrnoori (UT Austin), Marcos V. B. Machado (Petrobras S/A), and Lisa S. Lun (ExxonMobil Technology and Engineering Company).

# Acknowledgements (funding)

**University of Bergen:** The work was supported in part by the PoroTwin2 project, funded by Harbour Energy Norge AS.

**Calgary:** The work was partially supported by NSERC (Natural Science and Engineering Research Council)/Energi Simulation and Alberta Innovates Chairs; funding number: 365863-17.

**CAU-Kiel:** The work was supported by Kiel University by providing high-performance computing resources at the Kiel University Computing Centre.

**GEOS:** The work was funded by TotalEnergies and Chevron through the FC-Maelstrom project, a collaborative effort between Lawrence Livermore National Laboratory, Total Energies, Chevron, and Stanford University. Portions of this work were performed under the auspices of the U.S. Department of Energy by Lawrence Livermore National Laboratory under Contract DE-AC52-07NA27344.

**OPM:** The work was partially supported by Gassnova through the Climit Demo program and by Equinor ASA (622059). Sandve and Landa-Marbán acknowledge additional funding from the Centre of Sustainable Subsurface Resources (CSSR), grant no. 331841, supported by the Research Council of Norway, research partners NORCE (Norwegian Research Centre) and the University of Bergen, and user partners Equinor ASA, Harbour Energy, Sumitomo Corporation, Earth Science Analytics, GCE Ocean Technology, and SLB Scandinavia.

**Pau-Inria:** This project was provided with computer and storage resources by GENCI at CINES thanks to the grant AD010610019R3 on the GENOA partition of the Adastra supercomputer.

**PFLOTRAN:** The work was supported by Pacific Northwest National Laboratory's Laboratory-Directed Research and Development (LDRD) program, Award No. 211622. PNNL is operated for the DOE by Battelle Memorial Institute under contract DE-AC05-76RL01830. This paper describes objective technical results and analysis. Any subjective views or opinions that might be expressed in the paper do not necessarily represent the views of the U.S. Department of Energy or the United States Government.

**Rice:** The work was supported by member companies of our research consortium.

**Stuttgart:** The work was financially supported by the Collaborative Research Cluster CRC 1313 (DFG – German Research Foundation, Project Number 32754368-SFB 1313, where Wendel received three months of funding.

**TetraTech:** We thank Rock Flow Dynamics for their support and license provision.

# Appendix A: Detailed group descriptions

The group descriptions are provided by the participating groups and have been edited for clarity and consistency by the organizers. Individual group descriptions may contain statements and wording that have not been approved by all authors.

## A1.1 Calgary

**Main contributors:** Chaojie Di, Zhangxing Chen

We used our PRSI-CGCS Version 1.0 (Parallel Reservoir Simulator Infrastructure-Compositional Geological $CO_2$ Storage) simulator to generate the submitted results. This simulator is a non-isothermal parallel compositional reservoir simulator for large-scale geological $CO_2$ storage (Di et al. 2024). A standard finite-difference (volume) method is applied to discretize the multiphase flow equations, and efficient nonlinear and linear solvers are used to solve the discretized equations.

Gas solubility and properties of the water and $CO_2$-rich phases are obtained through linear interpolation of precalculated tables. The temperature and pressure intervals of these tables are set to 1°C and 1 bar for SPE11B and SPE11C and 0.01 bar for SPE11A to ensure accuracy of the linear interpolation. Pressure, saturation, and composition were solved implicitly, while temperature in the SPE11B and SPE11C cases was solved explicitly. Only thermal conduction and convection are considered in this model, since they are primary heat-transfer mechanisms within an underground formation. The effects of thermal radiation and mechanical work are minimal and thus neglected. SPE11A and SPE11B used uniform Cartesian grids while SPE11C used a corner-point grid derived from the SPE11A grid provided by the organizer. SPE11B and SPE11C consider the effects of diffusion and dispersion.

## A1.2 CAU-Kiel

**Main contributors:** Firdovsi Gasanzade, Sebastian Bauer

We used the OPM Flow open-source simulator, applying adaptive time-stepping and using parallelized computations on the high-performance Linux cluster of Kiel University. The simulator version used was 2024.04 (built time 2024-08-07 at 16:37). SPE11A was solved using only 14 logical cores, 32 cores were used for SPE11B and SPE11C. Mesh generation and parameterization were based on a structured grid, with facies distributions derived from the GitHub collaborative resource and remapped for each problem using in-house scripts (documented under github.com/fgasa/11thSPE-CSP-Kiel). Alternative to the official SPE11 description, the CoolProp thermodynamic library was used instead of the NIST database. For SPE11A, a 30-minute initialization period before injection start was added to the simulation to achieve numerical stability. SPE11B allowed for an extensive sensitivity study on, e.g., grid resolution, timestep size, and boundary effects. The main issue observed was the impact of grid size on temperature effects, since large grid cells could lead to unphysical temperature oscillations in the vicinity of Well 1. Therefore, the grid size in the K direction is half the size of the reporting grid. The computational grid for SPE11C was generated using a Gaussian function to mimic the anticline shape, keeping the crest point and dip angle according to the benchmark description. The resulting mesh is fully symmetric, as shown by Okoroafor et al. (2023). Although extensive sensitivity analysis was performed for each case, only the single set of model results seen as most representatives was submitted.

## A1.3 CSIRO

**Main contributors:** Christopher Green, Mohammad Sayyafzadeh

We used MOOSE (mooseframework.inl.gov), an open-source simulation framework with which we have developed compositional multiphase flow capability (Wilkins, Green and Ennis-King, 2021). A finite-volume discretization and fully implicit backward Euler time-stepping was used to solve the governing equations (formulated as component mass balances).

The grids were generated using the Gmsh and python scripts supplied by the organizers. Water and $CO_2$ properties, as well as mutual solubility of the two fluid components, were computed using the models specified in the description. All physical processes, boundary conditions, and wells were implemented as specified in the project description.

We submitted one case for both SPE11B and SPE11C, respectively, using a Cartesian grid with the same resolution as the reporting grid. We submitted two results for SPE11A: one using the base reporting-grid resolution (uniform 10 mm) and another with a single level of uniform refinement (uniform 5 mm). The results demonstrate a grid dependency, where an additional $CO_2$ plume is formed in Box B in the coarse simulation which is absent in the finer case. This discrepancy likely results from volume averaging in the cell-centered TPFA formulation. In coarse grids, the averaging artificially increases the pressure difference at the interface/vertices of cells. This can lead to a higher probability of exceeding the capillary entry pressure in adjacent facies, allowing the non-wetting phase to penetrate.

## A1.4 CTC-CNE

**Main contributors:** Yousef Ghomian, Nicolas Ruby

We assembled a diverse team, consisting of technical contributors from both the Chevron Technical Center (CTC) and Chevron New Energies (CNE), to address the three challenges presented by the SPE11 project. We utilized a thermal-compositional simulation modeling approach with INTERSECT™ (IX 2023.4). Petrel was employed to construct the static grid, incorporating the provided geometry, heterogeneity, thermal capacities/conductivities, and boundary conditions.

A single-component ($CO_2$) equation-of-state model was tuned to match pure gas-phase properties derived from NIST's chemistry webbook, along with the Crookston correlation for phase partitioning. Pressure and temperature-dependent K-value tables were used to determine component solubility in brine, utilizing in-house tools and data, in addition to Ezrokhi correlations for mixture density and viscosity calculations. For component enthalpy calculations, NIST data was extracted and tabulated at the given ranges of pressure and temperature.

All three cases were successfully simulated, with the team conducting additional sensitivities beyond the project's scope in results not submitted. This included convective mixing, a major focus of this study. This phenomenon was specifically amplified by both $CO_2$ dissolution and density differences due to the temperature contrast between the injected $CO_2$ and aquifer brine.

## A1.5 DARTS

**Main contributors:** George Hadjisotiriou, Denis V. Voskov

The Delft Advanced Research Terra Simulator (open-DARTS 1.1.4) is an open-source simulation framework designed for forward and inverse modeling and uncertainty quantification. Open-DARTS employs a unified thermal-compositional formulation and operator-based linearization (OBL), which makes it possible to adjust terms of the PDE and simulate various physics (Khait and Voskov 2017). OBL identifies operators in the conservation equations that depend on state variables (e.g., pressure, composition, temperature) and tabulates these over the parameter space of the problem. During simulation, operators and derivatives are evaluated for each control volume with an interpolator. This linearization approach is robust and flexible, as it allows for a control on accuracy and performance.

SPE11B was executed and achieved numerical convergence at a high resolution of 604,800 grid cells ($\Delta x$ = 5 m, $\Delta z$ = 3.3 m) using a structured mesh, two-point flux approximation, and variable time stepping. A hybrid fugacity-activity equation-of-state was used while phase properties are evaluated with the DARTS-flash thermodynamic library and validated against the NIST library (Wapperom et al., 2024). Velocity is reconstructed using a least-squares solution of fluxes across all cell interfaces (Tripuraneni et al., 2023) and subsequently explicitly incorporated into the numerical approximation of dispersion.

Computationally expensive parts of open-DARTS are implemented in C++ with OpenMP and GPU parallelization, while the framework provides a Python interface and is installed as a Python module. The linear equations were solved using an iterative GMRES solver with a CPR preconditioner. The submitted results were deployed on an NVIDIA A100 GPU, achieving a final runtime of 11 hours. For all DARTS-related code and the SPE11B Jupyter notebook, please refer to the open-DARTS repository (gitlab.com/open-darts/darts-models).

## A1.6 GEOS

**Main contributors:** Jacques Franc, Dickson Kachuma

GEOS (v1.1.0) is an open-source framework for simulating thermo-hydro-mechanical processes, with a focus on subsurface applications (Settgast et al. 2024). A particular focus of the effort is platform portability, providing the flexibility to run simulations on a variety of CPU- and GPU-based architectures in an efficient and scalable way.

A thermal-compositional flow solver employing a global variable formulation was used. This solver is based on a fully implicit finite-volume discretization with two-point flux approximation. Nonlinear systems were solved using Newton's method with line search, and adaptive time stepping was applied based on observed convergence behavior. Linear systems were solved using an iterative solver with a Multigrid Reduction (MGR) preconditioner from the *hypre* library (Bui et al. 2021).

Our setup deviates from the official description in the following ways: For SPE11B and SPE11C, the partitioning follows Duan & Sun (2003). Liquid enthalpy was determined using the correlation provided by Michaelides (1981). Gas viscosity follows Fenghour & Wakeham (1998). For SPE11C, a Brooks-Corey analytic function was used for the capillary pressure giving access to the full derivatives though discarding the renormalization prescribed in the description. Simulations were run using only diffusive fluxes and no dispersive fluxes. The injection conditions were modeled using source terms without a wellbore model.

The sets of results are differentiated by grid size: SPE11A used a regular mesh with facies 7 removed and grid-blocks size 1 cm × 1 cm for Result 1 and 1.25 mm × 1.25 mm for Result 2. SPE11B used a regular mesh with grid blocks of size 10 m × 10 m for Result 1 and 2.5 m × 2.5 m for Result 2. SPE11C used a regular mesh restricted to the simulation region with grid blocks of size 50 m × 50 m ×10 m for Result 1 and 25 m × 25 m × 5 m for Result 2.

## A1.7 IFPEN

**Main contributors:** Didier Ding, Eric Flauraud

Geoxim/CooresFlow is a research software developed by IFPEN. It is a non-isothermal compositional multiphase flow simulator that uses the variable switching formulation based on natural unknowns (pressure, temperature, phase saturation, and species molar fractions). To discretize the set of equations, standard cell-centered finite-volume method was used in space and a standard fully implicit scheme with dynamic time stepping was used for the time discretization.

Although unstructured grids can be used in Geoxim/CooresFlow, we only use Cartesian grids herein. A two-component table (in P, T) was used to describe pure densities, viscosities and K-values. Diffusion and dispersion were considered in our simulator. A custom function for $CO_2$–brine mixing density has been implemented specifically for this project. Well and boundary conditions were adapted to align with the benchmark.

We have provided two sets of results for SPE11B. The difference is the grid block size. For SPE11B, 840×120 cells with size of 10 m × 10 m are used on the uniform Cartesian grid for Result 1 and 3360×480 cells with size of 2.5 m × 2.5 m for Result 2. For SPE11A and SPE11C, the reporting grid was used.

## A1.8 KFUPM

**Main contributor:** AbdAllah A. Youssef

We employed ECLIPSE-300 2024.1 to simulate SPE11B, using a uniform block-centered structured mesh with 840×1×120 cells. Thermodynamic parameters were evaluated using the default CO2STORE model. A fully implicit discretization scheme was implemented alongside the JALS-tuned linear solver, which was crucial in maintaining mass conservation. To mitigate computation cost associated with the non-convergence of the non-linear solver, capillary pressure curves of flowing layers 2 to 5 were modified near and below the residual water saturation, $s_{wr}$, by imposing semi-log extension (Webb, 2000) to prevent huge pressure drop near $s_{wr}$, which causes reduction in time step. Additionally, the gas relative permeability of sealing facies 1 was set to zero, except at $s_g = 1$, to ensure that gas flow was restricted in this layer.

Two distinct approaches were researched to model the buffer region. First, the buffer was modeled using numerical aquifers connected to the main domain. However, this method introduced some deviations in estimating $CO_2$ mass within the buffer, as ECLIPSE limits the migration of components, allowing only $H_2O$ to move into the numerical aquifers. Our submitted result adopted a pore-volume multiplier approach for the buffer region, which enabled the transfer of all species between the buffer and the main aquifer, providing a more comprehensive solution.

Dirichlet temperature condition was applied at the top and bottom faces by connecting the model to cap and base rocks. These rocks, possessing high heat capacities, were initialized at temperatures of 40 °C and 70 °C, respectively. Both dispersion and diffusion were neglected as the “THERMAL” option in ECLISPE is not compatible with them. The sources were modeled using standard well model.

## A1.9 OpenGoSim

**Main contributors:** Pablo Salinas, Paolo Orsini

The simulator used for this study is the open-source PFLOTRAN-OGS 1.8 (opengosim.com/), which will be renamed Cirrus in future releases. PFLOTRAN-OGS is a finite-volume, fully implicit, and parallel reservoir simulator focused on $CO_2$ sequestration. The linear system of equations is solved using a constrained pressure residual method (CPR) with an algebraic multigrid (AMG) pressure solution step.

For this study, we have used a two-phase (aqueous and vapor), two-component ($CO_2$ and water) approach, both components can be present in both phases. The model was discretized using a structured grid; the time-step size is dynamically selected based on the performance of the non-linear solver.

Well models were used for the $CO_2$ injection, and for the thermal cases, we considered the Joule–Thomson effect not only in the reservoir but also within the well. Boundary conditions were applied to exchange heat but not flow.

The only difference between the different results presented is the grid resolution used.

## A1.10 OPM

**Main contributors:** Tor Harald Sandve, David Landa-Marbán, Kjetil Olsen Lye, Jakob Torben

OPM Flow (v. 2024.10, opm-project.org) is an open-source, fully implicit, black-oil simulator parallelized using MPI (Rasmussen et al. 2021). The CO2STORE option used relies on the simulator's $CO_2$–brine PVT model, which was adapted to the SPE11 description; see the simulator's technical manual (Baxendale et al. 2004) for details. Sources, boundary conditions, diffusion, and dispersion were also adapted to align with the benchmark. A Python tool, `pyopmspe11`, was developed to generate input decks, including corner-point grids, saturation-function tables, well/source locations, and injection schedules, as well as to produce reporting data in the required format. Along with the 12 configuration files (OPM1 to OPM4 submissions for the three SPE11 subcases), these resources are openly available on GitHub (github.com/OPM/pyopmspe11).

For all three subcases, the Result 1 submissions were generated with the grid size specified for the data reporting and standard simulation choices. Results 2 and 3 aim to show the impact of solver and parameter choices, grid orientation effects, and subgrid modeling for convective mixing, while Result 4 submissions aim to provide results of high resolution. Specifically:

- SPE11A: Maximum capillary pressure of 2500 Pa instead of 95000 Pa following the remarks in the benchmark description. Result 2 used the original uniform grid, Result 3 used a corner-point of approximately the same size, while Result 4 used a uniform grid with cell size of 1 mm × 1 mm.
- SPE11B: Results 2 and 3 both used a corner-point grid with cells of approximately the same size as the uniform grid used in OPM1 (10 m × 10 m). Result 3 applied a subgrid model for convective mixing for facies 2 and 5 (Mykkeltvedt et al, 2025), while Result 4 used a uniform grid with cell size 1 m × 1 m.
- SPE11C: Same differences as for SPE11B, but with grid size of approximately 50 m × 50 m × 8 m for Results 2 and 3 and 8 m × 8 m × 8 m for Result 4.

Using the maximum value of 2500 Pa (Result 2) did not significantly impact the results for SPE11A, but reduced simulation time. For SPE11B, results with the subgrid model (Result 3) for convective mixing compare reasonably well to the fine-scale simulations (Result 4). The 100-million cell case submitted for SPE11C demonstrates OPM Flow's scalability. For additional information about the different submitted cases, we refer to opm.github.io/pyopmspe11/benchmark.html.

## A1.11 Pau-Inria

**Main contributors:** Etienne Ahusborde, Michel Kern.

A compositional model from version 3.8 of the open-source platform DuMuX (Koch et al. 2021, dumux.org) was used to simulate a two-phase, two-component flow (water and $CO_2$). Phase transitions (appearance and disappearance) are handled in DuMuX by a variable-switching mechanism, using gas pressure and either liquid saturation or $CO_2$ molar/mass fraction, depending on the number and type of phases present.

All three subcases were simulated on structured orthogonal grids. For SPE11A, it was essential to employ a highly non-uniform grid. We used the cell-centered two-point flux approximation (TPFA) scheme and fully implicit temporal discretization, with grids managed through the DUNE-ALUGrid module (Alkämper et al. 2015). We only considered molecular diffusion and dispersion was not included. $CO_2$ injection was modeled using point-source terms in the relevant cells rather than wells. DuMuX uses the Spycher–Pruess EOS, in conjunction with the NIST database, for the thermodynamic data.

The nonlinear system is solved by the Newton–Raphson algorithm, with the Jacobian matrix approximated via numerical differentiation. Additionally, an adaptive time-stepping strategy is implemented, where the time step is adjusted depending on the number of iterations required by Newton to achieve convergence in the previous time step. The linear systems were solved using BiCGSTAB, preconditioned with an algebraic multigrid (AMG) solver. It was necessary to adjust the tolerance for the Newton–Raphson algorithm (1e-5 instead of 1e-8) to enable convergence for all time steps.

## A1.12 PFLOTRAN

**Main contributors:** Michael Nole, Glenn Hammond

The SCO2 mode in version 6.0 of the open-source massively parallel reactive multiphase flow simulator PFLOTRAN (pflotran.org) is specifically designed to solve conservation equations for $CO_2$ mass, water mass, salt mass, and energy, including miscibility and capillary effects. The SCO2 mode uses a finite-volume spatial discretization, is backward Euler in time with adaptive time stepping, and solves the fully implicit system of equations using Newton–Raphson.

Simulations were conducted on structured grids at resolutions roughly equal to the reporting grids for SPE11A/B. SPE11C employed higher resolution and interpolated to the reporting grid. Facies were generated using a custom set of scripts to convert to a PFLOTRAN-readable HDF5 format using Gmsh output provided by the SPE problem description. Dispersion was set to zero for all problems. A fully coupled well model was used for SPE11B; the other two cases modeled $CO_2$ injections using source terms. $CO_2$ thermodynamic properties were interpolated from a database compiled using the Span–Wagner equation of state.

## A1.13 Rice

**Main contributors:** Jakub Solovský, Abbas Firoozabadi

The simulations were performed using the research code Higher-Order Reservoir Simulations Engine (HORSE) advanced for dynamic adaptive gridding. The key features of the numerical scheme are:

- Fully unstructured grids in 2D and 3D.
- Flow equations formulated as total-volume balance and employing capillary potentials.
- Discretization of the pressure equation by the mixed-hybrid finite element method.
- Discretization of transport equations by the discontinuous Galerkin (dG) method with a slope limiter.

- Two-phase behavior description, compressibility, and density computations by the cubic-plus association (CPA) equation-of-state.

The simulator is fully compositional, with dynamic adaptive gridding used for both the mixed-hybrid finite element and discontinuous Galerkin discretizations. The criterion for dynamic refinement is based on gradient of composition of carbon dioxide (Solovský & Firoozabadi 2025a). SPE11B was modelled using an isothermal assumption; results for a non-isothermal formulation are presented in (Solovský & Firoozabadi 2025b).

Two sets of results are reported: coarse triangular grid with approximately 10,000 elements and one level of refinement with original elements divided into four. Simulations using two- and six-level refinement are presented in (Solovský, J., and Firoozabadi, A. 2025b). The six-level is for high resolution around the injection well.

**Acknowledgments:** We thank the member companies of our research consortium for supporting this work.

## A1.14 SINTEF

**Main contributor:** Olav Møyner

JutulDarcy (Møyner 2024, github.com/sintefmath/JutulDarcy.jl) is a general high-performance automatic differentiation reservoir simulator that has support for non-isothermal compositional flow and consistent spatial discretization schemes. There is significant overlap in the development of JutulDarcy and OPM Flow, which was used by two other teams. To complement the results of the OPM team, with whom we collaborated closely, we chose to focus on investigating the effect of facies-adapted meshes and errors from inconsistent discretizations.

The open-source MRST (mrst.no) software was used to mesh the facies model as a partially unstructured cut-cell type grid that accurately represents the given geometry while still retaining a Cartesian background grid in regions where the rock is uniform. Our submissions compare a consistent discretization scheme (AvgMPFA) with the standard and potentially inconsistent TPFA scheme. A fully implicit formulation with dynamic time stepping was used with both schemes. Custom functions for $CO_2$-brine mixing density were implemented. Differences are observed, even for our cut-cell meshes that should in principle minimize grid orientation effects.

Deviating from the official description, we used the NIST tables for injection enthalpy of $CO_2$ at well-cell conditions but used the tabulated component heat capacities together with mass fractions pressure and density to calculate mixture enthalpy in the reservoir. This inconsistency resulted in negative temperatures in degrees Celsius close to the well. The property tables were not extrapolated into negative temperatures. Uniform tabular grid spacing and explicit solutions for two-component Rachford–Rice were used to minimize the cost of property evaluations.

## A1.15 SLB

**Main contributor:** Marie Ann Giddins

Two general-purpose commercial reservoir simulators were used for the submissions: Eclipse Compositional (version 2024.2) for SPE11A (Result 1) and Intersect (version 2024.2) for SPE11B/C and Result 2 for SPE11A. We used inbuilt simulator options for $CO_2$ storage in aquifers, representing models that could be used in typical reservoir engineering practice, as a starting point for more detailed $CO_2$ studies.

The submissions used cell-centered TPFA discretization, on Cartesian grids corresponding to the reporting grid formats, with facies mapped from the provided mesh file. Fluid properties were calculated using methods from Spycher and Preuss (2005, 2009), with Ezrokhi's method for brine density calculations (Zaytsev and Aseyev, 1996). The diffusion model was adjusted to approximate combine diffusion and dispersion effects.

The simulators' standard well models were used for $CO_2$ injection. Thermal submissions assumed constant enthalpy for the fluid injection streams, defined at temperature 10°C and the initial pressure at Well 1. External boundaries were treated as no-flow, with heat boundary conditions represented by a semi-analytical heat loss model (Vinsome and Westerveld, 1980).

All runs were fully implicit, with CPR preconditioning and adaptive time stepping. Intersect models were run in parallel using ParMETIS-based unstructured parallel partitioning. Time steps were constrained to match the reporting times specified in the SPE description.

## A1.16 Stuttgart

**Main contributors:** Kai Wendel, Holger Class and Timo Koch

The open-source code DuMu$^X$ 3.9 (Koch et al. 2021, dumux.org) uses cell-centered finite volumes with two-point flux approximation for the spatial discretization and implicit Euler for the time discretization.

Stuttgart only submitted results for SPE11B, where a constrained-pressure-residual (CPR) and AMG-based solver was used as preconditioner and computations were done on the reporting grid. Sources, boundary conditions, and diffusion were adjusted to match the SPE11 problem description. In difference to the description, due to a setup error, the pressure at the top boundary was fixed to the initial condition instead of prescribing a no-flow boundary condition. Capillary-pressure and relative-permeability curves were approximated by piecewise-linear functions with 1000 support points, denser for small saturation values and computed by the transformation $s = (e^k - 1)(e^5 - 1)^{-1}$, where the k values are equidistant in [0, 5]. An application module to run the benchmark cases with DuMux is developed at https://git.iws.uni-stuttgart.de/dumux-appl/dumux-spe11.

Three different models were used, dispersion was not accounted for, and diffusion only in the 2p2cni model, as described. The models are sequentially coupled as explained in Darcis et al. (2011):

- 1pni: One-phase non-isothermal model with a pure water phase.
- 2pni: Two-phase non-isothermal immiscible model with pure water phase (wetting) and pure $CO_2$ phase (nonwetting).
- 2p2cni: Two-phase compositional model with symmetric switch based on the DuMu$^X$ core $CO_2$ model.

These were combined to generate four results:

- Result 1: Using a 2p2cni model for the whole simulation time.

- Result 2: The 1pni model was used for initialization, followed by the 2pni model for the injection period. After the injection, the 2p2cni model accounted for compositional effects. This sequential coupling follows the approach described in Darcis et al. (2001).
- Results 3: Same as Result 2, but with the 2pni until 25 years after the end of injection.
- Results 4: 1pni for the initialization and a 2pni model until the end of the simulation.

In all models, aqueous phase pressure [Pa] and temperature [K] are primary variables. In the 2pni model, gas phase saturation (Sg) was added to close the system. The 2p2cni model used a variable-switching approach depending on the local state.

## A1.17 TetraTech

**Main contributors:** Adam Turner, Hai Huang, David Element

The Tetra Tech / RPS group used two software packages: Rock Flow Dynamics' tNavigator simulator (v24.2) was used to model SPE11B/C and the STOMP-CO2 code developed at Pacific Northwest National Laboratory (PNNL) was used for SPE11A/B/C.

Both simulators run using a Cartesian grid with the same dimensions as for dense data reporting. For SPE11C, STOMP-CO2 used a curvilinear grid that follows the formation horizon topology in the Y-direction and remains rectilinear in the xy-plane and for tNavigator a similar corner-point grid was generated using the `pyopmspe11` library (github.com/OPM/pyopmspe11).

Both simulators adopt a fully implicit time integration approach with dynamic, adaptive time-stepping scheme, and use Newton's method for solving coupled multiphase flow and thermal/solute transport equations. Simulations were run in compositional mode. The simulator timesteps did not necessarily align with the reference temporal resolution and therefore required linear interpolation to generate the reported data at the frequency requested.

Both simulators applied Spycher–Pruess equation-of-state (EOS) to compute fluid PVT behaviors, and the relevant thermodynamic properties were taken from data published by NIST or from the problem description. For SPE11C, STOMP-CO2 ignored the thermal transport process, for the purposes of both computational efficiency and assessing the thermal effect on spatial-temporal distributions of the injected $CO_2$ at reservoir scales.

During post-processing of results, the convective mixing term, M(t), was calculated at the same reporting times as used for the dense data. For this calculation, the maximum solubility term was calculated using bilinear interpolation of a table of solubilities as a function of temperature and pressure constructed using the Spycher–Pruess equations.

For all three SPE11A/B/C, STOMP-CO2 ignored the calculations of dispersive fluxes and only considered the diffusive fluxes with diffusion constants specified in the problem descriptions. No molecular diffusion was modelled in the tNavigator simulations. STOMP-CO2 applied a pore-volume multiplier approach to "augment" the pore volumes of lateral boundary grid cells in SPE 11B/C in the way as specified by SPE11 problem descriptions. For all three problems, STOMP-CO2 simply treated injection wells as constant-rate pure $CO_2$ sources at specified injection temperature.

The tNavigator simulations for SPE11B/C also employed pore-volume multipliers to the model edge cells to account for the augmented boundary volumes. The constant temperature boundary conditions for the top and bottom were approximated by specifying high thermal conductivity and heat capacity for overburden and underburden.

## A1.18 UT-CSEE

**Main contributors:** Bruno R. B. Fernandes, Prasanna Krishnamurthy

We used two commercial reservoir simulators: SLB Intersect (version 2023.1) and CMG GEM (version 2023.30). Both simulators rely on traditional two-point methods with upstream-mobility weighting and a fully implicit formulation. We used a Cartesian grid for SPE11A/B and a corner-point grid geometry for SPE11C.

SLB-PETREL (version 2023) was used to generate Cartesian grids and properties for Intersect, as well as to define boundary conditions through boundary wells (SPE11A) or pore volume multipliers. Additionally, it was employed to generate different grid refinement levels, though only one grid result was submitted per benchmark case. A custom script was written to create the corner-point grid for SPE11C.

For SPE11B/C, Intersect used the thermal model with all the physics required by the problem description, but the aqueous phase density was computed with the Ezhorki equation. $CO_2$ enthalpy was provided in a table format, and the phase equilibria were calculated with K-value tables for various pressures and temperatures. The $CO_2$ gas phase density was computed with the Peng–Robinson equation-of-state. Three different fluid models were considered for SPE11A using Intersect. Result 1 considered the solubility of $CO_2$ in the aqueous phase using solubility tables and water vaporization was neglected. For Result 2, the molecular diffusion and physical dispersion were disabled to observe the impact of these mechanisms. Finally, Result 3 considered the black-oil model, which, even though it was quite efficient computationally, did not model molecular diffusion and physical dispersion. All runs performed with Intersect used an FGMRES solver with a Quasi-IMPES CPR (ILU+AMG) preconditioner.

Results for CMG GEM were submitted only for SPE11B. The Peng–Robinson equation-of-state was used for computing gas phase density and enthalpy, while phase equilibrium was fully based on K-values. Fluid properties and component parameters were calibrated using CMG WINPROP. Physical dispersion was not considered. The simulator utilized a GMRES solver with ILU preconditioner. Standard well models were used to represent the sources in both CMG-GEM and SLB-Intersect.

# Appendix B: Performance data for submitted results

A subset of the submissions to each of the three SPE11 cases also provided the requested performance data. The overall performance depends on many factors, many of which can only be partially assessed between different submissions. Assessable factors include grid resolution, time resolution, and, to some extent, degree of parallelization and efficient utilization of hardware (e.g., not all groups reported the number of processors). Unassessable factors, on the other hand, include convergence tolerances for linear and nonlinear iterative solvers, software architectures, implementation efficiency.

Due to these limitations, drawing strong conclusions from the performance data is challenging. However, we believe that experienced computational scientists will nevertheless find valuable insights, which justifies inclusion of performance data in this appendix as Figures B.1, B.2 and B.3, respectively.

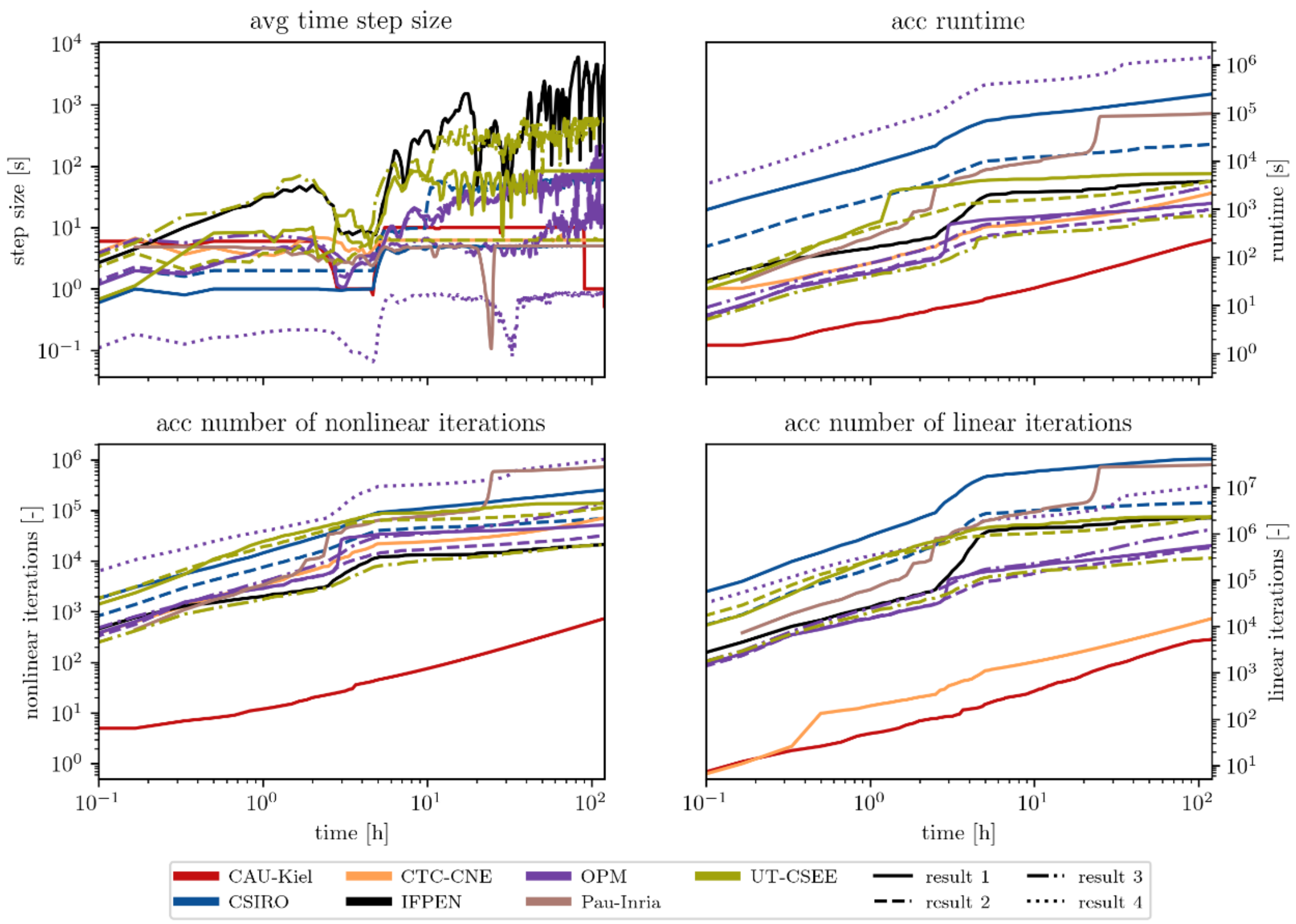

**Figure B.1**: Plots of sparse performance data for SPE11A.

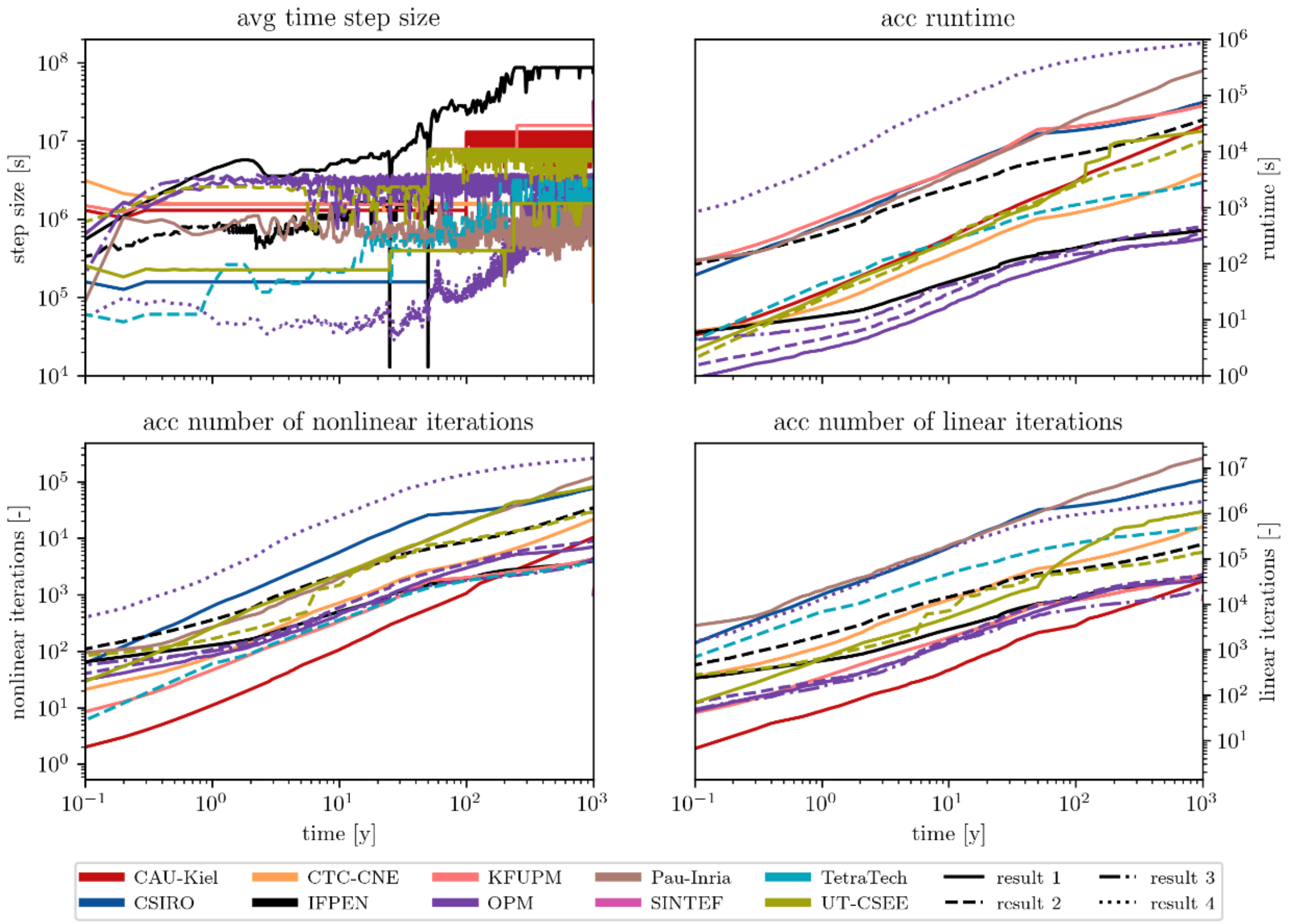

**Figure B.2**: Plots of sparse performance data for SPE11B.

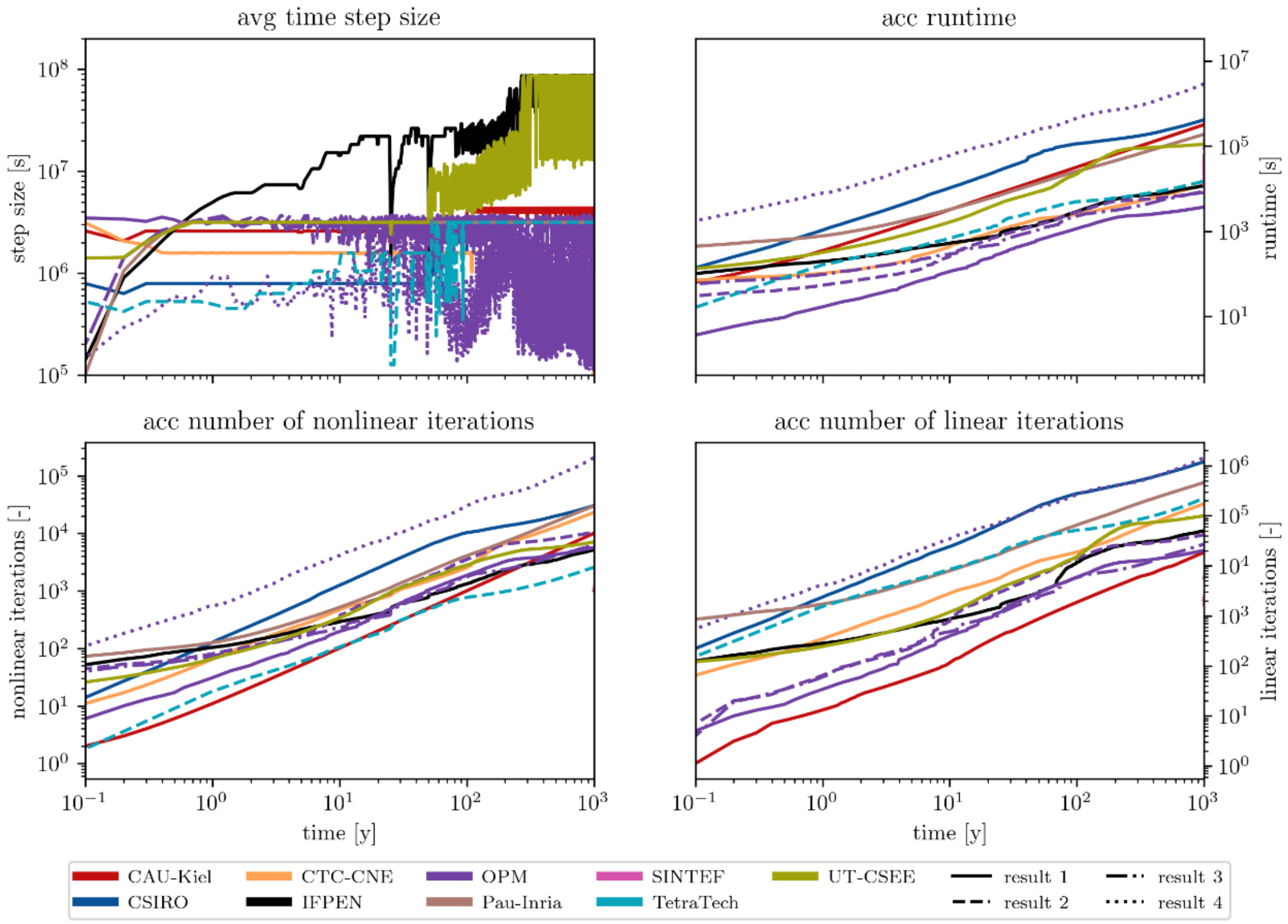

**Figure B.3**: Plots of sparse performance data for SPE11C.

# Appendix C: Editing of reported sparse data

For several results, some participants did not report the sparse data in a way where all measurables could be used directly for the visualization and analysis. Firstly, this concerns primarily the sparse data measurables #3, 4, 7 and 8, regarding the amount of mobile and immobile free-phase $CO_2$ in Boxes A and B. The definition of "immobile free-phase $CO_2$" is stated in the CSP description (Nordbotten et al., 2024) as $CO_2$ at saturations for which the nonwetting relative permeability equals zero. Nevertheless, for the final submission, three participants (Calgary, CTC-CNE and OpenGoSim) chose to also include in regions of mobile free-phase $CO_2$, the part of $CO_2$ below residual saturation. Secondly, other measurables were simply omitted in some submissions. This concerns mostly the sparse data measurable #11 on the amount of convective mixing in Box C. Seven participants chose not to report this quantity (Calgary, CAU-Kiel, CTC-CNE, PFLOTRAN, SINTEF, SLB and Stuttgart). Finally, sparse data #6 and 10, the amount of $CO_2$ in seal facies 1 in Boxes A and B, was not reported by PFLOTRAN, and only partially reported by two groups (Calgary and Pau-Inria).

In the instances enlisted above, the relevant sparse data was approximated by the organizers based on post-processing the submitted dense data, using linear interpolation in time to achieve the temporal resolution required. The post-processing of dense data is straightforward summation for sparse data measurables #3 through 10, while for measurable #11, a standard second-order central differencing stencil with respect to the reporting grid has been employed. In all instances, this necessarily disregards any potentially finer-granular information encoded in a higher grid resolution or particular spatial discretization method.

# Appendix D: Spatial norms used for SPE11

Comparing dense data on the reporting grid between two submissions at a single snapshot in time requires the selection of a meaningful metric. Different dense data metrics serve different purposes, and their choice must be adapted to the specific physical field being compared. As part of the SPE11 distance, three spatial norms are used:

- The $L^2$ norm between two submissions, $\gamma_i$ and $\gamma_j$, is given by the sum of spatially weighted squared (absolute) distances $D_\gamma^{i,j}(t) = \sqrt{\int_\Omega |\gamma_i(t) - \gamma_j(t)|^2 dx}$. This norm represents a measure of the overall magnitude of a spatial field and can thus be interpreted as an energy-like measure, which is particularly suitable for quantities such as pressures and temperatures. In practice, the integral is approximated using a summation on the reporting grid.
- The $L^2$ semi-norm between two submissions, $\gamma_i$ and $\gamma_j$, is defined by the sum of spatially weighted squared (relative) distances $D_\gamma^{i,j}(t) = \sqrt{\int_\Omega |(\gamma_i(t) - \bar{\gamma}_i(t)) - (\gamma_j(t) - \bar{\gamma}_j(t))|^2 dx}$, where $\bar{\gamma}_i$ denotes the spatial average of $\gamma_i$. This formulation of the $L^2$ semi-norm normalizes the data to zero mean value, thereby measuring relative variation rather than absolute distance. For field data like pressure and temperature, the $L^2$ semi-norm is less sensitive to structural differences in boundary values than the full $L^2$ norm, which minimizes the influence due to different implementations of boundary conditions.

- The weighted Wasserstein distance between two density fields, $\gamma_i$ and $\gamma_j$, with the same total mass, is defined by the minimization $D_\gamma^{i,j}(t) = \min\{\int_\Omega |\bar{\kappa} q| dx : \nabla \cdot q = \gamma_i(t) - \gamma_j(t) \; in \; \Omega, \; q \cdot n = 0 \; on \; \partial\Omega\}$ for a given spatially-varying parameter $\bar{\kappa}$. This distance quantifies the cumulative measure of mass-conservative transportation required to transform $\gamma_i$ into $\gamma_j$. Unlike the $L^2$ norm, which provides a merely volumetric comparison, the Wasserstein distance considers both shapes and spatial distributions. This makes it particularly suited for comparing $CO_2$ mass in the presence of density-driven convective mixing. The parameter $\bar{\kappa}$ is used to penalize the transport cost, thus encoding physical information about the geometrical structure and characteristic flow properties. In the case of the SPE11 distance, the weight used is the inverse of the absolute permeability tensor, globally rescaled by the highest permeability value. This approach associates low-resistance unitary weight with the high-permeable layer, while penalizing transportation through seals, reflecting the flow resistance of the reservoir. The minimization problem is approximated and solved using the open-source package `DarSIA` (Nordbotten et al., 2024b).

# Appendix E: SPE11 distances for SPE11A and SPE11C

The SPE11 distances for SPE11A and SPE11C are summarized and analyzed in the main part of the manuscript. Here, we provide the complete tables of pairwise SPE11 distances for these two cases in Figures E.1 and E.2, respectively. Table E.1. shows the correlation between the SPE11 distance and the individual metrics it is composed of.

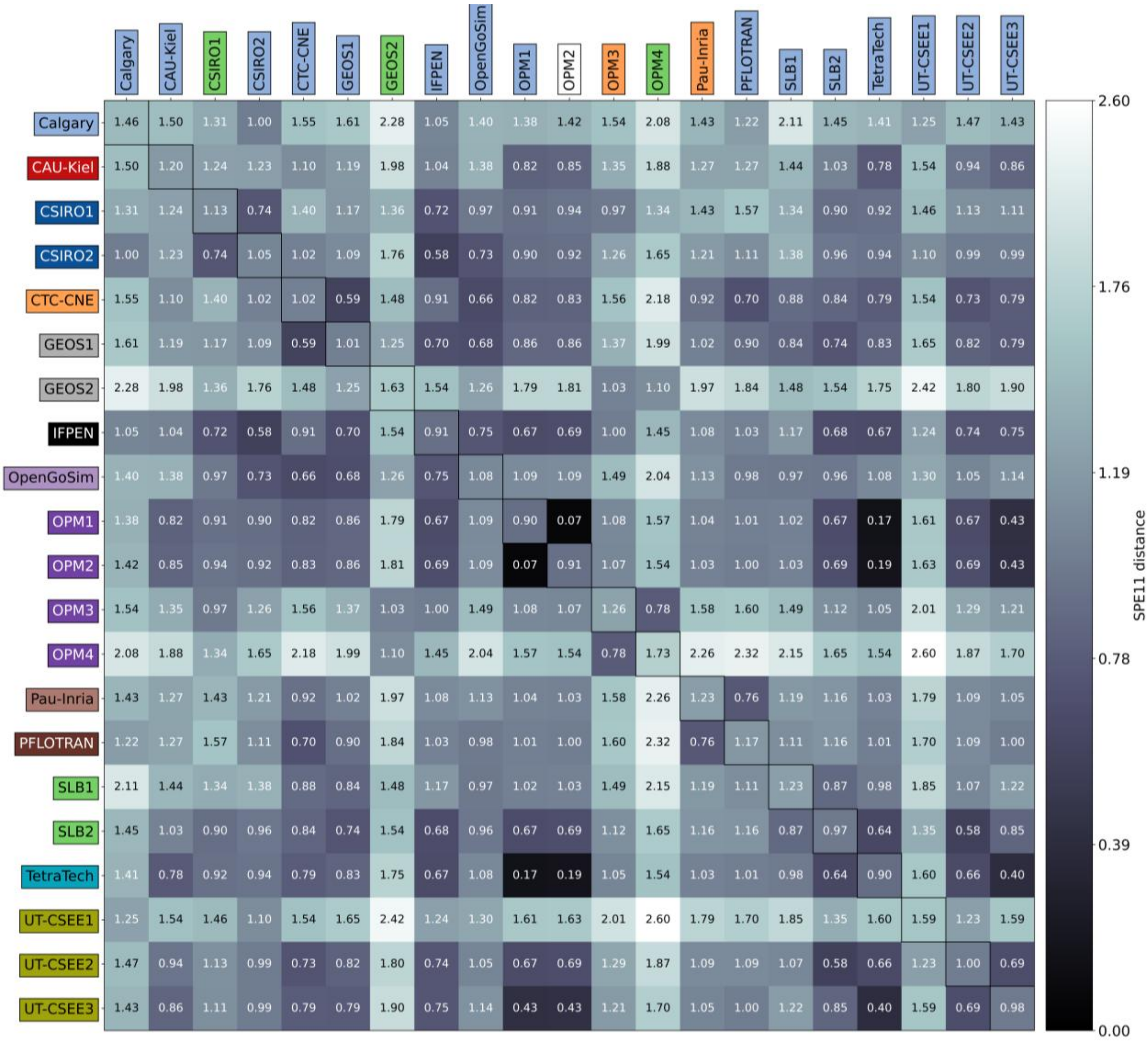

**Figure E.1**: Pairwise distances for all SPE11A submissions, measured in the SPE11 distance. On the diagonal, the mean distance of the respective group towards all submissions is provided.

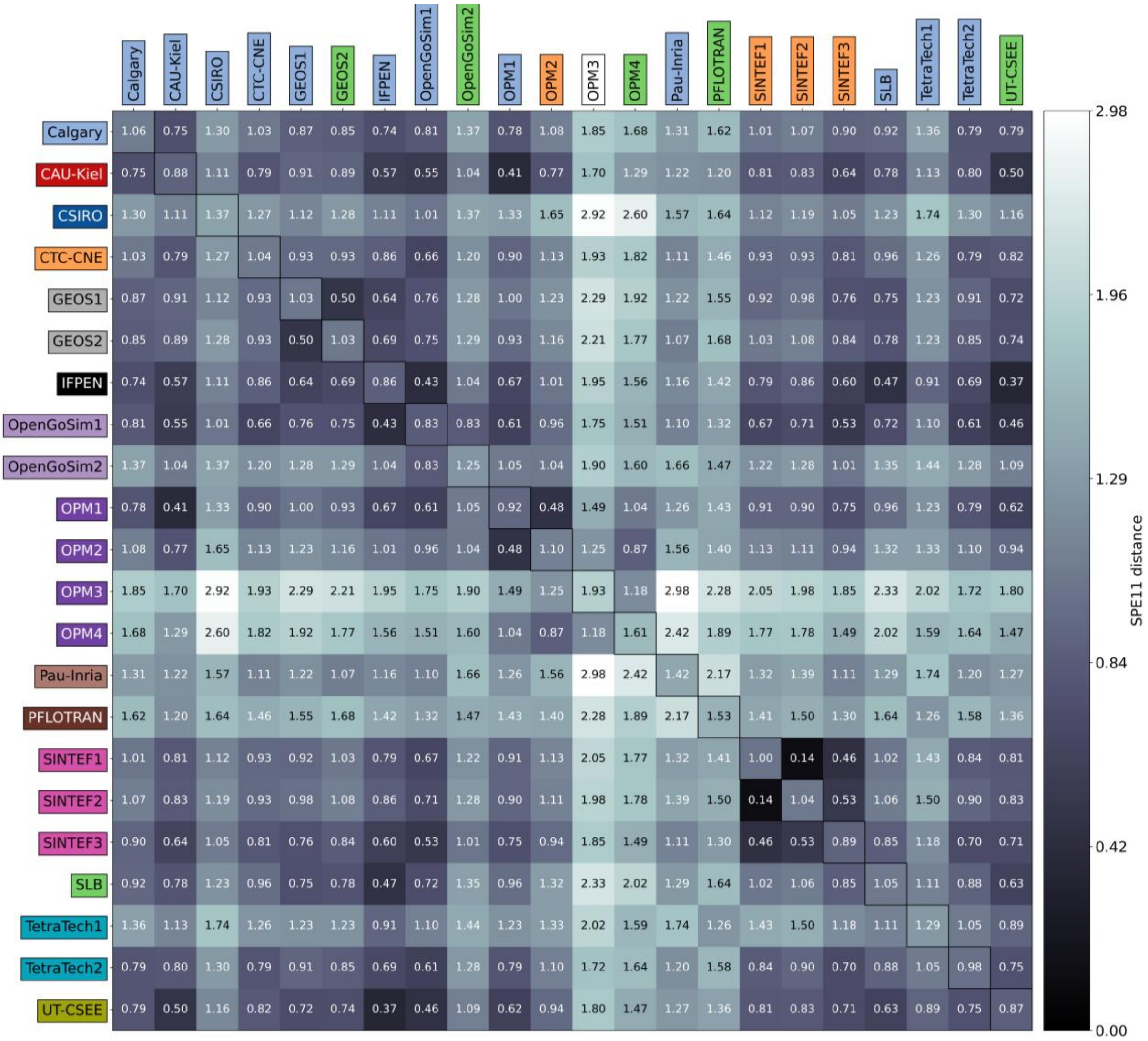

**Figure E.2**: Pairwise distances for all SPE11C submissions, measured in the SPE11 distance. On the diagonal, the mean distance of the respective group towards all submissions is provided.

| | # | $\bar{\mathcal{D}}_\gamma$ (11A) | PCC (11A) | $\bar{\mathcal{D}}_\gamma$(11C) | PCC(11C) |
|---|---|---|---|---|---|
| **Sparse data** | 3 | 7.59e-2 | 0.39 | 2.17e14 | 0.28 |
| | 4 | 3.19e-4 | 0.25 | 8.43e12 | 0.42 |
| | 5 | 7.27e-2 | 0.36 | 1.34e14 | 0.38 |
| | 7 | 5.54e-3 | 0.70 | 3.91e11 | 0.70 |
| | 8 | 8.56e-4 | 0.59 | 3.04e11 | 0.72 |
| | 9 | 4.01e-2 | 0.76 | 3.65e10 | 0.70 |
| | 11 | 1.17e2 | 0.48 | 1.24e11 | 0.42 |
| | 12 | 2.48e-2 | 0.47 | 1.85e13 | 0.28 |
| | 13 | - | - | 4.71e13 | -0.07 |
| **Dense data** | 14.e | 3.66e2 | 0.23 | 1.27e11 | 0.16 |
| | 14.l | 5.53e3 | 0.21 | 9.43e11 | 0.32 |
| | 20.e | 3.17e-3 | 0.55 | 9.37e15 | 0.36 |
| | 20.l | 1.00e-2 | 0.54 | 2.76e16 | 0.21 |
| | 21.e | - | - | 2.20e6 | 0.26 |
| | 21.l | - | - | 2.70e6 | 0.25 |

**Table E.1**: Median and PCC for SPE11A and SPE11C submissions.